\documentclass{article}%
\usepackage{amsmath}
\usepackage{amsfonts}
\usepackage{amssymb}
\usepackage{graphicx}%
\providecommand{\U}[1]{\protect\rule{.1in}{.1in}}
\newtheorem{theorem}{Theorem}

\begin{document}

\title{String theory deconstructed\\{\small Dedicated to the memory of Klaus Pohlmeyer}}
\author{Bert Schroer\\CBPF, Rua Dr.\ Xavier Sigaud 150 \\22290-180 Rio de Janeiro, Brazil\\and Institut f\"ur Theoretische Physik der FU Berlin, Germany}
\date{November 2010}
\maketitle

\begin{abstract}
This essay presents a critical evaluation of the concepts of string theory and
its impact on particle physics. The point of departure is a historical review
of four decades of string theory within the broader context of six decades of
failed attempts at an autonomous S-matrix approach to particle theory.

The central message, contained in sections 5 and 6, is that string theory is
not what its name suggests, namely a theory of of objects in spacetime whose
localization is string- instead of point-like. The result is corroborated by
the failure of the conformal embedding interpretation. Whereas the "target
space" of the chiral theory (higher dimensional vector or spinor indices of
currents) defined by the inner symmetry indices of an abelian current is
redefined as the spacetime of string theory, the one-dimensional chiral
conformal "source" theory is not embedded as a string in target space but
rather the oscillators of the would-be string go into an infinite dimensional
inner Hilbert space pictured as an inner space over a localization point in
target space. Hence string theory solves a problem which enjoyed some
popularity in the 60s namely the construction of \textit{dynamical}
\textit{infinite component fields}. This mean in particular that there are no
word-sheets which can be subjected to interactions in form of rulrd for tubes
in analogy to Feynman graphical rules.

The present work also subjects ideas to a critical test which arose in the
wake of string theory, as dimensional curling up, the quantum theoretical
aspects of branes and the Maldacena conjecture.

The essay is dedicated to the memory of Klaus Pohlmeyer since he studied the
intrinsic physical content of the Nambu-Goto Lagrangian as an autonomous
quantum system according to Faddeev's ideas concerning integrable systems.
This leads to a system which, as QFT, exists in every spacetime dimension and
is inequivalent to the above infinite component field of string theory.

\end{abstract}

\section{Preface}

Tracing ideas, which have been dominating for decades large parts of particle
theory, back to their historical origin is usually considered interesting
worthwhile only in case there is a happy end, which in particle theory means a
theory whose success is universally recognized. This is certainly the case
with renormalized quantum field theory (QFT) and in particular quantum
electrodynamics (QED) which gloriously confirmed the ideas about quantization
of relativistic matter waves at the cradle of quantum field theory. This essay
presents a critical look at a less successful but already almost 5 decades
lasting project which, since the 80s, became popular under the name of string theory.

Particle theory, being an extremely speculative science at its frontiers,
comes with a large list of failed projects which is much longer than that of
successfull ones. Most have been left on the wayside of the great particle
physics caravan, in fortunate cases after having received critical attention
and an appropriate closure. Who still remembers the frenzy about combining
inner symmetries with the relativistic spin under the name "SU(6)", the
"peratization" of quantum field theory (the precursor of "effective QFT"), or
the Lee-Wick model (complex poles finally shown to lead to acausal precursors)?

Failed projects which managed to hang on for a decade, as the S-matrix
bootstrap in its role of a unique theory of strong interactions, are more rare
and hence they stay longer in our memory, especially if some of their content
is reprocessed and together with other ingredients used as raw material of a
new theory, as it was the case with string theory.

In this area of tension between successful and failed ideas, string theory
plays a special role. The opinion on this issue is still split and if issues
in particle physics could be decided by a democratic vote, string theory would
most probably continue in a limbo. But, as the history of particle physics
amply demonstrates, popularity is not always coupled to the scientific
solidity of a theory; the vernacular "so many people cannot err" is in fact
turned on its head in particle theory by the observation that the larger the
community around a fashionable failed topic, the lesser the possibility of an
internal critical view and the more fades the chance to overcome the
derailment. This point is especially relevant since the old "Streitkultur"
which in the 50s and 60s provided a critical counterbalance to speculative new
ideas dicontinued in the 80s and many those who in earlier times were, as the
result of their scientific status, the natural pundits for analyzing new
ideas, are now their fiercest defenders, if not to say their salesmen.

It is impossible to criticise string theory in a profound scientific manner by
only looking at the present situation, and a reconstruction of the sudden
changes in the 5 decades old history and its Zeitgeist is not really
achievable without having lived through at least part of that time and in this
way having acquired a first hand knowledge. After all the conceptual errors
have been committed at the beginning (in the transition from the S-matrix
bootstrap to the dual model) and then propagated into the present problems.
These problems will be in the focus of this essay.

\section{How it all began, a historical sketch of the path from QFT to
S-matrix theories and back}

Nowadays there is agreement that the history of QFT started with Pascual
Jordan, one of Max Born's young collaborators. But simply stating this does
not explain Jordan's radical viewpoint about field quantization which
certainly was not shared by his friendly adversary Paul Adrien Maurice Dirac,
at least not up to 1950 when Dirac came around to embrace field quantization
as a general principle. At the time when Jordan pleaded to apply the
quantization formalism also to matter (de Broglie) waves, Dirac limited field
quantization to light, whereas (massive) matter was subject to the
Born-Heisenberg rules of quantum mechanics \cite{Darrigol}.

The radicality of Jordan's view has its explanation in an episode 1925 when
Einstein roughened Jordan's feathers about some points in his 1924 thesis "Zur
Theorie der Quantenstrahlung". Shortly afterwards Einstein wrote a
counter-article against Jordan's claim that in getting thermal equilibrium
between electromagnetic radiation and material oscillators one does not have
to make Einstein's "Nadelstrahlung" (needle radiation, his way of getting
corpuscular photons into the picture) assumption of a radiation recoil $hc$;
an assumption which naturally fitted into his 1905 idea of what received later
the name photons.

Einstein showed that Jordan's claim, although mathematically correct, came
into a contradicition with physical facts because such a theory could not
explain the absorption coefficient. The impact on the young Jordan was
profound, he not only convinced himself that the quantization of
electromagnetic radiation led to Einstein's corpuscular view of light, but he
also took this as a starting point to elevate wave quantization of matter to a
general principle; such giving de Broglie's ideas the missing
conceptual-mathematical underpinning.

Dirac on the other hand defended his view that, although classical waves as
electromagnetic radiation must be quantized, the classical-quantum
parallellism called appropriately "quantization" is best served by using
quantum mechanics for the description of massive quantum matter and reserve
field quantization to electromagnetism. Many successful ideas appeared to
support his viewpoint; in particular because it led Dirac to his seminal
relativistic equation, which in turn opened the path to a deep understanding
of the relativistic spin equation and the hole theory of the electron/positron
system. The first textbooks on relativistic particle theory by Heitler and
Wenzel were based on the Dirac's hole theory of particles; and as long as the
perturbative order was low enough in order to avoid vacuum polarization
contributions (as in those books), hole theory led to consistent and
observationally verified results.

But hole theory becomes inconsistent in the presence of vacuum polarization;
renormalized QED cannot be formulated and performed in Dirac's hole theory
setting. Dirac's long lasting detour to QFT was probably the conceptually
richest odyssey in particle theory. The conceptual and physical differences
between QM and QFT (historically represented by Dirac and Jordan) are most
pronounced in the ubiquitous presence of vacuum polarization in all processes
which deal with localized objects (as distinct from global quantities as the
scattering matrix). To be more precise, the only local objects which are free
of vacuum polarization are free fields themselves. But as Heisenberg had shown
\cite{Hei}, sharply localized (Wick-ordered) functions of free fields
(composite fields), in particular the quadratic function representing a
partial charge,%
\begin{equation}
Q_{R,\Delta R}="\int_{\left\vert \mathbf{x}\right\vert <R+\Delta R}%
j_{0}(x)f_{R,\Delta R}(x)d^{3}x",~~f_{R,\Delta R}(x)=\left\{
\begin{array}
[c]{c}%
1,~\left\vert \mathbf{x}\right\vert <R\\
0,~\left\vert \mathbf{x}\right\vert >R+\Delta R
\end{array}
\right.
\end{equation}
where the quotation mark is an alert that the current-charge connection in QFT
requires the definition of the current as a "normal product" (Wick product in
the free field case) and a definition of the integral (in view of the fact
that the integrand is an operator valued distribution). The test function
$f_{R,\Delta R}$ localizes spatially in a smooth way. In case one is
interested in the global charge limit $R\rightarrow\infty,$ the specification
in what sense of operator convergence the integral converges (weakly, strongly).

Heisenberg showed that if one integrates the charge density over a sphere of
radius $R,$ which corresponds to the limit $\Delta R\rightarrow0$ and fixed
$R,$ the integral diverges as a result of vacuum polarization caused by the
current density acting on the vacuum. The discovery of vacuum polarization in
composites of free field operators and the subsequent observation by
Oppenheimer and Furry that the presence of interactions leads to infinite
vacuum polarization clouds (finite in every finite perturbative order,
increasing with order), are in a retrospective evaluation the conceptually
most important contributions to the early phase of QFT.

For the first time it delineated QFT clearly from any form of relativistic QM.
In principle a profound understanding for the ubiquitous presence of vacuum
polarization as a consequence of causal localization could have saved QFT from
its first big crisis, the ultraviolet divergence crisis, which however,
different from the later crisis which will be the subject of this essay, did
receive a proper closure.

Part of this crisis was overcome by mathematics, namely the recognition that,
very different from classical theories and also different from quantum
mechanics, quantum fields are singular by their very nature and manipulations
with them go astray if one does not keep the Laurant Schwartz distribution
theory at least in the back of one's head in dealing with quantum fields. This
lead to a flurry of articles around the beginning of the 50s in which the new
message was that quantum fields are really operator-valued distributions.
These mathematical structures, which made QFT so different from QM, had their
origin in the phenomenon of vacuum polarization (and different concept of
localization from the Born localization of Schroedinger wave functions) which
renders quantum fields as objects with a short distance behavior which was
never seen in QM.

In the case of the above partial charge this connection can be nicely
exemplified. A sharp localization of e.g. the charge denslty within a radius
$R$ causes infinite vacuum polarization clouds which manifest themselves in
the contributions of infinitely many intermediate vacuum polarization
formfactors
\begin{equation}
\left\langle p_{1},...,p_{n}\left\vert j_{0}\right\vert 0\right\rangle
\end{equation}
which enter the calculation of the $\Delta R\rightarrow0$ singular behavior of
the norm of the partial charge applied to the vacuum which for dimensional
reasons is (modulo logarithmic corrections) proportional to ($R/\Delta R)^{2}$
since the region of the vacuum polarization grows with the area of the shell
around the sphere of radius $R$ and $R$ and $\Delta R$ are the only
dimensionful quantities\footnote{Although this argument is only valid in
conformal models, it is believed that the value of the mass does not enter the
leading behavior.}. This dependence on the "thickness" of the vacuum
polarization cloud hold also for other dimensionless quantities which diverge
if one compresses the fuzzy localization of the vacuum polarization cloud in
the attenuation shell by sending the attenuation collar $\Delta R\rightarrow
0.$

One such quantity which has been the object of recent studies is the
"localization entropy" \cite{BMS} which, like the charge, is also
dimensionless and diverges in the same way. In the limit $R\rightarrow\infty$
the vacuum fluctuations disappear and the value of the charge or the entropy
in the global vacuum vanish. The partial charge on the vacuum and the
localization entropy, being both dimenionless, do not only have the same short
distance behavior, but they also converge both to zero in the global limit
$R\rightarrow\infty.$

The distribution-theoretical cure together with the principles of QFT allows
to separate problems with genuine infinities, as for the partial charge or the
localization entropy, from those which are "man-made" i.e. have been caused by
using inappropriate physical concepts of QFT during perturbative calculations.
All divergence problems at the time of the ultraviolet crisis in the decade
before renormalization were of this kind. Even renormalization was often
understood as the consistent removal of cutoffs or regulators; this is
perfectly reasonable as long as one does not attribute any physical
significance to such computational tricks. This way of looking at dynamical
formulations of QFT in the old says came from QM, where the main dynamical
problem was to convert formally defined operators into mathematically well
defined selfadjoint or unitary operators and to compute their spectrum and
eigenstates. The most important operator of this setting is the Hamiltonian.

The conceptual setting of QFT is radically different. Here the main tool for
the classification and construction of models is the \textit{realization of
the locality principle}. This is particularily evident in the formulation of
Epstein and Glaser. One starts with an invariant interaction polynomial formed
from covariant free fields which in turn are obtained from the
covariantization of the Wigner representation. There are many free fields for
a given $(m,s)$, whether they are of the Euler-Lagrange kind or not is
irrelevant at this point\footnote{What is relevant for passing the
renormalizability test is that the fields have their snallest possible short
distance dimension which is 1 for Bosons and 3/2 for Fermions.} since strictly
speaking this is not a Lagrangian approach. The Epstein-Glaser iteration step
consists in showing that the n-fold time ordered operator products in the next
perturbative order are determined from the previous order, apart from the
spacetime diaginal. These freedom to add "counterterms" on the total diaginal
is restricted by a scaling requirement: do not add counterterms with higher
scaling degree than that before the addition of counterterms.

Models are called (perturbatively) renormalizable if the scaling degree
subjected to those rules does not increase with perturbative order. In d=1+3
the number of renormalizable models turns out to be finite, only free fields
with scaling degree one whose interaction polynomial has a scaling degree
which stays within the power counting limit of 4 lead to renormalizable
models. The latter form finite parametric islands among all couplings and all
fields (the universal Bogoliubov generating $S$ operator) which are stable
under application of the renormalization group and with only couping pointlike
fields, their number is finite.

From the vacuum expectation values of these time ordered operators one obtains
via the so-called KMS construction a Hilbert space representation which
contains all operators which are useful for the model, in particular the
Hamiltonian. Cutoffs and regulators maybe convenient computational tricks in
certain situations, but there is no conceptual necessity to work with them.

The described renormalization setting contains strictly speaking only
$s=0,\frac{1}{2}$ fields; $s=1$ fields remain outside since the scaling
dimension for the lowest dimensional pointlike field of spin s is $d_{sc}=s+1$
and $d_{sc}=2$ is to high for the power counting limit; the prerequisite for
renormalizability is $d_{sc}=1$ which is the lowest value permitted by
positivity (unitarity) in 4 spacetime dimensions. The saving grace came from
an analogy with classical gauge theory where the vectorpotential is an
important calculational tool. The prize for using such a
\textit{vectorpotential in QFT} is however quite high: in order to save its
pointlike field formalism one has to overcome a clash between quantum
localization and the fundamental Hilbert space setting of quantum theory, a
clash which has no analog in the classical setting.

The way to do this is well-known: \textit{quantum gauge theory}, formulated
either in the older (and more limited) Gupta Bleuler setting or by using the
more general and meanwhile better known BRST formalism. The indefinite metric
lowers the short distance scale dimension down to the power counting
range\footnote{The improved short distance behavior from negative
probabilities was used in innumerous forgotten short distance improving
calculations.} and at the end of the day, after all calculations had been done
in the unphysical metric, the BRST "symmetry" allowed the return to QFT in
form of finding a subalgebra pointwise symmetric under BRST transformation. In
this way one can enlarge the range of renormalization theory to $s=1.$

We owe all our insights into the standard model (the culmination of gauge
theory which started with renormalized QED) to the formalism of gauge theory.
But does this mean that gauge theory is the closure of an development which
started with QED ? Certainly not, the clash between pointlike localization of
vectorpotentials (and more generally $(m=0,s\geq1)$ tensor-potentials) and the
positivity principles of QT should not be solved by a technical trick but
rather at its conceptual roots.

Even staying with $s=1$ There are eminent practical reasons why one should not
view the present gauge formulation to be the last word. Gauge theory only
covers the very small subalgebra of local observables; all operators in QED
which carry electric charge remain outside its formalism; their presence is
only symbolically indicated in the form of Dirac-Jordan-Mandelstam exponential
semiinfinite line integrals whose perturbative definition and construction is
not part of the QED perturbative renormalization formalism \cite{charge1}. The
Dirac spinors in the field equation or in the interaction Lagrangian have no
direct physical significance, in fact it is known that there can be no
nontrivial electric charge in an indefinite metric space \cite{Pic}. The
situation worsens for Yang Mills theories and QCD; in that case all spacetime
dependent correlations have apparently incorrigible infrared divergencies and
those few spacetime independent quantities which one can compute, as the
beta-function, remain without those physical objects whose short distance
behavior they are supposed to describe.

The way out consists of two steps \cite{charge1}. The first step is a return
to the Wigner representation theory in order to resolve the clash between
localization and positivity right there. For the electromagnetic fields
$(m=0,s=1)$ this is quite easy: although there is no pointlike covariant
vectorpotential in the Wigner representation space, by relaxing the
localization requirement one easily obtains a semiinfinite stringlike
localized vectorpotential $A_{\mu}(x,e)$ which is causally localized on the
line $x+\mathbb{R}_{+}e.$ This construction can immediately be generalized to
$s>1.$ The use of such string-localized potentials avoids fallacies of the
indefinite metric pointlike vectorpotential e.g. the conclusion that the
application of the Stokes theorem to the magnetic flux leads to a an object
which is really localized on the boundary \cite{Roberts}\cite{charge1}.

The second step consists in the use of the stringlike physical
vectorpotentials in interactions. The power counting hurdle is easily passed
since one can always construct string-localized potentials whose scale
dimension is one, even for s%
%TCIMACRO{\TEXTsymbol{>}}%
%BeginExpansion
$>$%
%EndExpansion
1. In fact these potentials were not totally unknown in the gauge setting,
they are apart from important differences in interpretation identical to the
axial gauge potentials and it was also known that they live in a Hilbert
space. What was however overlooked was their stringlike localization. To be
more precise, the main consequence of this localization, namely the
incorrigible infrared divergencies (even in the QED correlations for charged
fields) were noticed and led finally to the dismissal of this gauge.

But reinterpreting the alleged gauge parameter as a spacelike localization
direction of a string $e$ (a point in a lower dimensional de Sitter space) the
fluctuations in $e$ and the resulting distributional character are is hard to
be overlooked. This mathematical situation asks for a smearing in the
$e^{\prime}s$ and a calculation of coincidence limits pretty much as one goes
about composite fields in $x$. There is also the problem of how to formulate
the iterative Epstein-Glaser renormalization for strings \cite{charge1}. The
lower short distance dimension of string-localized potentials also carries
over to the massive case which is of great relevance to the improved
understanding of the Schwinger-Higgs screening mechanism. Needless to add that
strings in QFT, gauge theory and the standard model have in content nothing to
do with string theory; it is however one of our conclusions in this essay that
the long period of stagnation in those areas is connected to the ascend of
popularity of theories of everything (TOE) and string theory in particular.

The perturbative approach based on the local coupling of free fields is not a
good starting point for mathematically rigorous model constructions, because
even under optimal circumstances the renormalized perturbative series is known
to diverge. In that case there exists a different strategy which is based on
the philosophy that algebras with larger localization regions admit generators
which have a lesser complicated vacuum polarization structure. For a certain
family of two-dimensional models it was possible to find generators for
wedge-localized algebras which are rather simple \cite{Annals}. The compactly
localized double cone algebras can then be shown to exist in a nontrivial way,
which securs the nontrivial existence of the theory with a clear path for the
construction of pointlike generating fields.

Quantum field theorists and mathematical physicists in the old days also
discovered a relation of a certain type of QFT (including those polynomial
field couplings investigated at that time) with Euclidean QFT and Statistical
Mechanics. In fact the contributions of E. Nelson, K. Symanzik, F. Guerra, K.
Osterwalder, R. Schrader, J Glimm and A. Jaffe did not only form the focus of
mathematical particle physics of the 60s, but they were also indispensible for
the linkage with statistical mechanics and the formulation of Wilson's
renormalization group project.

The connecting link of the particle theory based on QFT and string theory,
which will be ln the focus of our critical analysis in the present essay, is
S-matrix theory. By this we do not mean Dyson's presentation for the
perturbative scattering matrix which presented the crowning finish of the
first attempt at renormalized perturbation theory in the 50s, but rather the
partially misled attempts to escape the perceived ultraviolet catastrophy in
pre-renormalization times by exclusively dealing with operators which are free
of vacuum polarization clouds.

These attempts led around 1942/43 to Heisenberg's proposal to study autonomous
theories of S-matrices i.e. models which do not use localized interacting
fields and are therefore free of vacuum polarizations. In order to make some
more specific remarks about these unitary and Poincar\'{e}-invariant models it
is helpful to introduce some notation (the original notation, apart from
Wick-products, which did not yet exist in the 40s) about these oldest S-matrix
models%
\begin{equation}
S=e^{i\eta},~\eta=\int\eta(x_{1},..\eta_{4}):A(x_{1})...A(x_{4}):dx_{1}%
..dx_{4}%
\end{equation}
where $A(x)$ is a scalar free field and $\eta(...)$ is a connected
Poincar\'{e} invariant function.

Unitary and Poincar\'{e} invariance are obvious whereas the validity of the
cluster factorization property for asymptotic spacelike distances
$a\rightarrow\infty$
\[
lim_{a\rightarrow\infty}S\left\vert f_{1},.f_{k},g_{1}^{a},..g_{l}%
^{a}\right\rangle =S\left\vert f_{1},..f_{k}\right\rangle \otimes S\left\vert
g_{1},..g_{l}\right\rangle
\]
follows from the connectness of the Poincare invariant function $\eta(...)$

The spacelike clustering is the easy part of "macrocausality"\footnote{The
Heisenberg ansatz happens to fulfill this indispensable S-matrix property in
an accidental tacid way, at least it is not mentioned in any of the old
S-matrix papers.}. St\"{u}ckelberg criticised Heisenberg's approach on the
lack of a timelike manifestation of macrocausality called "causal
rescattering". The simplest illustration of this requirement is in terms of
the 3 to 3 particle scattering which should contain as an asymptotic
subprocess the scattering of first two particles and then at a much later
time, the scattering of one of the outgoing particles with the third one. He
showed that the asymptotic line which connects the two scattering events is
described by (what was later called) the Feynman propagator at the value of an
momentum which corresponds to the classical mass and velocity associated with
the two events. This timelike asymptotic cascading structure, which must be
contained in the S-matrix, together with the spacelike cluster-factorization,
constitute the most important manifestation of macrocausality. This expression
actually stands for all spacetime aspects which can be formulated in terms of
particles only i.e. without the intervention of interpolating fields i.e. it
does not include the crossing property of the S-matrix which cannot be
understood without the microcausality-carrying fields.

St\"{u}ckelberg used the asymptotic propagator expression for all distances
and idealized the particle interaction region by a spacetime point. In this
way he arrived prior to Feynman, but in a less legitimate way, to the Feynman
rules. As a result it is almost impossible to answer the question: what did
Stueckelberg really intend with his criticism of Heisenberg? Did he want to
point out that Heisenberg's S-matrix setting is incomplete from the viewpoint
of macro-causality or did he want to say that Heisenberg is barking up the
wrong tree i.e. that a pure particle-based S-matrix approach is not the right
way to counteract the ultraviolet catastrophe. The second view is probably the
correct one because Stueckelberg's later contributions to QFT show that he
wanted S-matrix ideas to be subservient to QFT. In principle he could have
discovered an iterative construction of a unitary S-matrix by complementing
his macrocausality requirement with an iterative unitarization. The analogy of
Stueckelberg's S-matrix problem and the problem of unitarization of the dual
model will be the issue which will be taken up later on.

This early S-matrix project moved out of sight for more than a decade, and a
QFT liberated to a considerable degree from its ultraviolet divergence problem
in the form of renormalized QED returned to the forefront. The crowning
achievement up to the present time of this project of perturbative
renormalization theory is the Standard Model. Although the SM only appeared on
the scene after the failure of the second return of S-matrix theory in form of
the S-matrix bootstrap, its concepts extended those underlying the pre
S-matrix QED renormalization theory and had no connection with S-matrix ideas
and the later string theory. The fact is that almost all\footnote{The new
aspects as gluon/quark confinement which were absent in QED and the closely
related asymptotic freedom have remained open problems.} the concepts of the
SM were preempted in renormalized QED and enriched with important new
computational technology; in this way the latter became part of the former.
The return of S-matrix theory between the QED renormalization and the later
S.M. had specific reasons. Since their understanding is important for our
deconstruction project, we will check them out more carefully in the following.

In the 50s there was a giant step taken towards a better understanding between
fields and particles. The omnipresence of localization-caused vacuum
polarization made it desirable to understand better how particle concepts as
the S-matrix, which are free of such vacuum polarization problems, can be
reconciled with fields. The key to progress in that area was the formulation
of the time-dependent LSZ scattering theory, the derivation of the reduction
formalism and the foundation of both in localization and spectral (positive
energy) requirements. In this way it became possible to derive properties of
the S-Matrix which, unlike the aforementioned macrocausality, were genuine
imprints of foundational properties of field theory. This opened the
possibility of experimentally checking the foundational property of QFT namely
the causal localization principle (micro-causal propagation) through the QFT
adaptation of the optical Kramer-Kronig dispersion relation.

In this context a new basic structure linking fields with particles arose: the
crossing property. It has its name from an obvious property of Feynman graphs:
a contribution to say, two-particle scattering, has an intermediate one
particle contribution which if viewed in the "crossed" channel looks like a
process in which a particle is exchanged. But Feynman diagrams are on-shell
(the mass-shell) and its not entirely trivial to show that there is an
on-shell analyticity which allows to recast the graphical crossing into an
analytic identity between a scattering process and its analytic continued
crossed version. For the dispersion theorist the crossing for 2-2 particle
scattering was sufficient, it was proven in \cite{BEG}. For many decades this
property was used outside the special established cases; most people
considered it as self-evident. Only more recently with the advent of the of
"modular localization" its conceptual subtlety (related to an extension of the
KMS property) has been finally appreciated and its correct foundational
position recognized \cite{founcor}.

The dispersion theoretical approach is the only successful concluded project
in an S-matrix setting. It is in fact the only successfully finished project
in particle theory: all its theoretical aims were achieved and experimentally
verified, so that the individuals who have been involved in this project could
turn towards other problems. Such a project would have lost its value if the
structural properties (e.g. the Jost-Lehmann-Dyson representation) used in its
derivation were based on guesswork and conjectures rather then mathematical
proofs. This may have been the reason why Jost, Lehmann and K\"{a}ll\'{e}n
were so unhappy about the Mandelstam representation which has remained
unproven and in retrospect defines the beginning of a more metaphoric
discourse which led to the derailment of a large part of particle physics.

All the later S-matrix-based projects ended in a cascade of failures: the
S-matrix bootstrap, the dual model and string theory. The common (but not the
only) reason for the failures is the incorrect understanding of the meaning of
crossing; on very recently the conceptual origin of the crossing property for
formfactors (of which the S-matrix crossing is a consequence) was
understood\cite{founcor}. The bootstrap proposal was based on requirements
which, as a result of their generality and vagueness did not have any
constructive power. The belief that one was in the pocession of a unique
particle theory namely a kind of theory of everything (without gravity)
remained a grand fata morgana; instead of a theory of something it turned out
to be a theory of anything in the worst possible meaning: each S-matrix
derived from a QFT fulfills the bootstrap structure and as in d=1+1 one finds
infinitely many explicit bootstrap solutions\footnote{This is not the result
of the bootstrap setting but rather that of the \textit{factorization
property} together with the bootstrap, which has no counterpart in d=1+3.}
even within the family of factorizing theories (purely elastic S-matrices),
every QFT in any spacetime dimension which has an S-matrix which fulfills
Poincar\'{e} invariance, unitarity and crossing which are the defining
properties of the bootstrap; so the boostrap project, with the exception of
factorizing models, was not wrong but empty.

The case of the dual model and string theory is much more concrete; as a
result the misunderstanding and conceptual errors become more visible; it is
our aim in this essay to expose them explicitely. This ends the constructive
part of the essay which mainly serves to get better platform for the
deconstruction of the conceptual basis of string theory.

The next section contains a critical evaluation of sociological and
philosophical reactions to string theory. This is however not the kind of
criticism supported by the author because it does not expose its scientific
roots. A sociological and philosophical critique dangles in the void unless it
is preceded by a scientific deconstruction. Certainly string theory and what
its defenders write about it \cite{Schele} arouses in many people the feeling
of looking at something bizarre, but it is not so easy to identify the
scientific reasons from which these feeling emanates.

It is therefore not surprizing that our analysis of string theory is preceeded
by a rejection of existing criticism of ST on the grounds that e.g. it did not
lead to observational verifications. If a correct theory which is an extension
of QFT does not lead to experimental verification this would be still
sensational from a conceptual viewpoint since to supersede a highly consistent
theory in a mathematical consistent way would be an incredible achievement.
The greater problem is however to confront a conceptually wrong theory which
over a long time agrees with observations (viz. phlogiston). Since we are
claiming that ST belongs to the second kind, but up to now without
observational agreement, we do not have to loose time on the first possibility.

The criticism starts in the third section in which the development of string
theory from the dual model through the Nambu-Goto Lagrangian is critically is
analyzed from the viewpoint of localization which is central for all
relativistic quantum theories.

As a contrapoint we explain in the fourth section the intrinsic meaning of
string localization. After this constructive interlude we pass again to the
destructive project in section 5 where the consistency of dimensional
reduction (Klein-Kaluza) will be critically analyzed.

The metaphoric picture about particle theory in the wake of string theory is
the topic of section 6, whereas section 7 analyses the impact of ST on the
situation in Germany in the aftermath of LSZ, a topic of living history for
the author. The last section tries to counteract the belief that
metaphor-based theories as string theory are the only "game in town".

\section{The aim}

...\textit{The history concludes with an unexpected and glorious success: the
so-called standard model. \ The way in which this structural classification
fell into place, and the great leaps of imagination involved, \ justifies a
degree of hubris among the few dozens truly extraordinary individuals who
discovered it. However both this hubris, and the complexity of the result, fed
the temptation to go on leaping, and to forget that these earlier leaps,
without exception, had taken off from some feature of the solid experimental
facts laboriously gathered over the years.... }

\textit{Philip Anderson, in }\ \textit{"Loose ends and Gordian knots of the
string cult" \cite{Ander}}

There is a widespread consensus among particle physicists that particle theory
is in the midst of a crisis. Even string theorists, who feature in this
statement of Phil Anderson as bearing responsibility for this situation, quite
readily concede that particle theory has seen better times. They however
propose a quite different remedy from that which would follow from Anderson's
diagnosis, namely they plead for the application of a stronger doses of the
same medicine, because in their mind "there is no other game in
town"\footnote{As well known to every string theorist, this phrase was used by
David Gross on several occasions.}.

Anderson's quite devastating indictment about nearly four decades of post
standard model domination of particle physics by the noisy but scientifically
unsuccessful "string cult" expresses an opinion of an eminent condensed matter
physicist which is shared by an increasing number of particle physicists.

Most of the criticism directed against string theory (ST) has been focussed
towards its lack of observational success and more general the absence of any
predictive power \cite{Woit}. Others are worried about the dominance ST has
even outside of popular science. Indeed its metaphoric discourse about an
alleged theory of everything (TOE) at universities and research institutions
is a serious problem and concerns about its suffocating influence on other
theoretical directions which are based on the less reductionistic ideas of a
theory of something instead of a TOE are certainly well-founded \cite{Smolin}.
There are also those who are irritated by the strange philosophical opaqueness
of ST \cite{Hed}. Meanwhile string theorists got used to this kind of
sociological and philosophical criticism which does not reach the scientific
foundations of their theory. On the other hand sociological ST-bashing has
become an activity by which one can build a reputation, so that the survival
of ST is not only important for string theorists but also to some of its critics.

It is noticeable that in all the sociological contributions the authors take a
critical look at the dominant position of ST and explain very well the
sociological reasons why younger people uncritically internalize its
catechism. \textit{But they never explain why respectable older people, who
are under no such career pressures (especially those who are the main string
proselytizers mentioned before) believe in the validity of the theory.} It is
of course common practice to blame the foot-soldiers (in the present context,
the young partisans of ST) and perhaps the propaganda division (as Brian
Greens and others), but spare the generals; in this respect these critical
contributions are not different from critique about many other human
activities where critical pundits have become interested in the survival of
target of their criticism, especially if their only fame or profile depends on
its continued existence.

What has been totally missing is a conceptual criticism of the physical
content of ST. If the string theorists view of their theory as extending QFT
is correct, than this would be \ a remarkable achievment independent of
whether nature realises it or not. One should not prematurely dismiss it on
the grounds of problems it has in leading to observational test. Even if it
contradicts observational facts, the mere observation that there is a
consistent theoretical setting which extends QFT would be a remarkable feat
which is bound to give new insights into the ongoing development of QFT
itself. This would create a similar situation as the one which Einstein
confronted with his purely theoretical discovery of General Relativity, when
he thought about its possible observational failure as a lost chance for the
Dear Lord. In the absence of observational support, the criticism of the
theoretical consistency is the only aspect which really counts, and here
Einstein was his own demanding critic.

In this note I will show that ST fails on two aspects. The first one is that
the claim that a string is a string-localized object in spacetime comes from a
misunderstanding of the intrinsic meaning of localization. Rather a string of
ST is an infinite component field which represent a infinite mass/spin tower
over one point and additional operators which transform between the levels of
the towers. The misunderstanding about localization can be traced back to the
dual model which shortly before its reformulation into the string theoretical
setting was incorrectly interpreted as defining an embedding of a
multicomponent chiral current "source" theory into the "target" space defined
by its inner symmetry components; the metaphoric error consisted in
attributing a string extension (a sheet in the graphical spacetime setting of
Feynman) in target space to this object. But a relativistic quantum theory
with causal localization cannot fulfill such an embedding idea: no lower
dimensional theory can be embedded into a higher dimensional one; a spacetime
which has to accomodate causally localized quantum matter (with the vacuum
polarization caused by localization) is simply too holistic. When people have
such ideas, they usually think about QM where this is possible. A chain of
quantum mechanical oscillators can be placed anywhere (including into an
internal space) since localization in QM, different from modular localization,
is not an intrinsic concept. The only way of having 1-dimensional causally
localized objects in higher dimensional spacetime is through string-localized
covariant fields.

The claim that the inner symmetry acting on the components of the chiral
current or rather its multicomponent potential in its exponential form (the
operator presentation of the dual model) can be a noncompact group is
surprising, since inner symmetry spaces in d%
%TCIMACRO{\TEXTsymbol{>}}%
%BeginExpansion
$>$%
%EndExpansion
1+1 are known allow only compact group actions; and that the requirement of it
being the Lorentz group which together with positive energy translations leads
to a finite number (connected by M-theory) of 10 parametric superstring
representation involving fermionic degrees of freedom is even more remarkable,
but nevertheless true. But why should a mathematical attention causing
statement on a chiral model be the starting point for a new theory of
spacetime? Isn't the more logical procedure to look for special aspects of
\textit{chiral theories which are known to admit different realizations of
spin, statistics and inner symmetries} (including noncompact structures in the
case of irrational chiral models)?

From this first error resulted a second problem: how can that strange
pointlike embedding of a chiral model (the source) into the "target space"
(the space spanned by its internal symmetry indices) be linked to Feynman
rules for strings in terms of world-sheets (tubes) instead of world lines?
Recipes for higher order scattering amplitudes were abstracted on the basis of
such world sheet pictures which were interpreted as higher order
approximations and up to second order explicit calculations were performed
(perturbation in the genus in the geometric string setting). Wheras Feynman
diagrams in QFT represent amplitudes in the operator/state setting of quantum
theory, the conceptual status of the worldsheet rules for the interaction of
an infinite component pointlike field remained mysterious. They become even
more mysterious if, as will be shown later on, the "dual model" starting
amplitudes are nothing else then the Mellin transforms of conformal QFT. A
mysterious new connection of a particle S-matrix with a conformal QFT ? Can
unitarization recipes on a conceptually doubtful lowest order save the day?

The waiting for a consistent theory in which the tube rules represent the
perturbation theory of an operationally defined problem (so that one can
forget the unfortunate start from pointlike localized objects or the incorrect
embedding picture which erronously were misred as stringlike and start with
the pictures) for almost 5 decades has had no success. No consistent
formulation in terms of states and operators was ever found, even though it
attracted the attention of the best minds in ST. The present situation is
psychlogically reminiscent of Samuel Beckett's "Waiting for Godot" where Godot
stands for "quantum theoretical rules" i.e. rules in terms of states and
operators as in Feynman's case which would give the graphical rules for
splitting and combining tubes an operational meaning. In view of the fact that
the zero order input in form of the Mellin transform (unlike a Fourier
transform) has no autonomous operational meaning this looks like a "mission
impossible". In this case the quantum theory associated with the worldsheet
rules could have been taken as a start of a new theory, and the not very
useful infinite component pointlike theory which is behind the source-target
reading of the dual model or the canonical quantization of the bilinearized
Nambu-Goto Lagrangian could have been ditched.

One can also see this situation of a missing quantum theoretical setting for
the tube rules in a more antropomorphic manner as the reaction of particle
theory against the imposition of an incorrect metaphor about localization
which is the central theme of any relativistically causal QT.

Our presentation of the mathematical-conceptual flaws on which ST erected its
claim of hegemony as a TOE may generate the impression that relativistic
quantum theory, after reaching such heights as QED, renormalization theory and
the standard model, was suddenly taken over by a new generation of half-wits
who rammed particle theory into the ground and in this way caused more than 4
decades of standstill. In order to avoid creating such an oversimplified and
incorrect view it is necessary to present the scientific criticism in a
historical context in which the different ideas leading to superST appeared.
Understanding why whole communities made conceptual mistakes and took an
erroneous path is as interesting and important as understanding what led to
important discoveries, especially if these errors were made about the most
subtle issue of localization which already led to many other misunderstandings
unrelated to ST \cite{inter}.

It would be very difficult to accomplish this without having lived through
those times with an open critical eye. Recent historical accounts as
\cite{Vech} are of some help as long as one does not expect any critical
viewpoint in a commemorative contribution.

Accepting that the dual model- and string-theorists were in no way less
competent to their field theory predecessors, the question arises what was
different after all, and why did a derailment of particle physics occur with
them and not before?

There were at least two situations which could have caused an erroneous trend:
the ultraviolet crisis which begun in the 30s and found its end at the time of
the renormalization theory of QED of the late 40s, and the S-matrix bootstrap
project which was supposed to lead from some S-matrix principles directly to a
unique solution and which started in the late 50s and ended in late 60s around
the time of the appearance of the standard model. The main difference to the
situation which led to the modern ST is that the leading figures before the
dominance of ST (starting in the beginning of the 80s) were much more aware of
the importance of balancing the speculative aspects of particle theory with a
critical analysis which goes to the bottom of its conceptual structure.

The discovery of a theory which generalizes quantum field theory (QFT) not
only on formal grounds but also on a conceptual level would of course have
been a remarkable achievement. Most attempts over 5 decades to supersede QFT,
be it through nonlocal/noncommutative changes or a pure S-matrix theory have
ended without useful conclusions, but not as a result of an
conceptual-mathematical inconsistency. ST is a special case; it is the only
project for which one can pinpoint to a conceptual error at the most important
concept which devides relativistic quantum mechanics from relativistic
theories which have a causal propagation at all distances (sections 4-7). This
is the central issue of localization which devides QM and QFT/ST more than any
other property: Born-Newton-Wigner localization for QM and and modular
localization for causally propagating theories (QFT, ST).

All the other sociological or philosophical arguments given against ST are
secondary, even the question whether ST during its 5 decades of existence has
explained any observed phenomenon. There is no time limit on exploration of
deep ideas beyond QFT, but the question is whether the 5 decades old ST
represents such a consistent extrapolation. Even if such a theory is
mathematically too complex to make new predictions or if a prediction comes
into contradiction with observations, it is still important to pursue its
consequences; the mere existence of a conceptually consistent theory which
contains QFT as a limiting case requires this. An experiment can decide
whether a theory is useful or not, but not whether it is consistent as a
conceptual-mathematical construct and whether it can be viewed as an extension
of an established theory and from the old phlogiston theory of burning we know
that a theory can explain phenomena and make prediction and yet be wrong.

A historical illustration about the importance of inclusive consistency is
Einstein's reaction when the gravitational deflection of light on the sun was
observed and he exclaimed that in the case of a negative result the Dear Lord
would have missed an interesting chance. The history of ST proceeded quite
differently, the theory had to undergo several metamorphoses passing through
real laboratory particle physics before ending at the secure Planck length and
bringing tenure to one of the protagonsists. before the ST community placed it
at the center of a new ultrareductionist TOE.

The concrete illustration of how ST was able to introduce misleading metaphors
into particle physics, and disturb the equilibrium between innovative
speculative ideas and their critical assessment, will be the main aim of this
essay. As regards to sociological aspects of the crisis, I will stay within
the boundaries of my personal experience and focus on how ST was able to get a
foothold in German particle physics.

The central metaphor of string theorists\footnote{We follow time-honored
traditions of respecting well-established terminology even in case when new
findings show that the chosen name is metaphoric and without an autonomous
significance}, based on the form of the classical Nambu-Goto Lagrangian
\cite{Nambu}\cite{Goto} which describes \textit{classical} strings, is the
credo that ST has to do with quantum objects which are string-like localized
in spacetime. It may be expected that its verification is not an easy matter
since the concept of \textit{quantum localization} is not only the most subtle
in QT, but there is also a big distinction between localization in QM which is
best be expressed in terms of wave functions on the spectrum of a position
operator and the modular localization\footnote{This is the the causal
localization (spacelike commutance of observables and their causal timelike
determination in the causal shadow region) but formulated independent of
"field-coordinatizations".} of quantum theories with a finite propagation
speed at finite distances as causal relativistic theories.

The conceptual differences are considerable: from a mathematical viewpoint
localized algebras in QM are von Neumann factors of type I, together with the
complementary localized algebra they tensor factorize which leads to the
notion of entanglement and the connected rich information theoretical concepts
\cite{inter}. On the other hand the local subalgebras in a theory with modular
localization are factors of hyperfinite type $III_{1}$ with very different
physical manifestations in that that there is no tensor-factorization for
causal subalgebras with respect to their causal disjoint because the vacuum
polarization at sharp localization boundaries prevents this.

Even if one creates a split distance $\varepsilon$ between the two which tames
the vacuum polarization and generates a fuzzy localized type I intermediate
factor which can be used for tensor-factorization, the restriction of the
vacuum to this tensor-factor is a thermal KMS state which gives rise to
\textit{localization entropy \cite{BMS}}. Actually causally localizable
theories have both; the frame-dependent BNW localization coming with the Born
probability and covariant modular localization mainly for the operators. Both
approach each other asymptotically in the scattering limit of large times
which is absolutely crucial for the correct understanding of the
particle-field relation.

Rephrased in a way which does not refer explicitly to quantization of
classical data, the ST credo relies on the possibility of embedding a chiral
conformal field theory as a stringlike subtheory of a higher dimensional
relativistic theory whose living space is referred as the \textit{target
space}. In contrast to the totally vague and useless general bootstrap
project, the stringy object is very specific; it is a model in a 10
dimensional dimensional spacetime. As in standard QFT in curved space-time to
each CST there is another (or many) reference state which is as similar as
possible to the Minkowski vacuum.

But the story of the source-target embedding is a fairy tale, since in
contrast to QM, where one can place a chain of quantum oscillators anywhere,
it is not possible to do this in theories with a causal localization. The
spacetime and the quantum matter in it are too \textit{holistic} as the result
of the omnipresence of vacuum polarization from localization. Even in doing
the opposite, namely restricting such a theory to a lower dimensional
spacetime (brane) there is a problem\footnote{It is however perfectly normal
to have together with pointlike fields also semiinfinite stringlike localized
field \cite{charge1} $\psi(x,e)$ in a theory (see section 7). Fields carrying
a Maxwell charge are of this kind where Field strengths remain pointlike.
Stringlike fields, despite their intuitive proximity having nothing to do with
source-target image.}, although it can be done, the brane theory exists
mathematically but not physically. A similar problem is encountered with the
Kaluza-Klein compactification.

The impact of these statements on actual calculations are easily seen in
concrete cases; in both cases, the canonical quantization of the Nambu-Goto
model and the dual model embedding of the chiral current theory (the operator
representation of the dual model) both lead to pointlike structure. The
infinitely many oscillators do not lead to a stringy extension in spacetime
because they enrich the internal Hilbert space over a point in a similar was
as the spin components.

One could of course dispose of localization aspects of dual models and string
models and directly ask whether the unitarization prescriptions for a
tentative string S-matrix which the string theorists came up with are
consistent and whether they justify the name "stringy". The first question is
clearly defined; one must show that the recipe can be formulated in terms of
operators and states. This is a problem which through all the S-matrix history
was never solved for any theory. The most famous attempt was that of
Stueckelberg mentioned previously. Unless one knows already that there is an
off-shell operator theory behind, there is no chance to guess one. Despite
more than 30 year research on this problem, including the most experienced
individuals as Witten, all these attempts have been in vain.

The second problem, the stringyness of an S-matrix, seems to be a
contradictory requirement. To the extent that "stringyness" is a spacetime
notion (what else can it be) the correct question is: how does off-shell
stringyness manifest itself on-shell? The best example of a field which cannot
be point-localized, but whose best possible localization is semiinfinite
string-like, is a Dirac particle which carries a Mawell charge. In that case
the LSZ limits vanish and as a result of perturbative infrared divergences
also the pertubative S-matrix is ill-defined. The inclusive cross section with
finite photon resolution is finite but one can neither abstract an S-matrix
nor relate it to a spacetime limiting process. A stringy S-matrix is simply to
slippery a notion.

In the next section it will be shown that the quantum counterpart of the
classical Nambu-Goto string (or its supersymmetric extension) describes higher
dimensional infinite component pointlike wave functions which can be
associated with infinite component free fields i.e. \textit{the mass/spin
tower sits over one point and is not carried by a spacetime string}. The
recipes about splitting and recombining of tube by which string theorists
assign a "would be" S-matrix, constitute an ad hoc attempt to use the (as it
turns out incorrect) metaphor of strings in spacetime in order to assign
transition probabilities between infinite component wave functions. Short of a
large time limit process by which the interacting operators are connected with
the in- and out- going free particle configurations, string theorists do not
derive the S-matrix from local quantum physics but rather \textit{defines} it
by assigning transition amplitudes to interaction-free wave functions with the
help of tube pictures, which are then translated into analytic formulas
(taking some inspiration from Feynman tules). So in QFT the S-matrix is
derived whereas in ST it is postulate.

The unraveling of the string metaphor will be the principle aim of the next
section. In the third section the string theorists incorrect idea about
localization in relativistically covariant theories will be retracted to the
very beginning of ST namely to the Veneziano-Virasoro-Dolen-Horn-Schmidt
duality and its interpretation in terms of an alleged string-like embedding of
a special chiral conformal QFT into the physical spacetime. For a previous
more detailed description of string theories historical roots we refer to
\cite{crisis}

At the root of these metaphoric disorientation there is a widespread
fundamental misunderstandings of the concept of causal localization. Therefore
section 5 is a reminder of "modular localization" (causal localization of QFT
after liberating it from the fortuitous field coordinatizations). Although
modular localization has rich group theoretical and geometric aspects and is
certainly quite deep as a mathematical theory with strong relations to
operator algebras, it has little relation to the kind of differential geometry
and topology in particle physics as used particularly in gauge theory as it
enjoyed popularity starting in the 70s. We will give several illustrations
which show that certain geometric interpretations are not supported by
intrinsic localization properties\footnote{A prior critical account similar to
the present one, but which more emphasis on the sociological aspects of the
present crisis can be found in \cite{crisis}.}.

In section 6 we will show that some ideas which originated from ST (but which
afterwards developed a life of their own), as extra dimensions, have to be
taken with a grain of salt. The only way they start to make physical sense is
if an operational meaning can be attached. It will be shown that the high
temperature limit coupled with the operational connection between real and
imaginary time local operator algebra is the necessary prerequisite for a
dimensional reduction which maintains the Hilbert space setting and the
positive energy condition of the 10 dimensional superstring. A quantum
generalization of the classical Kaluza-Klein setting which is consistent with
local quantum physics remains uncharted territory.

It is indicative that foundational work on QT which has been able to
incorporate even the counter-intuitive structure related to the quantum
reality has capitulated vis-a-vis ST. A framework of principles as it is
available for QM or QFT does not exist for ST, it would not make much sense in
a theory which claims to be a TOE. Who gets involved with ST and asks for
conceptual guidelines has to be prepared to confront the surreal. A
presentation of a theory as a realization of physical principles is only
possible after a theory has been known beyond its metaphors. For a theory as
ST, for which its main thesis, namely that it deals with string-localized
objects in spacetime, turns out to be a metaphoric illusion, any foundational
placement which overlooks this fact would be illusory.

Metaphoric ideas around a new theory are sometimes quite helpful, as long as
one conscientiously uses them as temporary placeholders for not yet precisely
understood facts. The complementarity principle and the uncertainty relation
of QM which later were derived from noncommuting operators are examples of
such a positive transitory role in particular at the beginning of a new
theory. However the situation of the string metaphor in ST is quite different.
It is a misleading metaphor which is still believed by all members of the
string community and even most particle physicists outside who do not find it useful.

The string oscillators are not objects in spacetime but are part of the
internal space. In particular the quantum object corresponding to the
classical Nambu-Goto string is an infinite component wave function
respectively an infinite component free field describing a mass/spin tower
over one point in spacetime. The search for infinite component fields (in
analogy to the group theoretical $SO(4,2)$ construction of the hydrogen
spectrum) which could produce an interesting particle spectrum without having
to go through tedious dynamical computations, was a popular research topic a
decade before ST, but it remained without success.

The existence of the solution of the quantum N-G model shows that such
infinite component objects do arise (but only in10 dimensions) if one uses
instead quantum oscillators as they arise from the Fourier decomposition of
closed/open strings. The oscillator degrees of freedom (respectively their
supersymmetric counterparts) set the mass/spin tower as well as their relative
strength in the infinite component wave function; but this has nothing to do
with extended objects in spacetime. String theorists were so spellbound by
their mind games, that up to this date they did not realize that even though
their construction has nothing to do with spacetime strings, it did solve the
older problem of infinite component wave equations, which apart from the high
dimensions probably would have pleased their older predecessors of the
infinite component project.

The present attempt of "deconstruction" of ST (in its contemporary
philosophical use it means subjecting something to critical analysis by
carrying it to the breaking point) is an attempt to fill this gap. Given the
present Zeitgeist in particle theory it is not very probable that this nor any
other criticism of the dominance of metaphoric arguments in particle theory
will have any short range effect, but it may facilitate the task of historians
of physics to understand what happened to the millennium project of a TOE.

\section{A critical review: from the dual model through the Nambu-Goto model
to the superstring}

In this section we will show that ST does not describe string-like extended
object in spacetime, but rather constitutes a construction of nontrivial
models of infinite component wave functions. It is well known that infinite
component generating (distribution-valued) covariant wave functions are in a
one-to one relation to multi-component free fields. In case the reducible
unitary representation of the covering of the Poincar\'{e} group has no
irreducible components which fall into the class of Wigner's "infinite
component massless representations" (see next section) both the generating
wave functions as well as their quantum field counterparts are pointlike
localized objects. By an infinite component field one does not simply mean a
direct sum of free fields. The theory should contain operators which
"communicate" between the different levels of the infinite mass/spin tower
over one point\footnote{Sine the objects are distribution-valued, this needs
an appropriate mathematical formulation.} in the sense that they do not
commute with the mass operator; if this would not be the case, there would be
no mechanism by which one could generate a "dynamical" spectrum and the
concept of infinite dimensional fields would be without physical content.

It is precisely in this dynamical sense that the terminology "infinite
component fields" entered the discussion some years before ST was proposed.
Attempts to work with infinite component wave functions date back to Majorana;
in the early 60s they underwent a revival\footnote{The guiding principle was
to emulate the O(4,2) description of the hydrogen spectrum (Barut-Kleinert,
Ruegg, Fronsdal, Budini,...) in the covariant relativistic context
\cite{Barut}\cite{Frons}.}. The increasing particle zoo made it seem advisable
to look for "dynamical symmetries" which generate a rich mass/spin spectrum in
analogy to the spectrum-fixing SO(4,2) symmetry for the hydrogen atom; the
problem of how this can be reconciled with a Lagrangian picture of
interactions was deferred to after a successful construction. These attempts
where all based on noncompact groups containing the Lorentz groups. This
project remained without success and was soon suspended.

More interesting possibilities arise if one goes outside noncompact groups and
permits a full quantum mechanics or more generally a chiral conformal field
theory at its place. Although this is not the way string theorists see their
subject, it is the best metaphor-free way to approach it.

Instead of using the above structural argument which says that every
representation space of a unitary ray representation of the Poincar\'{e} group
not containing massless infinite spin components is necessarily pointlike, it
is more revealing to use the pedestrian method of quantizing the bilinearized
Nambu-Goto string Lagrangian. A mathematical physics presentation can be found
in \cite{Dimock} and less rigorous but still correct computations have been
presented before in articles written by string theorists \cite{Mar}\cite{Lowe}.

Leaving temporarily the physical interpretation aside, one may ask the
mathematical question whether there exist quantum mechanical models which,
attached to a localization point in a particular way, leads to an infinite
component local free field with a nontrivial mass/spin tower. Since such free
fields are characterized by their (graded) commutator functions it is easy to
see that the problem of local field is equivalent to the construction of an
infinite component wave function. The infinite component one-particle Hilbert
space of such a model must be a subspace in a tensor product
\begin{equation}
H_{sub}\subset H=L^{2}(p,\rho(\kappa^{2})d\kappa^{2}\frac{d^{D-1}\mathbf{p}%
}{\sqrt{\mathbf{p}^{2}+\kappa^{2}}})\otimes H_{QM}%
\end{equation}
where, in order to obtain a unitary representation of the covering of Poincare
group with a spin spectrum (the first factor is scalar) the operators in
$H_{QM}$ must have a vectorial/spinorial multiplicity index i.e. they must be
vector or dotted/undotted spinor-valued.

At this point the setting is completely general and does even contain the
generating wave functions of so called generalized free fields (having a
continuous mass spectrum). In particular the Wigner representation theory of
irreducible positive energy representations is of this form. In order to
describe a massive s=1 vector representation one starts from $H_{QM}%
=\mathbb{R}^{4}$ with the generating 4 component unit vectors and sets
$\rho(\kappa^{2})=\delta(\kappa^{2}-m^{2}).$ The invariant metric is always
indefinite at the start, reflecting the fact that finite dimensional
vectorial/spinorial representations are not unitary. The construction of a
unitary subrepresentation is done with the help of intertwiners i.e.
vector/spinor valued functions $u(p,s),s~$=physical spin (and their charge
conjugate $v(p,s)$) which project the formal tensor-factorized states in
$\not H  $ into $H_{sub}~$\footnote{This procedure is described in the first
volume of Weinberg's book \cite{Wei}.}$.$ For D=4 and $dimH_{QM}=4$ one
obtains the s=1 representation in the vector potential description which
fulfills the Klein-Gordon as well as the vanishing divergence equation. A
generalization to reducible representations with several masses and spin would
be unproductive unless one finds a dynamical mechanism which generates a
specific mass/spin spectrum. Such a mechanism must provide additional
operators which connect the different irreducible levels and taken together
with the Poincar\'{e} generators act irreducibly on $H_{sub}.$

This was the idea behind the infinite component field projects in the
beginning of the 60s \cite{Tod}. Those authors tried to implement it with the
help of group theory by extending the Lorentz group to a larger noncompact
group and by looking for unitary representations of these groups. This turned
out to be a blind allay and the idea was given up after 5 years. The main
issue of this section is to point out that the project is successful with
infinitely many vector/spinor valued oscillators as they arise as Fourier
components from $D-$component abelian chiral currents on a circle and its
unique result is the 10 dim. superstring. The terminology "string" for the
resulting pointlike infinite component field is the most unfortunate
terminology in the history of particle physics.

Some renown physicists lend support to ST but had some second thoughts on the
issue of localization. Steven Weinberg thought the ultimate irony is if ST
would turn out to be equivalent to some unusual model of QFT \cite{W97}. The
dynamical infinite component field called 10-dimensional superstring is indeed
a pointlike object and therefore answers Weinberg's ironical question.

If one ignores the positivity of energy requirement one finds another solution
with D=26 which is somewhat simpler to present: the bosonic Nambu-Goto
model.\ One starts with the oscillators of a quantum mechanical string and
defines as $H_{QM}$ the Hilbert space generated by the n-component oscillator
Fourier coefficients leaving out the n-component zero mode which is identified
with the D-dimensional momentum $p$ of the first tensor factor. In this case
the construction of the positive definite subspace in which the Poincar\'{e}
representation becomes unitary is more subtle then in the Wigner case massive
particles or massless representations with finite helicity. The trick which
leads to simplification consists in remembering that the original nonlinear
formulation of Nambu-Goto was reparametrization invariant, which in turn leads
to subsidary conditions on the oscillator Fourier components. Their use leads
for D=26 to a subspace without negative norm states which still contains
null-states. The remaining canonical step consists in "dividing out"
null-states. Allowing spinorial valued oscillator Fourier component one can
remove the last flaw namely the absence of energy positivity and one arrives
at the D=10 superstring.

Admittedly this aspect of uniqueness is somewhat astonishing. But is it
sufficient reason for claiming a new era of understanding about spacetime?
Definitely not, as will become clearer in the sequel the solution must be
looked for in peculiar properties of inner symmetries of chiral models; under
special circumstances their inner symmetry space can support a Poincar\'{e}
group symmetry\footnote{The DHR superselection sector theory derived from
representation theory of local observables \cite{Haag} leads to compact
internal symmetry groups for $D\geq4$ but does exclude noncompact
("nonrational") structures in chiral models.}. This will become more explicit
in the setting of the dual model below.

All these calculation have been done a long time ago \cite{Lowe}\cite{Mar} but
there is a kind of schizophrenic gap between the correct calculation leading
to the pointlike nature of the resulting infinite component field on the one
hand and the metaphoric claim that the objects are extended "wiggling little
strings" (with the occasional admission that they are "invisible" apart from a
point). Not wanting to interrupt our presentation of ST and the dual model by
sociological explanations we desist from further comments.

For a scientific (partial) understanding of what may have been on the mind of
obviously intelligent young particle physicists it is necessary to look at
ST's predecessor which is the \textit{dual model}. This model as well as its
ST extension is a concrete model with explicitly given computational rules
which emerged from the legacy of the more abstract setting of the S-matrix
bootstrap. The latter was a framework of which the strong interactions
(actually described by QCD) were thought to be the unique solution. The
S-matrix bootstrap had incorporated one property which was missing in previous
S-matrix attempts namely the crossing property for scattering amplitudes. This
is perhaps the most subtle property in QFT concerning the particle-field
relation, and it is certainly not our intention to go in detail into its
conceptual setting for which we refer to \cite{founcor}.

In order to get a better understanding of crossing, Mandelstam proposed a
spectral representation for the elastic two particle scattering amplitude; but
neither did the understanding of crossing as a consequence of first principles
improve from it, nor was it possible to add the Mandelstam representation to
the various other rigorously derived spectral representations which one was
able to derive from the principles of QFT. It did however play an important
role in Veneziano's discovery of the dual model, which was followed with
extensions and improvements by several other particle physicists \cite{Vech}.
The physical motivation was purely phenomenological (related to Regge
trajectories); it was soon forgotten after new experiments disposed of it. But
the mathematical attraction of Veneziano's profound appearing observation on
the basis of pedestrian methods exerted a quite irresistible pull for many
workers. To many it looked like a deep observation whose physical realization
still had to be found.

The first operational presentation of the dual model arose in collaboration of
Fubini with various collaborators \cite{Fub}. Using the terminology of chiral
quantum field theory they showed how one can obtain the dual model from a
multi-component abelian chiral current. The important objects in this
construction are not the d-component currents themselves but rather the
exponential of their potentials.%

\begin{align}
&  \left[  j_{k}(x),j_{l}(y)\right]  =-\delta^{\prime}(x-y)\delta_{kl}%
,~~Q_{k}=\int j_{k}(x)dx,~\Phi_{k}(x)=\int_{-\infty}^{x}j_{k}(x)dx\\
~~  &  \psi_{\alpha}(x)=~"e^{i\alpha\cdot\Phi(x)}"\rightarrow
V(z,p)=e^{iP\cdot X(z)} \label{int}%
\end{align}

In the QFT setting the d-component vector $\alpha$ in the exponential of the
d-component potential $\Phi_{k}(x)$ is the multi-component charge carried by
the field $\psi_{\alpha}(x)$ which applied to the vacuum generates a state in
a new representation sector for the current algebras. Since the $\alpha$
spectrum is continuous, there is a non denumerable number of superselection
sectors and the direct integral Hilbert space is nonseperable. The indicator
for the emergence of new superselection sectors above the vacuum sector is the
infrared divergence\footnote{Infrared divergencies cannot be handled by
renormalization theory, their appearance points to basic physical changes, in
the case at hand that the charged fields $\psi$ are not composite fields in
the current in the vacuum sector.} in the integral representing the potential.
The first time this problem arose in QFT was in a paper by Pacual Jordan
entitled the "neutrino theory of light" which caused quite a confusion because
in reality it was about chiral bosonization and re-fermionization which in the
setting of states (Jordan only used commutators) shows the mentioned infrared
problem \cite{Jor}.

The mathematically and conceptional clearest presentation for a one component
current theory can be found in \cite{BMT}. If one wants to use chiral models
as a theoretical laboratory it is more interesting to convert this
"non-rational" chiral theory into a "rational" model; this done by enlarging
the algebra by declaring certain (those which commute for finite distances) as
members of the enlarged observable algebra. The maximally extended observable
algebras can be classified and are "rational" i,e, have a finite number of
charge sectors \cite{BMT}. This method can be extended to multi-component
currents \cite{Stas}\cite{Longo} where the maximal extended observables are
classified by even lattices and their superselected charge sectors correspond
to their dual lattices. The list of models contains some examples of high
mathematical attraction related to finite exceptional groups the moonshine.

Having spent some time on the QFT use of multi-component chiral current models
makes it easier to appreciate the way the dual model uses the multi-current
model. In that case the use of this formalism for setting up an operator
version is slightly different. By declaring the null-mode of the chiral
conformal potential to define an d-component quantum mechanics the infrared
divergence is removed and its exponential function is now acting in the
quantum mechanically extended vacuum sector which is the only one i.e. the
superselection structure including its maximal extension has been lost. One
now has the following re-interpretation%
\begin{equation}
\alpha\rightarrow p,\text{ }\dim\psi_{\alpha}=\alpha^{2}\rightarrow m^{2}%
\end{equation}

The d-component charge passes to the d-component particle momentum and the
dimension of the charge-carrying field goes to one of the masses in the
mass/spin tower of the dual model. The loss of the continuum of superselection
sectors is in accord with the fact that momenta can be superimposed.

In fact Mack's observation \cite{Mack} that any conformal theory in any
spacetime dimension always leads (through is Mellin transform) to a dual model
in the sense of a meromorphic function of Mandelstam variables with a pole
structure of a dual model with the correct factorization of the residua coming
from the locality of the conformal theory. Mack's most astonishing result is
based on the fact that the cause for the appearance of "particle poles" in the
Mellin transform is the Euclidean conformal operator product expansion
(OPE)\footnote{As a result of the covering structure of the physical conformal
group, the global conformal operator expansions in real time are more subtle
than their euclidean counterpart \cite{SSV}; they lead in fact to the
conformal block decompositions which reveal an algebraic commutation structure
in the timelike Huygens region which replaces the commutativity of
observables.}, which is a partially resummed Wilson expansion. It has its
rigorous backup in the \textit{conformal partial wave expansion} which also
establishes its convergence.

In more detail, the correct dual model factorization of the residua of the
particle poles in the Mellin transformed correlation function arises from the
conformal operator product expansion inside a conformal correlation function.
Mack shows that one can also use the dual model properties in order to define
a conformal QFT; in this inversion that fact that the field theoretic
positivity simplifies (only two-point function of fields have to be positive)
due to the validity of OPE is important. He also stresses that the dual model
does not depend on the spacetime dimension of the conformal QFT from which it
was constructed. On the other hand the recent proof of the crossing property
has shown that the latter has a completely different origin in which
scattering theory and not conformal structures play the main role

After the experimental situation went against Regge phenomenology and its
sophisticated extension by the dual model, the correspondence between dual
models and CFT via Mellin transformation also removed a good part of the
mathematical mystery. It will be shown later on the crossing property of
scattering amplitudes and formfactors in QFT has a totally different
conceptual origin than the duality of the dual model in particular duality is
not a special of the crossing property.

Whereas higher dimensional CFTs leads via the Mellin formalism to the duality
relations in terms of Mandelstam variable, it does not possess the d-component
continuous charge $\alpha$ of chiral current models which in the operator
presentation of the dual model is re-interpreted as the momentum $p.$ Although
this is not needed from the strict dual model viewpoint, the requirement of
"charge turning into momentum" is essential for the transition to ST. For this
passage one needs to construct a unitary positive energy representation of the
covering of the Poincar\'{e} group and, as in Wigner's approach one needs an
momentum space. The only step which causes a major conceptual headache is the
use of an inner symmetry space defined by the charge components of a chiral
current model. The reason is that the quantum origin of inner symmetries,
which finally was understood thanks to the Doplicher-Haag-Roberts theory of
superselection sectors \cite{Haag}, had revealed that there exist only
countably many superselection sectors on an observable algebra and the
associated symmetries are described by the action of compact groups on the
indices of fields.

The theorem excluding noncompact inner symmetry structures does not hold in
chiral conformal theories, and existence of the current model with its
continuous charge spectrum shows that. On the other hand all the interesting
chiral models (minimal models,...) have a compact internal structure in the
sense that they are rational (finite number of sectors), unlike the current
model which in this terminology would be "irrational". But the uncountable
nature of the charge sectors does not mean that (returning to the case at
hand) one can find a subspace on the linear space of all real charges on which
a unitary positive energy representation of the Poincar\'{e} can act. If
somebody would have asked me to bet on this, my answer would have been that
there is no chance. Well, statistically I would have been right because there
is precisely one case which I would have overlooked: the d=10 superstring,
which belongs to a representation in which the chiral current has also
spinorial indices in addition to vectorial ones.

The world of chiral CFT is full of surprises and one has to except that there
is a chiral objects whose internal symmetry structure mimics 10-dimensional
spacetime behavior. But should one turn a chiral mimicry into a metaphor for a
10 dimensional living space which then becomes the starting point of curling
away unwanted dimensions in order to explain our living space? All the
properties discovered in ST, i.e. the finitely many different versions of the
10 dim. superstring and the possible interrelation by the big Latin letter M
are clearly properties of a particular multicomponent chiral current model and
should be understood right there. To draw conclusions about physical spacetime
in the form of "M theory" carries particle physics into the muddy water of
metaphors and a kind of new age particle physics in that one cannot prove that
it is wrong, but one does not have the impression that it can right either..

But there is one aspect which actually led to a wrong conclusion. This is the
claim that chiral theory associated with a multicomponent current can be used
to define an embedding of the chiral current theory (the source theory) into
the 10 dimensional inner symmetry space re-interpreted as a unitary positive
energy representation space of the covering of the Poincar\'{e} group (the
target space) defines a string in target space (an extended string-localized
object). This metaphoric but unfortunately incorrect idea results from
picturing the embedding as a map of the 1-dim. chiral theory into a 1-dim.
object geometrically embedded into the 10 dim. target space. What really
happens is that apart from the zero mode, the oscillator degrees of freedom
resulting from the Fourier decomposition of the chiral conformal QFT in the
circular compactification form the infinite components of

The remaining problem is to characterize such a subspace. But this is
precisely what the $u$-intertwiner accomplish.

In the case of the infinite dimensional setting of the N-G model there two
conditions which these oscillators have to obey: the string boundary
conditions and the reparametrization invariance condition. The corresponding
quantum requirements are well-known. In the present tensor product setting
they mix the momentum of the sought object with its "internal" quantum
mechanical degrees of freedom and in this way one gets to the physical states.
The two conditions provide the additional knowledge for a master-intertwiner
which intertwines between the original covariant transformation law and a
positive metric subspace $H_{sub}$ on which the representation is
semi-unitary. As mentioned the formation of a factor space $\hat{H}_{sub}$
leads to a bona fide unitary representation.

Since all unitary representations are completely reducible and a free field in
a positive energy representation is fully determined by its two-point
function, it is clear that the resulting object is an infinite components
pointlike field and not a string in physical spacetime. The string has not
disappeared, but contrary to naive expectations it did not become embedded in
spacetime but rather is encoded into the mass/spin spectrum as well as in the
irreducible component $u^{(A\dot{B},s)}$ intertwiners. (more below).

If one wants to cling to a geometric picture of a string, it should be a
"vertical" string in the highly reducible "little Hilbert space" which is the
inner Hilbert space over each point on which the oscillator variables act.
Those transformations in the oscillator space $H_{QM}$ which leave the
subspace invariant and do not implement Poincar\'{e} transformation, mix the
irreducible components of the infinite component field. This corresponds to
the "wiggling" of the "inner" string, a totally unphysical process which
corresponds to the flipping of the masses and spins between different
multiplets. It is not the string in spacetime which wiggles but rather the
particles in the tower over one spacetime point; they wiggle in the sense as
the oscillator interpretation of pointlike fields always did to those who
think that this is an appropriate visualization of free fields.

There is full agreement with the computation in the cited papers, but there is
also total disagreement if it comes to interpretation. Whereas for string
theorists it is a spacetime-embedded string of which only its center of mass
is visible, our derivation shows that it is really the construction of an
infinite component wave function and the localization point is not the center
of mass of anything.

Results of calculations often need the help of interpretation. The most famous
illustration is the Lorentz-Einstein disagreement in the interpretation of the
formula for the Lorentz interpretations with Einstein%
%TCIMACRO{\U{b4}}%
%BeginExpansion
\'{}%
%EndExpansion
s radical ether eliminating proposal coming out as favoured by nature. Since
ST originated as a mind game, there does not seem to be any role for nature to
play. But in both cases the correct interpretation is selected by the
application of Ockham%
%TCIMACRO{\U{b4}}%
%BeginExpansion
\'{}%
%EndExpansion
s razor; in the ether-relativity interpretation it is the superfluous
invisible ether, whereas in ST the invisible string which has to conk out.

Another comparison within quantum theory designed to highlight the magnitude
of the conceptual disaster is to create an analogy between the illusionary
invisible strings and a (hypothetical) attempt to attribute a position as well
as a momentum on the basis of the integrand in Feynman%
%TCIMACRO{\U{b4}}%
%BeginExpansion
\'{}%
%EndExpansion
s path integral. This actually may have happened if Feynman%
%TCIMACRO{\U{b4}}%
%BeginExpansion
\'{}%
%EndExpansion
s path integral representation would have been discovered before Heisenberg%
%TCIMACRO{\U{b4}}%
%BeginExpansion
\'{}%
%EndExpansion
s matrix mechanics. The nonexistence of a material string in spacetime has its
exact counterpart in the impossibility to embed a chiral conformal field
theory as a stringy-localized subalgebra into a higher dimensional spacetime
in such a way that the latter is the "target space" of the former. This was
actually the interpretation of the dual resonance model before it became
re-packed into the Lagrangian setting of ST. The target space interpretation
with extra dimensions only generates the spacetime dimension but does not
embed the chiral theory into this target space. Analogous to the spin degrees
of freedom the degrees of the chiral theory go into the "inner" Hilbert space
over each spacetime point and there is no physical interaction which is
capable of transforming this into a spacetime string.

The above argument about the canonically quantized bilinear bosonic Nambu-Goto
Lagrangian has an analog for the various 10-dimensional superstrings. The main
difference to the Nambu-Goto case is the undotted/dotted spinorial valuedness
of the oscillator variables which replaces the vector valuedness.

Staying for a moment with the chiral embedding picture of ST, it is
interesting to note that this way of looking at ST helps to lend plausibility
to restriction of the embedding construction to 26 respectively 10 spacetime
dimensions. The target space of the chiral conformal field theory is (by
definition) the finite dimensional vector/spinor valued index space of the
multicomponent conformal current (whose circular Fourier decomposition lead to
the oscillators). With other words the target space is a visualization of the
internal symmetry space of the source theory. In higher dimensional QFT the
nature of internal symmetries can be derived from the localization structure
of the observables \cite{Haag}. The DHR theory, which leads to compact group
symmetries, is however not valid in d=1+1 so one should be prepared for
exceptions with noncompact inner symmetries. The fact that within the family
of chiral current models this only occurs with 26 respectively 10
supersymmetric components leading to d=25+1 and d=9+1 Lorentz symmetry in
target space has to be proven. The intuitive argument only reveals that its
existence goes against naive expectations. This stands out against two other
facts which harmonize with naive expectations, genuine spacetime strings exist
in every dimension $d>1+2$ (see next section) and a more intrinsic
quantization of the Nambu-Goto string (in the original square root
formulation) which is based on the fact that this system is completely
integrable \cite{Po} also does not lead to restrictions of spacetime
dimensions. It is only the string theorists strange use of this Lagrangian as
a vehicle to generate spacetime from the inner symmetry of a multicomponent
chiral current (the infinite oscillators on a circle) which leads to this
restriction. This picture and its consequences is identical to the
source-target embedding of its dual model predecessor. A positive energy
representation of the Nambu-Goto algebra in a Hilbert space is indeed unique
and equal to the 10-dimensional superstring representation, just as the string
theorists claim, but it is point- and not string-like. One can even be
surprized about the uniqueness of the answer to the target space
interpretation because the general expectation is that the inner symmetry
space of a QFT cannot be the localization spacetime of a higher dimensional
QFT\footnote{In dimension
%TCIMACRO{\TEXTsymbol{>} }%
%BeginExpansion
$>$
%EndExpansion
2+1 there is a theorem that the inner symmetry space of a QFT is always the
representation space of a compact group \cite{D-R}.}, but this theorem does
not hold in chiral theories. Although there is no proof that in chiral
theories one cannot represent a noncompact group as the Poincar\'{e} group the
one and only one case of the pointlike infinite component superstring field is
the only known counterexample. The problems starts if one wants to transfigure
the component space of a chiral model into our living spacetime after
descending from 10 to 4 dimensions by "curling up" 6 of the spatial
dimensions\footnote{The curling up generates a problem of its own: it does not
get rid of the degrees of freedom which are by far too many for a physical
4-dimensional QFT.}.

Comments about the introduction of adding interactions to infinite component
wave functions will be deferred to later.

The formula just describes the most general pointlike field in d=1+3. There
are corresponding formulas for every dimension but the number of Casimir
invariants increase with increasing spacetime dimensions for both the massive
as well as the massless components. Both, the unitary representation of the
Nambu-Goto string as well as its supersymmetric analog are pointlike
\cite{Dimock}; in fact a unitary representation of the covering of the
Poincare group which does not contain an irreducible infinite spin component
is automatically pointlike generated and this is inherited by the associated
free field. The application of the oscillator algebra to a pointlike string
state leaves it pointlike but changes the system of $c$-parameters i.e. the
application of the oscillator algebra does not change the localization but
modifies the relative admixture of component contributions beyond what is
already done by the action of the Lorentz group.

The question of whether all admissible pairs $(A,\dot{B};s)$ occur, and if not
which ones do appear can only be settled by a more detailed computation which
for obvious reasons we are not prepared to do since neither infinite component
wave functions not its ST metaphor justify such additional work.

An infinite component free field theory is still a QFT and hence the
introduction of interactions should follow the standard logic of coupling
pointlike fields. But instead the rules to deal with interactions rely on the
metaphor of a spacetime extended string. Indeed, the graphical tube rules,
which define the ST interaction as a string-like analog of the Feynman rules
for point-like objects, only make physical-intuitive sense for genuine
spacetime strings. With this not being the case, the tube rules loose their
physical support and the question arises whether the transition matrix
elements calculated according to those metaphoric rules fulfill the physical
properties of an S-matrix of particle scattering, as unitarity and
macrocausality \cite{crisis}. In the case of Feynman rules this is possible
even without using an operator description. For the tube rules this has not
been done; there exists not even a credible perturbative (in the genus
associated with the combining and splitting tubes) proof of unitarity.

The senseless metaphor of a spacetime string together with the rules for
interactions which precisely rely on the presence of such strings looks to me
like a particle physics analog of what J. Heller in his famous novel with the
same title calls a "catch 22" situation, a situation which offers no way out;
the quantization gives an infinite component pointlike field, but this is not
what one needs for implementing interactions in terms of worldsheets. This
mismatch explains why in the more than 40 year history of ST it was not
possible to find an operational definition of interacting strings. It is not
only in the banking sector where people deal with metaphoric bubbles.

\section{Genuinely string-localized objects in spacetime and gauge theories}

The localization concept of QM referred to as "Born-localization" is based on
the Born probability-density associated with x-space Schroedinger wave
functions. It is directly related to the spectral decomposition of the
Hermitian position operator; in fact the projectors $E(\mathcal{O})$ measure
directly the probability to find the particle (at fixed time) in the
\textit{spatial region} $\mathcal{O}$. This quantum mechanical localization
operator has a multiparticle generalization to (bosonic or fermionic) Fock
space, but it is incompatible with Poincar\'{e} covariance and causality
(propagation in a theory with a maximal velocity); there is no operator which
localizes in spacetime regions in a frame-independent way.

The problem to find a localization which is compatible with special relativity
and causality attracted Wigner's attention already in the early days of QFT;
after he discovered the first intrinsic classification of (noninteracting)
relativistic particles he hoped to find an autonomous path toward QFT via a
relativistic concept of localization. Here "intrinsic" or "autonomous" stands
for "properly quantum" i.e. without using any classical parallelism as one
does in quantizing a classical Lagrangian field theory. This is an important
point because any theory which claims to be more fundamental than its
predecessor, should be able to arrive at its main results without referring to
a less fundamental theory. To delegate the intrinsic understanding of the
localization underlying QFT to the one inherent in classical Lagrangians was
not acceptable from Wigner's foundational viewpoint

Together with his collaborator Newton, Wigner adjusted the Born localization
to the relativistically invariant inner product of relativistic wave
functions. In this way the violation of covariance and causality of the
"Born-Newton-Wigner" localization becomes manifest.

Of course Wigner knew that QFT comes with a relativistic and causal
localization which is inherent to pointlike quantum fields, but quantum fields
even within one specified model of QFT are highly nonunique. In fact the
existence of an ever increasing zoo of physically equivalent but different
looking free field equations during the 30s was his principle motivation for
the intrinsic representation theoretical approach over quantization methods
which created this zoo.

If one could find a unique covariant and causal localization on the level of
one particle states, then the functorial relation between causally localized
one particle subspaces and von Neumann subalgebras of the total Weyl (CCR) or
Dirac (CAR) algebra would secure (at least for theories without interactions)
an intrinsic new localization concept which is by construction independent of
which pointlike generating field one uses to describe the spacetime indexed
net of local algebras\footnote{The best way to think about the relation
between the net of spacetime indexed algebras and the infinite class of
pointlike field generators is that between the intrinsically defined geometry
and its infinitely many possibilities of its coordinatization.}. The fact that
for a given spin $s$ there are infinitely many admissible spinorial free
fields $\Phi^{(A,\dot{B})}$ and each one gives rise to infinitely many
Wick-ordered composites would be no problem as long as the causal localization
property itself is unique. Despite its impressive observational success,
Wigner always maintained a certain distance to QFT. The intrinsic localization
concept was only found in the last two decades long after Wigner's lifetime.
In the present context of representation theory it can be found in
\cite{B-G-L}\cite{M-S-Y}, and, in a more special context already in
\cite{Annals} and older publications cited therein.

Wigner's sceptical view about the imperfections of QFT resulting from a lack
of intrinsic understanding of its most central localization properties was
well justified because recent progress on existence proofs for a certain
nontrivial family of two-dimensional models uses modular localization
properties in an essential way \cite{founcor} and a mathematical control in
higher dimensional QFT is hardly conceivable without this new intrinsic
localization setting. Only if one is in the possession of sufficiently many
nontrivial models about which one has a mathematical control of their
existence and properties, one can claim that QFT can be called a theory on par
with all the other mature physical theories as mechanics, electrodynamics,
statistical mechanics and quantum mechanics. It is one of the main points of
this essay to show that \textit{modular localization} plays a crucial role in
this process.

A good understanding of the issue of localization also strengthens the
understanding of the content of the previous section; after learning that ST
leads to pointlike generated infinite component wave functions rather than to
vibrating strings in spacetime, it is interesting to know how genuine
string-localized objects really look like.

In view of the fact that even for such eminent physicists as Wigner the issue
of localization was fraught with problems and perils, the critique of ST and
the string community is obviously not a gleeful reprimand of a committed
conceptual error. What is really worrying and distressing is not so much the
error itself, but rather the absence of any profound criticism afterwards over
several generations. For almost four decades particle theory for a majority of
physicists consisted in executing calculational recipes, formal games with
functional integrals and occasionally using sophisticated geometry and
topology i.e. in activities which are not of much help in conceptual problems
related to localization.

This raises the question about the addressee of an article like this. Is it to
encourage those who know better but prefer to be silent (section 7), or to
loosen up the mind of the hardened string theorists, or is it to strengthen
the mind of the uncommitted sceptic? Even if it does not achieve anything of
this, it certainly will be helpful for future historians and philosophers
because it is part of their professional obligations to say something about
what was going on in particle theory in all these years and why even the
standard model and more specifically gauge theory went into stagnation shortly
after its impressive start.

The localization aspect of ST is the most accessible illustration its
misunderstanding caused by its geometrical appearance in the source target
relation as compared with its intrinsic physical meaning. There are many other
such misleading metaphors obtained from naively identifying geometry with
localization. One which is very close to that of ST is the idea that one can
embed a lower dimensional QFT into a higher dimensional one. This is not
possible, but what one can do is restrict a QFT on a spacetime manifold to a
submanifold. However if the submanifold contains the time axis (a "brane"),
the restricted theory has \textit{too many degrees of freedom} in order to
merit the name "physical", namely it contains as many as the unrestricted
\cite{Mack}; the naive idea that by using a subspace one only gets a fraction
of phase space degrees of freedom is a delusion, this can only happen if the
subspace does not contain a timelike line as for a null-surface (holographic
projection onto a horizon).

The \textit{geometric picture} of a string in terms of a multi-component
conformal field theory is (section) that of an embedding of an n-component
chiral theory into its n-dimensional component space (referred to as a target
space), which is certainly a string. But this is not what modular localization
reveals, rather those oscillatory degrees of freedom of the multicomponent
chiral current go into an \textit{infinite dimensional Hilbert space over one
localization point} and do not arrange themselves according according to the
geometric source-target idea. A theory of this kind is of course consistent
but ST is certainly a very misleading terminology for this state of affairs.
Any attempt to imitate Feynman rules by replacing word lines by word sheets
(of strings) may produce prescriptions for cooking up some mathematically
interesting functions, but those results can not be brought into the only form
which counts in a quantum theory, namely a perturbative approach in terms of
operators and states.

ST is by no means the only area in particle theory where geometry and modular
localization are at loggerheads. Closely related is the interpretation of the
Riemann surfaces, which result from the analytic continuation of chiral
theories on the lightray/circle, as the "living space" in the sense of
localization. The mathematical theory of Riemann surfaces does not specify how
it should be realized; if its refers to surfaces in an ambient space, a
distinguished subgroup of Fuchsian group or any other of the many possible
realizations is of no concern for a mathematician. But in the context of
chiral models it is important not to confuse the living space of a QFT with
its analytic continuation. For the case at hand this means that the chiral
model is in a KMS temperature state with respect to the conformal Hamiltonian.
The analyticity region is a torus and the boundary values taken on its two
cycles are the "living space" of two chiral theories which, apart from one
theory being at the dual temperature of the other, are the same theories i.e.
the situation is selfdual\footnote{This is a special case of the
Nelson-Symanzik duality for d=1+1 massive theories.
\par
{}}. The spacetime interpretation of this situation in the sense of a
worldsheet is a totally misleading metaphor. Again we are reminded that
localization is a much more holistic notion than geometry. In the case of the
torus, one cicle may be that of a the living space of a compactified chiral
theory, but the other creates a thermal aspect (a state with vacuum
polarization clouds). The simple picture of QM where particles occupy
pre-assigned energy levels is never valid in the presence of interactions.

Whereas geometry as a mathematical discipline does not care about how it is
concretely realized the geometrical aspects of modular localization in
spacetime has a very specific geometric content namely that which can be
encoded in subspaces (Reeh-Schlieder spaces) generated by operator subalgebras
acting onto the vacuum reference state. In other words the physically relevant
spacetime geometry and the symmetry group of the vacuum is contained in the
abstract positioning of certain subalgebras in a common Hilbert space and not
that which comes with classical theories.

The dominance of the geometric-mathematical point of view over the physical
intrinsic localization-based interpretation started at the time of the
Wess-Zumino-Witten-Novikov model. Its predecessors, the various prior
publications on the multicomponent Thirring models in the setting of current
algebras have a seemless relation to modern modular localization whereas the
functional W-Z-W-N action representation was invented because in those days
there was a prejudice that in order to really understand something, one has to
bring it into a geometric-topological form with the help of a functional
integral representation. This added a non-intrinsic topological element which
may is interesting for mathematicians but did not add anything to QFT;
concrete calculations for these soluble models are still done in the
representation theoretic setting. In fact no chiral model has ever been
constructed under mathematical control with geometric functional integral method.

Another illustration of this point are the recent constructions for
factorizing models rely on modular localization \cite{Lech}; it would be hard
to imagine results of a non-metaphoric kind to come from the geometric path
integral setting. As most metaphoric tools their main purpose is to facilitate
communications between workers with different backgrounds and not model constructions.

Something similar happened with 3-dimensional plektonic models (models with
braidgroup statistics). They were geometrically visualized in terms of
Chern-Simons actions, and the messages that a euclidean geometric action has
to pass a sophisticated Osterwalder-Schrader positivity test before it can be
called a QFT was largely ignored. The correct representation theoretical way
would consists in the use the 1+2-dimensional Bargman extension of the Wigner
representation theory (\cite{Mu2} and previous papers of the same author cited
therein) which combined with modular localization leads to string-localized
algebras on the covering space whose generators fulfiill braid group
commutation relations. The generators of the wedge algebra are generalized
Wigner creation/annihilation operators\footnote{This step still needs to be
carried out. Since the raison d'etre for modified commutation is the
implementation of braid group statistics and not of interactions, the
resulting theory which has very nontrivial vacuum polarization is the QFT of a
"free" plekton (anyon).} whose braidgroup commutation relation play a similar
role as the operators of the Zamolodchikov-Faddeev algebra in d=1+1.

The modern version of the old representation theoretical methods is the
modular localization method which shows its true power in generalizations of
Ising like QFTs \cite{L-R}. By noting that the Zamolodchikov-Faddeev algebra
admits an interpretation in terms of modular localization in the region of a
wedge \cite{Annals}, the Karowski-Weiss bootstrap-formfactor program was
enriched by modular method which are powerful enough to secure for the first
time the mathematical existence of interacting nontrivial models with strictly
renormalizable short distance behavior \cite{Lech}.

In order to dispel the impression that I am a post-dicting fault-finder who
looks gleefully at blind alleys in the past struggle for the correct path in
particle theory, I should mention that I enthusiatically embraced the arrival
of that perfect blend of geometry, analysis and topology known as the
Atiyah-Singer index theory. Suddenly the connection between zero modes of
matter fields and winding numbers of gauge configurations which was observed
in certain exactly soluble two-dimensional models was understood \cite{A-S}.
After a couple of years my enthusiasm faded when I realized that this quantum
mechanical method is limited to Euclidean field theory and there was no easy
passage through the Osterwalder Schrader positivity requirement in order to
find its real time localization counterpart. Ideas around the A-S index theory
were popular in the 80s when they were used to classify euclidean gravity
models according to their topological anomalies. This direction of research
petered out at the end of the 80s.

In QM, where path integral representations had a solid mathematical basis, it
was difficult to use them outside of quasiclassical approximations. There was
a famous problem at that time posed by L. Schulman, namely to show why the
summation over all quasiclassical saddlepoint contributions of the rigid top
action (and generalizations beyond SO(3)) leads to the correct result. The
analog of this problem in finite dimensions was covered by the
Duistermaat-Heckmann theory. I spent almost two years on its
infinite-dimensional generalization before giving up. The message I drew from
this futile attempt was that if this does not work for geometric problems in
QM, there is a fortiori no chance for geometrical QFTs as d=1+1 sigma models
to work along those lines. The conclusion was that the best bet for actually
constructing models consists in the use of operator methods. In the case of
the rigid top the problem has a simple solution in terms of group
representations and in d=1+1 the idea of modular localization applied to the
family of factorizing models \cite{Annals} does secure their mathematical
existence \cite{Lech}.

The main point of the present criticism, namely the localization problem of ST
(section2) was noticed even by some string theorists, but "massaged away"
under the influence of the TOE ideology and as a result the string metaphor
was preserved. In more concrete terms, the commutator of the alleged
stringfield associated with the Nambu-Goto Lagrangian was correctly computed
and its pointlike field nature was noticed, yet the field was declared to be
an "invisible" spacetime string, apart from one point on the string identified
with the c.m. of the invisible string.

In this way the hard-won and proudly cherished conceptual autonomy of quantum
theory, which started with profound theoretical considerations and
Gedankenexperiments testing the philosophical range of the new QM by Bohr,
Heisenberg and others became eroded. Their old message was to abandon any
non-observable metaphoric relicts from classical thinking, but this has
withered away in the modern setting of ST. Even if this theory enters the
dustbin of history in the future, the metaphoric fallout of its discourse will
linger on for a long time to come. It would be a big loss if ST fades away
because it did not lead to observational consequences and not because it is
conceptually untenable.

Getting back to the history of localization, the modern causal localization
concept called "modular localization" would probably have pleased Wigner, but
unfortunately it only appeared in the late 90s. Its predecessor, the
(Tomita-Takesaki) "modular theory of operator algebras" arose in the middle
60s, the inspiring idea on the physical side came from the formulation of
thermal statistical mechanics for open systems \cite{H-H-W}. Less than a
decade later also the relation between the modular theory and wedge
localization was understood \cite{Bi-Wi} which soon afterwards led to a
fundamental understanding of the thermal aspects of the Hawking Unruh
localization behind black hole horizons. The spatial modular theory goes back
to some unpublished remarks of Longo and entered the work in \cite{Annals}%
\cite{B-G-L}\cite{F-S} and \cite{M-S-Y}. As mentioned before, together with
the modular reformulated bootstrap-formfactor program, it finally led to the
first mathematical constructions of strictly renormalizable models (models
whose short distance behavior is worse than that of free QFTs).

To give a\ detailed account of modular localization would go beyond the scope
of an essay. In the following we will explain some of the concepts in the
context of the simplest particle representation: a scalar massive particle. In
this case the Wigner representation of the Poincar\'{e} group acts as
follows:
\begin{align}
&  H_{Wig}=\left\{  \psi(p)|\int\left\vert \psi(p)\right\vert ^{2}\frac
{d^{3}p}{2p^{0}}<\infty\right\} \\
&  \left(  \mathfrak{u}_{Wig}(a,\Lambda)\psi\right)  (p)=e^{ipa}\psi
(\Lambda^{-1}p)\nonumber
\end{align}
We now define a subspace which, as we will see later on, consists of wave
function localized in a wedge. We take the standard $t-x$ wedge $W_{0}%
=(x>\left\vert t\right\vert ,~x,y$ arbitrary) and use the associated Lorentz
boost $\Lambda_{x-t}(\chi)\equiv\Lambda_{W_{0}}(\chi)$%
\begin{equation}
\Lambda_{W_{0}}(\chi):\left(
\begin{array}
[c]{c}%
t\\
z
\end{array}
\right)  \rightarrow\left(
\begin{array}
[c]{cc}%
\cosh\chi & -\sinh\chi\\
-\sinh\chi & \cosh\chi
\end{array}
\right)  \left(
\begin{array}
[c]{c}%
t\\
z
\end{array}
\right)
\end{equation}
which acts on $H_{Wig}$ as a unitary group of operators $\mathfrak{u}%
(\chi)\equiv$ $\mathfrak{u}(0,\Lambda_{z-t}(\chi))$ and the $x$-$t$ reflection
$j:$ ($x,t)\rightarrow(-x$,$-t)$ which, since it involves time reflection, is
implemented on Wigner wave functions by an anti-unitary operator
$\mathfrak{u}(j).$ One then forms the unbounded\footnote{The unboundedness of
the $\mathfrak{s}$ involution is of crucial importance in the encoding of
geometry into domain properties.} \textquotedblleft analytic
continuation\textquotedblright\ in the rapidity $U_{Wig}(\chi\rightarrow
-i\pi\chi)$ which represents an unbounded positive operators. Using a notation
which harmonizes with that of the modular theory in mathematics \cite{Su}, we
define the following operators in $H_{Wig}$
\begin{align}
&  \delta^{it}=U_{Wig}(\chi=-2\pi t)\equiv e^{-2\pi iK}\label{pol}\\
\mathfrak{\ }  &  \mathfrak{s}=\mathfrak{\ \mathfrak{j}}\delta^{\frac{1}{2}%
},\mathfrak{\mathfrak{j}}=U_{Wig}(j),~\delta=\delta^{it}|_{t=-i}\nonumber\\
&  ~\left(  \mathfrak{s}\psi\right)  (p)=\psi(-p)^{\ast}\nonumber
\end{align}
Since the anti-unitary operator $\mathfrak{j}$ is bounded, the domain of
$\mathfrak{s}$ consists of all vectors which are in the domain of
$\delta^{\frac{1}{2}}.$ With other words the domain is completely determined
in terms of Wigner representation theory of the connected part of the
Poincar\'{e} group.

In order to highlight the relation between the geometry of the Poincar\'{e}
group and the causal notion of localization, it is helpful to introduce the
real subspace of $H_{Wig}$ (the closure refers to closure with real scalar coefficients).%

\begin{align}
\mathfrak{K}  &  =\overline{\left\{  \psi|~\mathfrak{s}\psi=\psi\right\}
}\label{K}\\
dom\mathfrak{s~}\mathfrak{=K}  &  +i\mathfrak{K},~\overline{\mathfrak{K}%
+i\mathfrak{K}}=H_{Wig},~\mathfrak{K}\cap i\mathfrak{K}=0\nonumber
\end{align}

The reader who is not familiar with modular theory should notice that these
modular concepts are somewhat unusual and very specific for the important
physical concept of causal localization; the fact that despite their physical
significance they have not entered the general mathematical physics literature
and remain unknown outside a tiny group of theorists underlines the
observation that the mainstream of particle physics has favored geometry and
neglected quantum localization.

One usually thinks that an \textit{unbounded} anti-unitary involutive
($\mathfrak{s}^{2}=1$ on $dom\mathfrak{s}$) operator which has two real
eigenspace associated to the eigenvalues $\pm1$ as something very peculiar,
but its ample existence is the essence of causal localization in QFT.

The second line (\ref{K}) defines a property of an abstract real subspace
which is called \textit{standardness} and the existence of such a subspace is
synonymous with the existence of an abstract $\mathfrak{s~}$operator
affiliated with that subspace.

The important analytic characterization of modular wedge localization whose
mathematical origin is the existence of a dense domain $dom\mathfrak{s}$ for
the unbounded involution $\mathfrak{s}$ consists in the strip analyticity of
the wave function in the momentum space rapidity $p=m(ch\chi,p_{\perp}%
,sh\chi).$ The requirement that such a wave function must be in the domain of
the positive operator $\delta^{\frac{1}{2}}$ (the operator responsible for the
unboundedness) is equivalent to the analyticity of the wave function
$\psi(p_{\perp},\chi)$ $\in dom\mathfrak{s}$ in the strip $0<\chi<i\pi$
together with the action of $\mathfrak{s}$ (\ref{pol}) which relates the
particle wave function on the lower boundary of the strip which is associated
to the antiparticle wave function on the negative mass shell. It is easy to
see that the dense subspace of such wave functions equipped with the graph
norm of $\mathfrak{s}$ becomes a Hilbert space in its own right.

This relation of particle to antiparticle wave functions is the conceptual
germ from which, after generalization to the interacting setting, most
fundamental properties of QFT, such as crossing, existence of antiparticles,
TCP theorem, spin-statistics connection and the thermal manifestation of
localization originate. Apart from special cases this fully quantum
localization concept cannot be reduced to support properties of classical test functions.

More precisely the modular localization structure of the Wigner representation
theory "magically" preempts properties of a full QFT on the level of the
Wigner representation theory; this follows from the realization that
scattering theory permits to extend these one-particle properties to the
interacting QFTs \cite{Mu}. The crossing relation is one these properties
resulting from modular localization. This brings this property into a sharp
contrast with understanding of crossing behind the Veneziano duality; most of
the ideas which came out of the S-matrix setting starting in the 60s was not
only observationally without success, but also needs careful critical
conceptual attention.

The sharper than wedge localization in causally closed subregions of a wedge
(spacelike cones, compact double cones) is obtained by intersecting
$\mathfrak{K}$ spaces for wedges in different positions \cite{B-G-L}. These
intersected $\mathfrak{K}$ spaces are again related to modular $\mathfrak{s}$
operators (with no direct relations to the representation theory of the
Poincar\'{e} group). Among the 3 families of Wigner representations (massive,
finite spin massless, infinite spin massless) the infinite spin family has
trivial compact localization subspaces with the tightest localized nontrivial
subspaces being semiinfinite string-like localized (the singular limit of
spacelike cones). Such quantum matter cannot be generated by pointlike fields.
The case of zero mass finite helicity is also quite interesting since certain
covariant potentials, unlike in the massive case, do not exist as pointlike
but rather only as semiinfinite stringlike objects. We will return to this
observation in more detail below.

From the spatial modular localization setting of Wigner%
%TCIMACRO{\U{b4}}%
%BeginExpansion
\'{}%
%EndExpansion
s representation theory one can directly pass to the of interaction-free net
of local algebras by exploiting the functorial relation between real subspaces
and von Neumann subalgebras \cite{B-G-L}\cite{M-S-Y}. On can also use the
Wigner representation data to construct the covariant fields. According to the
previous section one only needs to determine the $u^{\left(  A\dot
{B},s\right)  ,i}(s;p,s_{3})$ intertwiners and their conjugates. This can be
either done by group theory (covariance) as in Weinberg \cite{Wei} or by using
modular localization. The second method also works in the case of string-like localization.

It is helpful to rewrite the above result about the oscillator-driven infinite
dimensional representation of the Poincar\'{e} group into the more
conventional setting of irreducible free fields.

For this purpose it is helpful to recall the form of the most general
covariant free field of mass $m>0$ and spin s
\begin{align}
\Psi^{(A\dot{B},s)}(x,m)  &  =\frac{1}{\left(  2\pi\right)  ^{\frac{3}{2}}%
}\int(e^{-ipx}\sum_{s_{3}=-s}^{s}u^{\left(  A\dot{B},s\right)  }%
(s;p,s_{3})a(s;p;s_{3})+h.c.)\frac{d^{3}p}{2\sqrt{\vec{p}^{2}+m^{2}}%
}\label{spinorial}\\
&  \left[  \Psi^{(A\dot{B},s)}(x),\Psi^{(A\dot{B},s)\ast}(y)\right]
_{grad}=i\mathbf{\Delta}^{(A\dot{B},s)}(x-y;m)\nonumber
\end{align}
\ \ The meaning of the notation is as follows. The $A,\dot{B}~$are the Casimir
values which characterize the irreducible $\left(  2A+1\right)  $ component
undotted respectively the $(2\dot{B}+1)$ component dotted spinor
representations. For a given physical spin s there is an infinite supply of
covariant fields. They are described by admissible triples $(A,\dot{B};s)$
which are characterized by the inequality $\left\vert A-\dot{B}\right\vert
\leq s\leq A+\dot{B}.$ For a given such triple there is precisely one $\left(
2A+1\right)  (2\dot{B}+1)$ dimensional irreducible representation of the
two-fold covering $\widetilde{O(1,3)}$ of the Lorentz group but for a fixed
$s$ there is a denumerable infinity of spinorial descriptions i.e. free fields
which all belong to the same Wigner representation and therefore share the
same Wigner creation (annihilation) operators $a^{\ast}(s;p;s_{3});$ their
only difference is in the $u^{\left(  A\dot{B},s\right)  }$ system of
intertwiners and their conjugates (the $v$-intertwiners) \cite{Wei}. The
$\left(  2A+1\right)  (2\dot{B}+1)$ column indices of the intertwiners have
been suppressed. Needless to add that these objects are explicitly known. They
define the pointlike covariant wave functions. The (graded) commutator of two
covariant fields is a covariant polynomial in spacetime derivatives acting on
the spinless $\Delta(x-y)$ commutator function. All these objects have been
computed in the literature \cite{Wigh}\cite{Tod} in terms of known functions.

Besides the massive family there are two irreducible massless representation
namely the zero mass \textit{finite helicity} representation ($m=0,s$) and the
zero mass "\textit{infinite spin}" \ representations ($m=0,\kappa$) whose spin
analog is a continuous Casimir parameter $\kappa.$ The infinite dimensional
unitary representation of (the covering of the) Poincar\'{e} group, leading to
the infinite dimensional wave function of ST in the previous section, does not
contain this third kind of massless representation and hence is
point-localized and leads to pointlike free fields.

The covariant form of the zero mass finite helicity representations lead also
to a undotted/dotted spinorial calculus as above, however there is a small but
important distinction; instead of the full range of admissible ($A,\dot{B}$)
values one is only left with spinorial representations with $\left\vert
A-\dot{B}\right\vert =s$ i.e. the physical spin equals the absolute value of
the difference between the dimension of the undotted and dotted "Lorentz
spins". This considerably reduces the possibilities, but there remains still
an infinity of covariant fields for a fixed $s$. In the case $s=1$ the field
strength tensor is associated with (m=0, s=1) is an allowed covariant
pointlike field, but there is no pointlike covariant vectorpotential
consistent with the unitary representation theory of the Poincar\'{e} group.
It turns out that covariant vectorpotentials do exist if one lifts the
pointlike restriction in favor of semiinfinite stringlike localization. In
fact \textit{the full range of admissible possibilities of the massive case
can be restored in the massless family if one permits string localized
fields}\footnote{Modular localization never leads to finite strings as in ST,
semiinfinite spacelike strings appear because they are the core of the
simplest noncompact simply connected causally complete regions (spacelike
cones) as points are the core of the simplest compact simply connected
causally complete. This has nothing in common with closed or open wiggling
strings of ST.}. It turns out that for $(m=0,s\geq1)$ the covariant fields
with the lowest short distance dimension are always semiinfinite string-localized.

In the interacting case there is no such functorial relation, however in this
case the validity of the modular theory for wedges (the Bisognano-Wichmann
property) can be related to free case by invoking scattering theory \cite{Mu}.
The fact that the noncompact wedge algebra still permits affiliated operators
which applied to the vacuum create vacuum-polarization-free (PFG) states is of
crucial importance in recent existence proofs for factorizing models
\cite{Annals}\cite{Lech}.

Modular localization theory reveals that, contrary to the quantum mechanical
Born-Newton-Wigner localization which is related to position operators and a
family of space-dependent projectors $E(\mathcal{O}),$ the localization
subspace in the Hilbert space $H$ of a QFT is defined as the dense domain of
the algebraic modular (Tomita) involution $S~$associated with a "standard"
pair ($\mathcal{A(O}),\Omega$) where $S$ is defined as%
\begin{align*}
SA\Omega &  =A^{\ast}\Omega,\text{ }A\in\mathcal{A(O})\\
H(\mathcal{O})  &  \equiv domS=K(\mathcal{O})+iK(\mathcal{O}),\text{
}\overline{H(\mathcal{O})}=H
\end{align*}
Here the real subspaces $K(\mathcal{O}),~iK(\mathcal{O})$ are the $\pm$
eigenspaces of the Tomita $S$-operator. For an interaction-free theory the
$K(\mathcal{O})$ and their complexification are Fock space generalizations of
the subspaces (\ref{K}) of the Wigner representation space. It is not our
intention to explain the details of modular theory, the only point to which we
want to direct the attention of the reader is the that, contrary to the B-N-W
localization, which leads to bona fide subspaces and projectors, the
information of causal relativistic localization is encoded in \textit{dense
subspaces} $H(\mathcal{O})$ (or equivalently in the real subspace
$K(\mathcal{O})$) which change continuously with the spacetime region
$\mathcal{O~}$and are determined in terms of the representation of the
Poincar\'{e} group i.e. all models with the same particle content have the
same spacetime indexed net of dense subspaces $H(\mathcal{O}).$ The Tomita
$S$-operator contains more detailed information about the interaction. For the
case of the wedge region $\mathcal{O}=W$ the polar decomposition of $S$
contains the $S_{scat}$-matrix%
\[
S=J\Delta^{\frac{1}{2}},~J=J_{in}S_{scat}%
\]
Here $\Delta^{it}$ is (up to a rescaling) the unitary representation of the
W-preserving boost group, $J$ is the anti-unitary modular reflection operator
which transforms the algebra $\mathcal{A}(W)$ into its commutant
$\mathcal{A}(W)^{\prime}=\mathcal{A}(W^{\prime})$\footnote{The equality
between the commutant and the geometric opposite algebra follows from the
proof of the Bisognano-Wichmann theorem \cite{Mu}
\par
.}$.$ $J~$can be decomposed into the anti-unitary reflection operator of the
incoming interaction-free situation $J_{in}$ and the unitary scattering matrix
$S_{scat}$. Hence modular theory does not only define the intrinsic content of
causal localization but it also attributes a hitherto unknown role to the
scattering matrix: $S_{scat}$ is a relative modular invariant between the
interacting and the free wedge algebra; with other words $S_{scat}$ is not
only that global object which connects with QFT through the asymptotic LSZ
limit but it is also in a very deep way related to the semi-local wedge
algebra. In fact the recent existence proofs for factorizing models, which are
the first existence proofs for strictly renormalizable models in the history
of QFT, depend precisely on this modular. Such concepts and mathematical
objects one does not meet anywhere in QM. These contrasts can be traced back
to the local algebras: whereas local algebras in QM are always of the same
type as the global algebra namely type I factors, the local algebras of QFT
are all equivalent to the unique hyperfinite type III$_{1}$ factor (the
monad). This leads to the extraordinary result that the full content of a QFT
can be encoded into the positioning of a finite number of copies of the monad
into one Hilbert space, a potentially powerful new structural property of QFT
whose exploration is only beginning.

But the fact that the B-N-W localization violates covariance and causality
does not mean that it is of no use in QFT. For asymptotically large timelike
separation of B-N-W localization events the relation becomes causal and
covariant and this asymptotic covariance is sufficient to prove the
Poincar\'{e} invariance of the S-matrix; last and not least the cross
section\footnote{In fact Born in his famous paper did not link the probability
interpretation of QM with the absolute square of the Schroedinger wave
function but rather with the cross section as it arose in the Born
approximation.} is a probability and it would be a disaster for relativistic
QFT if B-N-W would not be at least asymptotically invariant ("effectively"
outside a Compton wave length).

For the rest of this section we will return to the matter of our principle
concern: string-like localization.

In a theory which is generated by pointlike fields it is always possible to
introduce string-like localized generators. There are two cogent reasons for
doing this. One is that for each pointlike covariant free field $\Psi(x)$ one
can construct an associated semiinfinite string-like localized spinorial
fields $\Psi(x,e)$ with%
\begin{align}
&  U(a,\Lambda)\Psi(x,e)U^{\ast}(a,\Lambda)=D(\Lambda^{-1})\Psi(\Lambda
x+a,\Lambda e)\label{string}\\
&  \left[  \Psi(x,e),\Psi(x^{\prime},e^{\prime})\right]  _{\operatorname{grad}%
}=0,\ ~x+\mathbb{R}_{+}e><x^{\prime}+\mathbb{R}_{+}e^{\prime}\nonumber
\end{align}
The first line expresses the covariant covariant transformation property of
the string-like free field which lives in the same Wigner-Fock space as the
pointlike field. The second line justifies the name "string-like localized"
since the graded commutator only vanishes if all points of the two linear
strings which start at x and x' are spacelike relative to all points on the
second string. In this way the string become visible if one enters the causal
dependenc region of the other; this kind of causal visibility is part of the definition.

The first cogent reason for passing from $\Psi(x)$ to $\Psi(x,e)$ is that the
string-localized counterparts have better short distance properties: instead
of the well-known increase of the short distance dimension (sdd) with spin
$s$, the string localized $\Psi(x,e)$ can be chosen such that%
\begin{equation}
sdd\Psi(x,e)=1~\forall~s
\end{equation}
which is the formal power counting prerequisite for renormalizability up to
quadrilinear interaction terms. Renormalizability is however more than power
counting. The standard approach uses properties of relative pointlike
localization in an essential way; this is borne out in the formulation of
Epstein and Glaser where the parametrization of the freedom in passing to the
next perturbative order depends on the pointlike localization in an essential
way. Hence one needs a very nontrivial adjustment to the new string
localization; this is presently being investigated. A E-G approach to string
like fields would certainly enlarge the range of renormalizable models;
whereas before the renormalizability stopped at s=1, the power-counting
requirement for renormalization is valid for all s.

Another cogent reason arises in connection with zero mass finite helicity
representions. In that case, as was briefly mentioned in the previous section,
the admissible covariant fields are severely restricted, namely instead of the
inequality between the formal spins $A,\dot{B}$ and $s$ one has the more
restrictive equality $s=\left\vert A-\dot{B}\right\vert $ \cite{Wei}$.$ This
restriction prohibits pointlike covariant vectorpotential for photons and
gluons and metric $g_{\mu\nu}$ tensors for s=2 (gravitons) and more general
pointlike objects for all ($m=0,s\geq1$) representations. The full range of
possibilities of spinorial descriptions is however restored if one allows
\textit{semiinfinite string localization}.

Among the string-localized fields with sdd=1 there are in particular the s=1
vectorpotential $A_{\mu}(x,e),$ and the s=2 tensorpotential $g_{\mu\nu}(x,e)$
which are linearly related to the pointlike field strength and respectively
linearized Riemann tensor. For the purpose of a compact terminology let us
call the massless pointlike localized fields with $s=\left\vert A-\dot
{B}\right\vert $ "field strengths" and the remaining string-localized
spinorial fields which saturate the inequalities (\ref{spinorial})
"potentials". \ So free potentials are always string-localized and by
appropriately differentiating potentials one obtains pointlike free field strengths.

Already in the absence of interactions the observable algebra generated by the
field strength has a subtle structural property which distinguishes it from
the corresponding free massive algebras and clearly points towards
string-localized objects. In the massive case the algebras fulfill the
unrestricted Haag duality namely the property that the local fields which
commute\footnote{For Fermions there is a formulation in terms of graded
commutators.} with the fields inside a causally complete region are precisely
those fields and their composites which are localized in the causal complement%
\begin{equation}
\mathcal{A}(\mathcal{O})^{\prime}=\mathcal{A}(\mathcal{O}^{\prime})
\end{equation}
With the exception of regions $\mathcal{O}$ which consist of disconnected
parts, this Haag duality relation holds for all massive theories even, if
$\mathcal{O}$ is multiply connected. However for ($m=0,s\geq1$) this is only
true for simply connected regions\footnote{The violation of Haag duality for
spacelike separated disconnected spacetime regions is related to the existence
of charge transporters and leads to the theory of locally generated
superselection rules and the concept of inner symmetries.}; for spacetime
regions with multiple connectivity (e.g. a toroidal spacetime region) Haag
duality is violated%
\begin{align}
\mathcal{A(O})  &  \varsubsetneq\mathcal{A(O}^{\prime})^{\prime}\\
\mathfrak{K}(\mathcal{O})  &  \subsetneqq\mathfrak{K}(\mathcal{O}^{\prime
})^{\prime}\nonumber
\end{align}
where the second line contains the spatial modular version of the Haag duality
violation in which the upper dash on the real subspace $\mathfrak{K}$ stands
for the symplectic complement within the Wigner space.

In this case there are additional operators which, although they commute with
all field strength in the causal complement, and are therefore expected to be
objects localized in $\mathcal{O}$, cannot be generated by field strength
localized in $\mathcal{O}$. The explicit calculation was done in a famous (but
unfortunately unpublished) paper by Leyland, Roberts and Testard \cite{LRT}
for a free Maxwell field strength i.e. s=1. For $\mathcal{O}$ the authors
chose spacelike separated tori $\mathcal{T}_{i}$ i=1,2 and construct electric
and magnetic fluxes through orthogonal disks $D_{i}$ whose circular boundaries
are the cores of the $\mathcal{T}_{i}$ whose which are interpenetrating in
such a way that the spacelike separation is respected. The crucial point is to
show that the fluxes through D$_{i}$ can be chosen in such a way that their
electric and magnetic wave function fulfill
\begin{align*}
e_{i},h_{i}  &  \in\mathfrak{K}(\mathcal{T}_{i}^{\prime})^{\prime}\\
Im\left(  e_{1},h_{2}\right)   &  \neq0\neq Im(e_{2},h_{1})
\end{align*}
If there would exist a pointlike vectorpotential $A_{\mu}(x)$ then the
imaginary part would vanish and there could be no violation of Haag duality.
This argument is perhaps the strongest argument against the indiscriminate use
of pointlike vectorpotential which lives necessarily in an indefinite space in
which localization has no physical significance. A stringlike vectorpotential
avoids this contradiction \cite{M-S-Y}. causal completion of a torus obtained
by regularizing the boundary of a disk a circle. The above result does not
contradict the Stokes theorem but only the classical geometric idea that the
magnetic potential on the boundary is localized there. The absence of a
pointlike magnetic vectorpotential has no classical counterpart since it is a
result of the subtle clash between pointlike localization of a vectorpotential
and the principles of quantum theory; the use of the stringlike potential
shows that there is a consistent operational derivation of the above result in
terms of vectorpotentials.

A classical counterpart (classical only with respect to the electromagnetic
field) of this phenomenon is the Aharonov-Bohm effect to be more precise we
will consider the violation of Haag duality as the quantum
\textit{Aharonov-Bohm effect} \cite{charge1}). As in most cases of passing
from the classical to the quantum realm, the quantum relation is more subtle
than its classical counterpart There can be no doubt that for s%
%TCIMACRO{\TEXTsymbol{>}}%
%BeginExpansion
$>$%
%EndExpansion
1 there will be extended A-B effects. Since the distance between the dimension
of the best (in the short distance sense) field strength and the sdd=1 of the
best potential increases with s there arises the curious question wether the
increase of the possible potentials with s could lead to a increased violation
of Haag duality via fluxes through higher genus tori.

The potentially most promising application of these ideas is to find a
substitute or rather an extension of the method of gauge theory. The gauge
theory approach uses pointlike potentials and overcomes the representation
theoretical $~$No-Go theorem against the existence of covariant pointlike
vectorpotentials through sacrificing the rules of quantum theory by allowing
an indefinite metric state spaces in intermediate perturbative computational
steps. This problem of positivity and unitarity is absent in the classical
gauge theory of electrodynamics and hence despite all formal similarities the
conceptual meaning of the word "gauge" differs considerably. In the quantum
setting it leads to the introduction of auxiliary ghost degrees of freedom,
but apart from this more violation of the rules of quantum theory and the
necessity to relate gauge invariance with the necessarily more subtle return
(or creation) of a Hilbert space setting, the quantum version of the gauge
theory formalism follows its classical counterpart. However this is only true
for the local observables; the description of necessarily nonlocal quantum
electric charges is outside the gauge formalism. Their nonlocality is the
result of a subtle quantum phenomenon which has no counterpart in the
classical setting.

In fact the sharpest localization which is possible for electrically charged
particles is the semiinfinite stringlike localization, there are no compactly
localized charge carrying operators. Since the time of Jordan, Dirac and
Mandelstam the simplest\footnote{In the Coulomb and other representation the
em flux is not forced through an infinitely thin semiinfinite line but is
permitted to spread over larger part of spacetime.} formal string
representation of a charged field has the form%

\begin{align}
&  \Psi(x,e)=~"\psi(x)e^{\int_{0}^{\infty}ie_{el}A^{\mu}(x+\lambda e)d\lambda
}"\label{DJM}\\
&  \Psi(x,e)\overset{\alpha(\Lambda,a)}{\rightarrow}D(\Lambda^{-1}%
)\Psi(\Lambda(x+a),\Lambda e)\nonumber
\end{align}
where the second line is the automorphic action of the Poincar\'{e} group. It
is prohibitively difficult to lift the quotation marks i.e. to give a precise
meaning to this formula within the setting of renormalized perturbation
theory. Whereas the unphysical fields $\psi(x)$ and $A^{\mu}(x)$ (which only
have a "ghostly" existence) are part of the perturbative gauge formalism,
physical charged fields defined in terms of nonpolynomial functions in the
unphysical fields as (\ref{DJM}) are outside this formalism and have to be
defined in terms of a special and hardly manageable new formalism
\cite{Stein}. One may say that gauge invariance requires charged fields to
have noncompact stringlike localization as in (\ref{DJM}) but this is not very
revealing since localization is the raison d'\^{e}tre of QFT whereas gauge
invariance is a technical tool which allows to treat massless s=1 interactions
the with the same renormalization formalism as that used for pointlike s=0,1/2
fields. The prize to pay is precisely that important string-localized fields
are outside this formalism and one is confronted with the aforementioned
problems. Knowing that gauge theory maintains the known formalism at the
expense of the most important principle of QFT, the better strategy seems to
be to confront the stringlike localization right from the beginning without
any technical crutch which permits the standard setting of renormalizable
perturbation theory and to look for a new perturbative renormalization theory
which incorporates stringlike localized potentials. The incorrect handling of
the localization issue and the transgression of quantum theory by indefinite
metric spaces are two sides of the same coin; correcting the first
automatically resolves the second \cite{charge1}.

The observation by Bloch and Nordsieck \cite{B-N} that the scattering of
electrons in QED leads to more radical changes from standard scattering theory
in the infrared regime than the quantum mechanical Coulomb scattering was made
at the same time as (\ref{DJM}), but the question whether there is a relation
between the momentum space treatment of the emission of infinitely many
infrared photon in a collision of charged particles, and the possibly
semiinfinite stringlike localization of their charge-carrying fields was not asked.

Only in the 60s the first model studies on this problem begun. It was shown
that two-dimensional Lagrangian couplings between massless mesons and massive
fermions lead to solutions of the form \cite{infra}
\begin{equation}
\psi(x)=\psi_{0}(x)e^{ig\int_{x}^{\infty}j(y)dy}%
\end{equation}
where $\psi_{0}$ is a massive free Fermi-field and $j$ a chiral current of a
massless Fermi-field. This leads to a an analytic momentum space behavior
similar to the one observed in \cite{B-N}\cite{YFS} so the notion of
"infraparticle" i.e. a particle-like entity which is inexorably tied to its
infrared photon cloud and string-localized charged fields appeared as two
manifestations of the same phenomenon.

Since in d=1+3 QED there is presently no mathematical control on the level of
operators, in particular on string-localized charged operator, one must rely
on structural arguments which disprove the possibility that there can be
compactly localizable generators of charged fields and in this way confirm
string-localization as the sharpest possible one. Such a rigorous structural
argument follows from a quantum adaptation of Gauss law which also leads to
two additional interrelated results namely that the Lorentzgroup on charged
states is spontaneausly broken and that the velocity direction of the
outgoing/incoming particles define superselection rules \cite{Haag}.

But these structural results unfortunately did not lead to a modified
perturbation theory for electrically charged fields. The actual computations
are still made in the same gauge theory setting as that used more than half a
century ago: a formally pointlike vectorpotential in an indefinite metric
space and a gauge invariant physical space of local observables which does not
contain the string-localized electrically charged objects. The gauge invariant
observables correspond in the new setting based on interactions of matter with
string-localized vectorpotentials are those which are independent of e; since
the e-dependence only enters through $A_{\mu}(x,e)$ the criterionj for
pointlike localization is invariance under the change
\begin{equation}
A_{\mu}(x,e)\rightarrow A_{\mu}(x,e^{\prime})=A_{\mu}(x,e)+\partial_{\mu}%
\Phi(x,e,e^{\prime}) \label{sub}%
\end{equation}
As expected from the gauge theoretic setting, all correlation functions
involving products of chargeless operators\footnote{The e-dependence of
external lines in Feynman graphs for time-ordered correlations is harmless
since it can be removed by passing to field strength which does not change the
physical content.} are independent of $e$ i.e. they are point- instead of string-localized.

In fact the string-localized potential $A_{\mu}(x,e)$ looks, apart from the
fact that a gauge parameter does not change under Lorentz transformations,
precisely like the axial gauge potential in that it obeys identical relations
$e^{\mu}A_{\mu}=0=\partial^{\mu}A_{\mu}.$ The difference is mainly one of
interpretation: $e$ is not a gauge parameter but as $x,$ a spacetime
coordinate (a point in a 3-dim. de Sitter space) in which the field
fluctuates. Although $sdd_{x}=1$ it has reached this lowest possible dimension
only by deflecting part of the fluctuations into directional fluctuations
which are then perceived as infrared divergences. In fact the axial gauge was
never used for higher order calculations because in order to deal with these
infrared singularities one first has to understand the distributional nature
of stringlike fields in $e$ in order to appreciate how to deal with coalescing
$e^{\prime}s$ and this was not possible as long as they were considered as
gauge fixing parameters.

So the string theoretic alternative to the gauge theory setting should be
clear by now; one wants to trade the pointlike indefinite metric potential
together with its unphysical Gupta-Bleuler or BRST formalism with a coupling
of conserved matter currents to a string-localized vectorpotential $A_{\mu
}(x,e)$ in a Hilbert space. Besides reproducing the gauge invariant subalgebra
as the e-independent algebra of local observables, the main aim is to
understand the nonremovable stringlike localization of charged fields as
resulting from the interactions with stringlike potentials. Whereas the
stringlike nature of the latter has no bearing on the charge neutral subspace
of the physical Hilbert space\footnote{The cyclic application of $A_{\mu
}(x,e)$ and that of $F_{\mu\nu}(x)$ generates the same Hilbert space.}, the
more severe\footnote{In distinction to the potentials there is no linear
process which removes the strings, only by passing to charge neutral
composites this can be achieved.} stringlike nature of charge generators in
the perturbative conceptualization results from the interaction with the
stringlike vectorpotentials. In setting up a renormalization theory in which
stringlike potentials participate there, are 3 points to be observed.

\begin{itemize}
\item The power-counting requirement on the interaction polynomial
$sdd\mathcal{L}_{int}\leq4$ otherwise the number of undetermined
renormalization parameters will be infinite. But since for arbitrary spin $s$
there is always a string-localized covariant potentials with sdd$\Psi
=1,~$there are infinitely many couplings which fulfill the power-counting
requirement and thus are potential candidates for representing renormalizable QFTs

\item The iterative n$^{th}$ order step $n\rightarrow n+1$ must be such that
the resulting ambiguities are not worse than string localized (the string
adapted Epstein-Glaser requirement).

\item There must be pointlike local equivalence subclasses of fields i.e.
fields whose correlations functions are independent of e. In the simplest case
of QED this is achieved by the independence of their correlation functions
under change of $e$-direction. In case of QED this means that the correlation
functions of charge neutral fields must be independent under a change of $e$
which amounts to the substitution (\ref{sub})
\end{itemize}

At this point also the weakness of the geometric/topological point of view,
which rose to prominence in particular with gauge theory, becomes exposed. It
offers answers to questions about fibre bundles and related mathematical
subjects\footnote{The Seiberg-Witten view of gauge theory which led to
invariants bearing the same name as well as duality structures in gauge
theories are (semi)classical observations i.e. their local quantum physical
content is not known. The same holds for similar result about results about
euclidean gravity invariants.}, but does not shed light on quantum physical
questions in which localization plays an important role. Since fibre bundles
and other classical constructs associated with the Lagrangian formalism
contain no reliable information about quantum localization since modular
localization has representation theoretic and algebraic origins, the
string-localization of charged fields remained outside the Lagrangian
perturbative formalism.

The lowest order perturbative expression for the two-point function of the
charged field can be computed in terms of the product of the free massive
Dirac two-point function with that of the $e$-dependent vectorpotential. As
expected, the result cannot return to pointlike localization by any linear
operation \cite{charge1}. The point 2 in the above list is presently under
intense investigation, the generalization of the E-G iteration to strings is a
very nontrivial problem. Higher orders are expected to strengthen\footnote{For
stringlike fields allow to speak about the strength of their stringlike
localization.} the string dependence.

In this setting of viewing localization as the central concept of QFT, the
Schwinger-Higgs screening of scalar QED is a process of "re-localization" from
the noncompact Maxwell charge to the pointlike "screened charge" i.e. the
Higgs matter and the pointlike massive s=1 vectormeson. Figuratively speaking
half of the degrees of freedom of the "pre-screened" charged field has hooked
onto the massless vectorpotential and converted it into a massive vectormeson.
In this way of looking at the problem it becomes the QFT counterpart of the
Debeye screening mode of a quantum mechanical Coulomb gas. Locality in the
context of QM becomes the "range of forces"; the concept of modular
localization needs vacuum polarization which is absent in QM. Quantum field
theoretic analoga tend to be more abstract and more radical; the intrinsic
meaning of S-H screening is that there exists a self coupling of real scalar
fields and a coupling with massive vectormesons which leads to a theory which
is pointlike generated. The use of a stringlike massive vectormeson in the
coupling to the scalar field would then only play the role of a
"renormalization catalyzer" for maintaining the power counting restriction.
Although the massive vectorpotential is string-localized, its associated field
strength and the screened real scalar field is pointlike, in other words the
model is pointluke generated.

Since Schwinger's role in this birth of this concept seemed to have been lost
in the maelstrom of time, here are some helpful remarks. Schwinger thought at
the beginning of the 70s that there could exist a screened mode of actual
(spinor) QED in which the photon changes to a massive vectormeson. When he
realized that 4-dimensional spinor QED was not suitable in order to
demonstrate his point, he proposed the 2-dimensional Schwinger model which
exists only in this massive screened mode. About the same time Higgs succeeded
in showing that 4-dimensional scalar QED poccesses such a perturbatively
accessible mode. If in addition one couples other spinor fields in addition to
the Higgs (= screened complex) field, their global charge will be unaffected,
only the Maxwell-charge of the former complex scalar field will be screened
and in this way become a real field $R$.

Charge screening means that the selection rule of the Maxwell charge breaks
down because%
\begin{equation}
"\int j_{0}(x)d^{3}x"=0
\end{equation}
Here the quotation mark indicates that the definition of charges from
conserved currents requires some mathematical care. The interaction with the
real part contains also a $R^{3}$ term \cite{charge1} so that even-odd
selection rule is also broken. Although most calculations about
Schwinger-Higgs screening are perturbative, the charge/screening issue is also
the subject of a theorem. Swieca's screening theorem \cite{scree} in which the
occurrence of a massive "photon" (i.e. a massive object which complies with
the Maxwell structure of the interaction) is linked to an increase of
analyticity of certain formfactors which indicates an better localization.

Whereas the nonabelian generalization of the screening mechanism has a
similarity with its abelian counterpart in that the physical content is that
of a system of massive vectormesons and hadrons without having to invoke
infraparticles and string-localized interpolating fields, the "unscreened"
Yang-Mills theory and QCD remained a mystery. As with any mystery it is the
source of folklore and metaphors. This is reflected even in the terminology
(gluon, quark) confinement which referrs to QM, whereby the use of appropriate
potentials can create a vault for particles. However this is not possible in
QFT where the only principle is locality; theories with pointlike generators
(or compactly localized algebras) can not justify this terminology. The very
strong infrared divergencies (stronger than in the abelian case) of
correlation functions point into the opposite direction: noncompact localization.

In this context it is interesting to take a closer look at the third class of
positive energy Wigner representations, the infinite spin representations.
They are string-localized but in a very different way from string-local
potentials of zero mass pointlocal field strength and also from states created
by a charge-carrying field. Whereas in those cases the state on which the
Poincar\'{e} group acts contains (under reduction into irreducible components)
only pointlike generated representations, this is remarkably different for the
infinite spin representation.

Even if one could generate such a positive energy state by applying
interacting finite spin fields (e.g. gluon fields) to the vacuum, it would be
impossible to perceive its presence by performing measurements on it. A
measurement apparatus in local quantum physics is represented by an operator
from a compactly localized subalgebra, but in view of the inexorable vacuum
polarization from localization with sharp boundaries it is customary to relax
the restriction from local to quasilocal \cite{Haag}. Such an apparatus can
only interact with a finite part of the infinite extended string. The latter
should then experience a local change, which is prevented by the
irreducibility of the string representation. Similar arguments would lead to
the statement that it is impossible to create such strings in collision
processes (which is consistent with the string-breaking mechanism).

In earlier times, when the attempts show that a pointlike covariantization of
the infinite spin representations failed, these representation were dismissed
as "not used by nature" \cite{Wei}. But knowing that these representation
describes irreducible string states and its field theoretic generalization a
new kind of string-localized free field theory (which most probably does not
even possess pointlike composite fields \cite{M-S-Y}), and being aware of the
need to accomodate the degree of freedom of interacting in the gluons as
fields and states but in contradistiction to their abelian relatives the
photons not as particles, one should perhaps not outright dismiss this very
large third Wigner representation class. It is hard to find a theoretical
reason why a representation class which fulfills the energy positivity should
be discarded for conceptual reasons in times where the string-localized
interacting gluons ask for a better understanding.

This infinite spin representation and its generating field is by itself not
very interesting. But if one could find an argument that the interacting
string-localized gluon field applied to the vacuum contains such a string
representation the interest would certainly increase. Unfortunately the wisdom
in perturbation theory seems to be that only those representations which
entered in the form of free fields can be found in the decomposition of
interacting states, but this argument may be to naive. It is also possible
that the physics of strong interactions is described by a form of QCD in which
the infinite spin representations enter in some other form. But in no way can
invisibility in QFT result from confinement into a vault.

Some hint about the severity of the infrared divergences caused by gluon
string fields should be contained in the infrared divergencies in the string
direction $e$ which are much stronger in the Yang-Mills case than in
correlation functions of charged fields in QED. Whereas the standard gauge
theory approach has no means to avoid these divergences, the string setting
treats the string directions $e$ on the same level as the $x$ i.e. one needs a
testfunction smearing. In this setting the calculation of coalescing e's is
expected to be similar to the calculation of composite fields. Whatever the
progress on Yang-Mills and QCD will lead to, the stringlike localization will
remain the important issue. The reason why this issue appears in an essay with
the title "ST deconstructed" is that ST has succeeded to create a tremendous
amount of confusion and derailment in a area which is central to particle
physics and which has to be regained in order to end the almost 4 decades
lasting stagnation on the crucial project of the standard model.

Even though this perturbative realization of the Schwinger-Higgs screening
picture works, it is always advantageous to ask whether a certain way of doing
things is still afflicted with residues of metaphors even if, as in this case,
the metaphors have been helpful. Since the observable massive vectormeson does
not reveal in any observational manner whether it is coming from a (fattened)
photon or not, an anti-metaphor cleansing is physically justified. Indeed, by
using the BRST formalism in a perturbation setting, starting with massive
vectormesons, one derives that, in order for the BRST formalism to lead to the
expected cohomological properties, one needs to introduce an additional
physical degree of freedom whose simplest realization is an interacting
massive scalar field (naturally without a vacuum "condensate") \cite{Du-Sch}%
\cite{Du} of the same kind as the real scalar field after the Schwinger-Higgs
screening mechanism has done its job. This looks as one has finally succeeded
to show that the presence of a renormalizable vectormeson requires to be
accompanied by a s=0 particle. If it would not be for the BRST indefinite
metric aspect, this would be the end. But one cannot base a general statement
like this on arguments using indefinite metric or to put it in more drastic
terms: one cannot extinguish fire with gasoline. Hence the last step would
consist of starting with massive string-localized (and therefore
"power-counting renormalizable") vectormesons and find out whether the
requirement that the perturbation theory is pointlike generated (or at least
admits pointlike subobservables) requires the presence of a scalar
"companion". The presence of lower spin objects as a result of the locality
requirement would be the "ultimate" step in the understanding of the
Schwinger-Higgs mechanism within a theory in which all properties are
realizations of the locality principle under different circumstances; it would
certainly be much deeper than the rather dull appearance of families of
particles with different spins as a result of supersymmetry.

The merger of gauge theory into the setting of string-localized fields is
quite satisfactory because the analogy to the classical gauge theory which
finds its mathematical setting in fibre bundles is not helpful; there is no
classical analog of unitarity and Hilbert space structure since a classical
vectorpotential is like any other pointlike classical field. The modular
localization\ is an intrinsic quantum attribute and any property one
encounters in QFT must be resolved in terms of localization properties, only
then one can claim to have understood its roots. This does not only apply to
string-localization of generators of electrically charged algebras and their
pointlike "re-localization" in terms of Schwinger-Higgs screening and the
still poorly understood invisible nature of gluons in Yang-Mills theories, but
includes also such well-understood issues as the existence of superselection
sectors and the associated internal symmetries. to state it in slightly
provocative terms, if mathematicians would not have discovered the theory of
compact groups a long time ago, it would have appeared as a consequence of the
theory of nets of local operator algebra where the DR refinement \cite{D-R} of
the DHR theory \cite{Haag} of superselection rules finds compact group theory
in the form of its dual. Of course this also applies to spontaneously broken
symmetries. Even the Minkowski space together with its Poincar\'{e} symmetry
is a consequence of "modular positioning" of a finite numbers of copies of a
monad\footnote{All the local algebras in QFT are of the same type and are for
brevety referred to as a monad.}.

Although QFT is very indigent about principles, it is quite abundant and
ingenious in the different realizations of its only principle: causal
localization. As the above illustrations show, it takes some skill to show
that a certain property in QFT is a manifestation of locality. Therefore the
incorrect understanding of the crossing property and the misleading belief
that duality is its one-particle approximation, as well as the later
misunderstanding in interpreting multicomponent chiral models as defining an
embedding of a string into the target (= component) space, each one of these
conceptual errors is in the realm of human weakness confronting the most
subtle physical principle on which already Wigner despaired. After all QM in
distinction to QFT has no intrinsic notion of localization; where a quantum
mechanical string is "localized" in spacetime or in an internal Hilbert space
is specified by the person who carries out the computation. Not so in QFT
which thanks to its intrinsic modular localization can very will distinguish
between a string in spacetime and a collection of oscillators in a "little
Hilbert space" which are able to change the relative admixture in an infinite
particle/spin tower "above" spacetime.

In this way the superstring is what we called a dynamical infinite component
pointlike field, the first and only successful construction of an earlier
research program which was restricted to group theory (noncompact groups which
generalize the Lorentz group \cite{Tod}) and ended in failure. So the
misunderstandings of the delicate causal relativistic localization is not
different from other conceptual misconceptions in the history of theoretical
physics. But never before had a misconceived idea that much success that it
remained in the headlines of physics journals for 5 decades and recently even
led to a new string-oriented IOP journal. From a scientific point of view it
is nearly impossible to understand how such an incorrect idea, which in 50
years was not able arrive at any observationally verifiable prediction (but
"postdicted" General Relativity, as their followers claim), but led to
thousands, if not ten-thousands of publications and is still supported by what
are considered the to be leading particle physicists \cite{Witten}. To shed
some light into what has been going on will be a herculean task for historians
and philosophers of science, and it is my hope that essays like this may give
some guidance in this conceptual slippery terrain.

\section{Quantum Kaluza-Klein reduction and the degrees of freedom problem}

The Kaluza-Klein proposal originated in the aftermath of Einstein%
%TCIMACRO{\U{b4}}%
%BeginExpansion
\'{}%
%EndExpansion
s general relativity as an attempt to obtain a geometric unification with the
classical Maxwell theory by passing to a 5-dimensional Einstein-Hilbert theory
and imposing an appropriate restriction on the 5$^{th}$ spatial dimension. The
restriction consisted in assuming that the coordinate is compact i.e. circular
and that in its Fourier transform only the lowest frequency enters
("dimensional reduction" via a shrinking circle).

As we learned in the previous section a geometrical unification may be
compatible with a classical setting, but there is a priori no reason to
believe that it fits into QFT with its modular localization. Even on a
classical level there arises the question what is really gained if the
reduction of 5 dimensional Einstein-Hilbert gravity by a special prescription
for the compactification of one spatial dimension leads to the 4-dimensional
coupled Einstein Maxwell system which of course one also could have obtained
directly by adding to the energy-stress tensor the Maxwell contribution. The
fact that the Kaluza-Klein (K-K) idea did not play any significant role in
classical general relativity could be interpreted as indicating that it does
not go beyond a nice metaphor which does not enrich or simplify any
calculation nor is there any conceptual gain.

The local quantum physical setting is completely intrinsic in the sense that
all physically meaningful properties do not depend on how the model was
produced (operator methods, functional integration), rather all its properties
are autonomous once its vacuum correlations or its characterization in terms
of operator algebras has been settled. Therefore the first question is: does
there exist a suitably defined quantum version of K-K with an autonomous
meaning? This removes from the start manipulations involving Lagrangians as in
\cite{Sei-Wi}.

There are two ways of formulating a QFT adaptation of K-K which would have an
autonomous meaning, this freezing of one spatial coordinate leading to a
"brane" and its compactification in the sense of "curling up" as envisaged by
string theorists.

The brane construction is mathematically well-defined. One may picture it as
the thinning of a slab region in the sense that the spatial thickness of the
slab goes to zero, whereas all the other infinite spatial extensions, as well
as the extension of time, remains preserved. It is well known that operator
algebra in an arbitrary thin tube around the time axis is equal to that of the
subtended double cone, so the full time axis determines the entire global
algebra. All the degrees of freedom of the original bulk description are also
contained in the slab, never mind that this goes against classical intuition.

At first one's instinctive reaction is that this looks fishy because the
geometric symmetry group of the brane is smaller. But taking notice that the
infinitely many appropriately defined fields with derivatives into the brane
transform within themselves under the symmetry generators which lead out of
the brane \cite{Mack}, one realizes that, in agreement with the preservation
of the phasespace degrees of freedom, the full spacetime symmetry group is
maintained\footnote{The action of the generators out of the brane resembles
the action of an inner symmetry. Degrees of freedom and spacetime symmetries
are inexorably linked (viz the AdS$_{5}$-CFT$_{4}$ correspondence). However
projections onto a horizon (a nullsurface) are degree of freedom reducing.}
and thus the strong relation between causality and covariance remains intact.

There is one case for which the spacetime symmetry group continues to act in
the conventional way on the field variables of the lower dimensional QFT, this
is the famous AdS$_{n+1}$-CFT$_{n}$ correspondence in which the CFT$_{n}$
represents a restriction to a brane at infinity. In that case the
\textit{preservation of degrees of freedom prevents to have a physical theory
on both sides of the correspondence}. Either the AdS side has a physical
cardinality of degrees of freedom which leads to an overpopulation on the CFT
side, or the CFT side represents a physical QFT in which case the AdS$_{n+1}$
side is "too anemic" in degrees of freedom in order to deserve the attribute "physical".

At this point it may be helpful to be more explicit about the degrees of
freedom issue and to explain the pathological properties which arise from the
presence of too many of them.

There is a second idea to reduce spatial dimensions which has been
predominantly used by string theorists, this is the \textit{compactification
of spatial dimensions} and their subsequent pointlike limit ("curling up").

In order for the compactification idea to have any intrinsic physical meaning
beyond a formal game played with classical actions, one needs a structural
argument which allows to execute the compactifying limiting process for a
given model in an intrinsic operational manner. As mentioned before this
excludes arguments based involving the Lagrangian setting of the kind used in
\cite{Sei-Wi}. The ideal method would be to start from the correlation
functions or the uniquely associated operator formulation in a Hilbert space
and to implement the compactification either directly on the correlation
functions or in the operator setting. An implementation via spatial boundary
condition has to argue against the mathematical fact that the change of the
compactification radius leads to unitarily inequivalent representations which
strictly speaking represent different theories and not an intrinsic process.

The good news is that there exists such an intrinsic implementation of the
compactification idea. This is possible with the help of the intermediate use
of thermal physics which amounts to an inequivalent global representation of
the same spacetime indexed net of operator algebra. It is well-known that the
introduction of a KMS temperature amounts to compactify one coordinate of a
euclideanized QFT which then becomes circular. The curling up process is
implemented by the idea of a sequence of mutually contained subtheories
localized inside a global ambient theory; this is the "open system" point of
view which is very different from the simpler but nonintrinsic "quantization
box" setting \cite{Haag}.

The execution of such open system limits is prohibitively difficult. Its
general setting is the Nelson-Symanzik temperature duality. The best studied
case (which recently has been placed into a modern context \cite{Jae}%
\cite{po-int}) is that of two-dimensional superrenormalizable models.

\begin{theorem}
(Nelson-Symanzik duality) The correlation functions of two-dimensional QFT
whose spatial coordinate is periodic (e.g. lives on a circle with radius R)
and whose time coordinate fulfills a $\beta$-KMS conditions state at imaginary
times\footnote{In terms of covariant pointlike fields the difference between a
spatial periodicity and an imaginary time-like periodicity is in the different
prescription of taking boundary values. In the Nelson-Symanzik duality these
different prescriptions become interchanged.} are equal to those in a dual
theory where R\ and $\beta$ are interchanged together with space and time.
\end{theorem}

Hence the classical idea of simply compactifying a noncompact coordinate
compact "by hand" (imposing boundary conditions) has a flaw since the
intermediate compactified quantum theories do not represent a sequence of
mutually included theories with a compactification radius contacting to zero.
\ In the other hand the \textit{formalism of open system,} in which the
smaller system is constructed as a genuine subsystem included into the bigger
one, is insufficiently understood.

To some of the readers this discussion may sound a bit like nit-picking. But
the maintenance of the hard fought-for representation theoretical consistency
conditions which led to the d=10 infinite component pointlike ST does require
the utmost care in descending in dimensions in order to avoid throwing out the
baby (the consistency) with the bath water. \ 

The obvious difficulty with any quantum compactification is the ubiquitous
occurrence of localization-induced vacuum polarization. The only way to handle
this problem is to find a multiplicative renormalization which keeps the
limiting correlation finite.

In chiral theories for which the fields obey more general exchange algebras
than fermionic/bosonic ones the Nelson-Symanzik temperature duality passes to
a much richer selfduality which mixes the vacuum sector with the charged
sectors via the Verlinde-Rehren matrix S. This suggests to implement the
"curling up" as an infinite temperature limit. \ 

The simplest QFT are higher dimensional conformal free fields For a scalar
neutral free d=3+1 dimensional conformal field at inverse temperature $\beta$
the two-point function is
\[
\left\langle \varphi(x)\varphi(y)\right\rangle _{\beta}=\frac{\sinh2\pi
\frac{\left\vert \mathbf{x-y}\right\vert }{\beta}}{8\pi\beta\left\vert
\mathbf{x-y}\right\vert }(\cosh2\pi\frac{\left\vert \mathbf{x-y}\right\vert
}{\beta}-\cosh2\pi\frac{\left\vert x^{0}-y^{0}\right\vert }{\beta})^{-1}%
\]
It is easy to see that for $\beta\rightarrow\infty$ this function approaches
the massless zero temperature function of a field with scale dimension sdd=1
whereas the small $\beta$ limit at vanishing real time values behaves as%
\[
\left\langle \varphi(x)\varphi(y)\right\rangle _{\beta}\overset{\beta
\rightarrow0}{=}\frac{1}{8\pi\beta\left\vert \mathbf{x-y}\right\vert }%
\]
which after liquidating the divergent $\beta^{-1}$ factor by multiplicative
field renormalization is a two point function of a euclidean d=3 two-point
function of scale dimension $sd=\frac{1}{2}$ of a massless free field. The
last step namely the passing to a d=2+1 real time free field is best known in
the form of analytic continuation, but there is also an intrinsic operational procedure.

Clearly this kind of dimensional reduction works for any dimension and it can
be applied iteratively. The supersymmetry, which plays a crucial role in ST,
is a symmetry unlike any other symmetry\footnote{The physical origin of inner
symmetries in QFT (a concept which does not exist in classical physics) is the
DHR superselection theory \cite{Haag}. All compact groups appear as potential
symmetries except supersymmetry. This makes SUSY appear as an artificial
junction of fermions and bosons by tuning couping strength at a spacial value;
it could be the explanation why a thermalization leads to a collapse of SUSY
and not to the expected spontaneous symmetry breaking as in all other cases.}.
Unlike the Lorentz symmetry (and most other symmetries) which is spontaneously
broken in thermal states, the supersymmetry is the only symmetry which
"collapses" in a heat bath \cite{Buch} showing that supersymmetry is more like
an instable tuning between Fermions and Bosons than a stable symmetry. This
raises the question whether supersymmetry is stable under dimensional reduction.

All this is applicable to ST which is described by a pointlike infinite
component field. But there is a prize. Reducing dimensions in this way does
destroy the delicate oscillator mechanism for generating a mass/spin tower
with oscillator bridges between the different mass/spin levels. With other
words the source-target picture which led to the 10 dimensional superstring is
destroyed in any process of dimensional reduction and one does not really know
why one first has to go to the 10 dimension superstring in order to arrive at
a 4-dimensional consistent theory. What is the intrinsic property one wants to
achieve in 4-dimensions which requires the detour?

An even bigger hurdle is the comprehension of the computational rules for
interactions beyond the "tree approximation" represented by the dual model.
These recipes have been conjectured by adjusting the Feynman rules to the
situation of world-sheets instead of world-lines. But the string field is
pointlike i.e. there are no world sheets in the target space so what should
one make of those recipes? One would be in for a serious conceptual trouble if
string amplitudes could be expressed in terms of operators and states as in
any other accepted quantum theoretical activity in particular in Feynman's
perturbation theory.

In this case the tube rules would have to be taken as the definition and the
dual model and the canonical quantization of the Nambu-Goto Lagrangian which
inevitably led to pointlike objects must be rejected, which would amount to a
clear position in a conceptually confused situation. But this is not the case;
during 4 decades the efforts of the most renown string theorists have been
unable to come up with an operational setting in a Hilbert space. Far from not
having solved these problems, the string theorists have not even asked the
questions because believing in an incorrect metaphor has prevented them to do so.

As we argued before QFT and with it its infinite component counterpart called
ST have difficulties with the idea to generate internal symmetries from
spacetime symmetries of QFT by a curling up process. We will argue that such a
possibility is excluded on philosophical grounds since it would go against the
very raison d`etre for the concept of internal symmetries whose purpose is to
describe the relation between (neutral) observable algebras and their higher
representations in charged sectors which combined together define the
symmetry-carrying field algebras. \ For this it is useful to have a quick look
at the history of the representation theoretical origin of internal symmetries
and their relation to the structure of local observable algebras.

The concept of internal symmetries was always clouded by mystery ever since
its inception via the SU(2) isospin of nuclear physics introduced by
Heisenberg. The symmetry of classical physics is restricted to spacetime
transformations\footnote{The concept of inner symmetry in classical field
theory does not exist in any natural way. But it is sometimes intoduced by
reading back properties from QFT into classical physics (similar to classical
Fermions = Grassmann valued fields).}. Since the concept of causal
localization is fundamental to QFT it is natural to inquire whether the
quantum adaptation of localization inherent in QFT could shed some light on
inner symmetries. This is precisely what Doplicher Haag and Roberts set out to
do in their research on the spacetime origin of the superselection rules of
local quantum physics which started at the end of the 60s and ended at the
beginning of the 90 with a complete answer \cite{D-R}. Not all programs which
require such a large intellectual investment end with such a beautiful clear
answer: by viewing QFT in terms of a dichotomy between (neutral) local
observables and charge-carrying superselected states one delegates the issue
of (particle) statistics, charge superselection rules and their incorporation
into a canonical construction of a larger field algebra\footnote{In this way
the input into the Lagrangian quantization is understood as the output of a
more basic autonomous theory.} on which the symmetry acts as a global gauge
group with the local observables re-emerging as the fixed point algebra under
the inner symmetry.

In general the field algebra does not permit an immediate intuitive access
since it involves unobservable (but extremely useful) operators, but
fortunately its conceptual and mathematical structure is preempted in the
structure of the local observable net of algebras indexed by spacetime regions
in a somewhat hidden manner. By a sequence of conceptually quite interesting
and profound steps this unique field algebra (including the concrete inner
symmetry group which acts on it) can be constructed solely from the properties
of its observable projection. This is similar in spirit (but much more subtle
in detail) to Marc Kac's famous saying \textquotedblleft how to hear the shape
of a drum\textquotedblright.

Although there is a deep connection between spacetime locality and inner
symmetries there is no support for a naive passing of curled up spatial
coordinates to inner symmetry indices. Certainly the above (only known)
intrinsic implementation via infinite temperature limits shows nothing in this direction.

Additional support against the string theorists view of spacetime and its
symmetries through the 10-dimensional target space of special class of chiral
theory comes from recent insights on space-time and the emergence of spacetime
symmetries in the algebraic setting. It turns out that all causally localized
subalgebras of the global observable algebra are indistinguishable (unitarily
equivalent) independent of the localization regions; in technical terms they
are copies of the unique hyperfinite type $III_{1}$ von Neumann factor
("monads"). \ The modular localization theory (section 3) provides a relative
positioning of monads in terms of modular concepts \cite{inter} and that from
an appropriate positioning of a finite number (two for d=1+1, six for d=1+3)
of monads in a common Hilbert space one can derive the full content of QFT
including the Poincar\'{e} symmetry \cite{Ka-Wi} which emerges from
algebraically defined modular groups. Whereas spacetime symmetries are related
to positioning of monads in a joint Hilbert space, inner symmetries correspond
to endomorphisms of operator algebra. Dimensional reduction as well as
supersymmetry is caught between two stools.

This casts considerable physical doubts on the \textit{target space
construction of spacetime symmetries}. To avoid any misunderstanding, it is
not the emergence of a 9+1 dimensional infinite component field free wave
function per se from such a target construction which is being questioned, but
rather the physical rational behind it. The motivation for the old failed
attempt for constructing dynamical infinite component theories using
generalizations of the Lorentz group representation theory is at least
comprehensible, whereas the conversion of an inner symmetry component space of
a chiral theory into a living space space of higher dimensional relaltivistic
theory is a weird requirement. Unfortunately mathematics allowed an
exceptional positive answer in form of the 10-dimensional superstring;
unfortunately because this was taken by the string community as a revelation
about the origin of spacetime instead of looking for an explanation within the
setting of chiral theories which are known to offer a much greater richness at
the place where in higher dimensional theories one finds compact group inner
symmetries. In this way we will probably never know what really is behind the
M-theory metaphor.

String theorists will probably retort that this choice is a fortiori justified
by the tube rules for the implementation of interactions. But since the tubes
refer to metaphoric and not genuine strings there is no reason whatsoever why
one should take for granted that transition amplitudes computed in this way
can reproduce those properties which any relativistic S-matrix describing
particle scattering (independent of whether it comes from QFT or any other
consistent particle scheme) must obey \cite{crisis}. If string theorists
succeed to derive these properties then they could start to claim that, even
though all the physical arguments do not extend beyond metaphors, they
nevertheless found by luck (but not as a gift of the 21 century!) an unusual
prescription which assigns a consistent S-matrix to a special kind of infinite
component field.

The prescription is very different from what one would obtain if one considers
the infinite component field as the limit of finite component fields and
applies the standard rules of causal perturbation theory. It is also different
from what one would obtain if one were to use the string-localized description
of section 5 (which each free field permits) for each irreducible component in
the infinite component field and an extension of the causal perturbation
theory to such objects.

\section{Particle physics under the spell of metaphors}

ST would never have come into being without the dual model and the latter
would probably not exist without the Mandelstam representation for the elastic
scattering amplitude. The fact that the Mandelstam representation is probably
inconsistent with QFT does however not affect the dual model since the latter
is a phenomenological tool (no connection to QFT required) whereas the former
was thought to follow from locality and spectral properties. At the time of
Veneziano's discovery of the dual model, the peculiar mixture of mathematics
of Gamma and Beta functions together with clever pedestrian tricks, as they
are only available to experienced physicists, led many people to think that
this model was unique in d=4. On the other hand the idea of uniqueness in the
realization of a structural idea inevitably led in all cases to metaphoric
thinking in the direction of TOE, starting with the S-matrix bootstrap. The
production machine for dual models via Mellin transforms of conformal theories
in any spacetime dimension should have turned down any mysticism in that
direction. The imposition of unitarity and positive energy leads to uniqueness
but, as we argued before, this is of doubtful value, since even if we were
spared a stringy existence by recognizing the pointlike naure, we still would
facte the problem of how to descend from 10 dimensions to our 4-dimensional flatland.

In order to make contact with reality one has to get rid of the excessive
spatial dimensions in analogy to the Kaluza Klein model of classical field
theory. It was shown in section 5 that neither of the two proposed methods
(branes, curling up a compactified dimension) works. In the first case the
cardinality of degrees of freedom stay the same a fact which is related to the
preservation of the original spacetime symmetry with the brane-changing part
of the group acting in an unfamiliar way \cite{Mack}. The physical cardinality
of the quantum matter in the bulk becomes unphysical in the brane (albeit
mathematically consistent) leading to causal poltergeist phenomena and similar
unwanted events. The curling up method on the other hand and the subsequent
shrinking to a point creates infinite vacuum polarization effects. There is
also the problem why ST think that the properties they have implemented with
so much effort (unitarity, positive energy) will remain unscathed; and if this
is really the case, why it would not be better to understand the resulting
object directly in 4 dimensions.

The most spectacular consequence of the metaphor about ST as a relativistic
theory dealing with spacetime strings is the Maldacena conjecture \cite{Mal}
about a correspondence of a certain 4-dimensional supersymmetric conformal
gauge theory with a ST in 5 spacetime dimensions in which 5 dimensions of the
original 10 dimensions were compactified in the sense of Kaluza-Klein and
became inner symmetries; following the standard parlance of string theorists
this is the "gravity-gauge correspondence". This rather special conjecture
became the focal point of the whole string community and led to an immense
number of publications as no particle physics paper before; but none of these
papers got anywhere near a proof or at least a plausibility argument.

What makes this sociological phenomenon appear similar to mass-psychosis as
commonly observed in other human activities, is the fact that while most of
these papers where written, there already existed a precise theorem on the
AdS-CFT correspondence \cite{Re}. The method of its proof is in the best
tradition of local quantum physics; instead of importing geometrical ideas it
uses those autonomous localization and spectral properties of QFT which led to
important structural theorems (TCP, spin-statistics, the DHR theory,.. ).

It shows in a mathematical rigorous way that the correspondence exists, but
there is a physical hitch to it. If one starts from a 5-dimensional physical
theory, the 4-dim. conformal theory is unphysical for the same reasons as the
QFT on a brane in the previous section. So the correspondence theorem cannot
not support the Maldacena conjecture\footnote{For a more detailed account see
\cite{crisis}} which was viewed as corner stone of ST. Rehren's theorem was
shrugged off by the string community. A single person would be reluctant in
ignoring such a theorem for fear of appearing ignorant, even if its
conclusions are uncongenial to a proposed idea. But inside a sufficiently
large community, such inhibitions are lost.

The saying "many people cannot err" is only valid in a restricted sense,
namely if an individual arrives at an interesting result he should try to get
the approval of some of his colleagues in order to minimize the possibility of
an error. This step Rehren certainly certainly passed.

To have the support of a worldwide community however diminishes the chances
for correctness, because the drag to add an approving contribution to an
ongoing woldwide fashion is much stronger than that for a critical assessment,
in addition it is extremely helpful for increasing the impact and advances the
career. This was already clearly visible at the time of the S-matrix bootstrap
(when I was Lehmann's assistent) except that in those days there still was a
critical reaction (Jost,..); the Streitkultur of old Europe was still going
strong and the S-matrix fashion disappeared in less than a decade. This is a
reasonable lifetime for a proposal which did not contain a real conceptual
error but was too vague in order to yield something useful. The almost 5
decade dominance of ST is to a large part the result of the disappearance of
an internal Streitkultur which was replaced with connecting the rise of
acceptance with the size of the community; this represents the new Zeitgeist
which finds its stronger and better known manifestation in the financial
capitalism and its markets.

This throws a somewhat different spotlight on Anderson's critical remarks
about ST in the introduction. Leaving aside whether it was hubris of some
particle theorists which led in this uncritical way right into a new kind of
particle theory without observational facts, even on purely theoretical
grounds the theory has nothing to do which what it pretends to describe. Many
particle physicists who have kept their common sense and tacitly rejected ST
because they had the feeling of entering a surreal counter-world of particle
physics, should feel vindicated. In fact this essay is intended to identify
the cause of these gut feelings and provide them with a
conceptual-mathematical expression. It is clear that it will not have an
effect on hard core string theorists who live in their self-spun cocoon of
reduced conceptual perception. A theory which substituted a metaphor for the
most central property of local quantum physics namely \textit{localization,}
should actually create \textit{discomfort with any particle theorist}.

To perceive the full extend of the philosophical aberration into which ST
drove a sizable part of the particle physics community for several decades,
the reader should take a look at the "paradigmatic" new insights and
"revolutionary" upheavals about the new fundamental physics which one finds in
books and essays as \cite{Suss}\cite{Schele}\cite{Teg}.

To close ranks between members of the community and to convince those who
still waver, the message some of the string community leaders is that there is
nothing to be gained looking outside ST; this is clearly the purpose behind
their slogan "there is no other game in town". For many gifted and innovative
non-stringy particle physicists who got the short side when competing with
string theorists about positions, this slogan took on a literal existential meaning.

Leaving the difficult problem of an in-depth understanding of the action of
the same Zeitgeist on different human endeavours as science, politics, money
economy and cultural activities to sociologists, philosophers and historians
of science, I will confine myself to some observations about the environment
which facilitated the ascent of the string metaphor in particle physics.

The idea that particle physics may permit a description in terms of a unique
theory (reductionist extremism), called a "Weltformel" or a TOE, certainly
preceded ST. Einstein was probably the first who proposed a generalized
(classical) field theory containing his own gravity theory and all of that
time known classical field theories. It had to fail because it was designed
against the at that time already well established QT. The next attempt at a
TOE, Heisenberg's nonlinear spinor theory, gained for a short time the support
of Pauli and had a strong local attraction around the MPI in Munich and . The
reason may be related to the traditional hero worship of Germans, because at
the first lecture tour of Pauli to the US, it did not survive the engaging
critique by Feynman. The third shot for a TOE, namely the S-matrix bootstrap,
emanated from Berkeley, California. It is hard to imagine that it could have
come from the east coast or middle west of the US since the number of
religions, ideologies and new age life styles, which it needed in order for a
TOE to flourish, is more limited there. A first hand inside into the cultural
surrounding of a TOE can be found in \cite{Capra}.

In fact the uniqueness of an appropriately formulated S-matrix theory was the
driving force behind the S-matrix bootstrap, i.e. the predecessor of the dual
model and ST, without which the latter would not have arisen. The bootstrap
setting was not a perfect TOE, since there was no place allocated for
gravitation, but on the other hand nobody worried about gravitation as a part
of particle theory at that time.

Actually there was nothing metaphoric about its underlying S-matrix
assumptions, the mysticism was solely in the expectation that these S-matrix
principles will nail down a theory uniquely. In the middle of the 70s it
became clear \cite{STW}\cite{KTTW} that in d=1+1 the bootstrap principles
produce an infinite set of elastic S-matrices which have unique field theories
associated with them; one of the most fascinating discoveries about the inner
workings of QFT. The ensuing important discovery of factorizing models of QFT
via the bootstrap-formfactor program \cite{KW} attracted quantum field
theorists (and plays a fundamental role in modern nonperturbative
constructions of QFT), but was not even noted by the S-matrix bootstrap
followers; for them the alternative seemed to have been either a TOE or nothing.

The same underlying algebraic structure appeared almost simultaneous in
separate areas of mathematical physics. It became known under the name
\textit{Yang-Baxter relation} and already these names stand for two very
different areas, Yang for two-dimensional scattering of a nonrelativistic
n-particle system interacting via delta-functions and Baxter for the
statistical mechanis of integrable 2-dimensional lattice systems. Even the
field theoretic use at different places was not the same, in Saint Petersburg
the main interest was integrability, whereas at the Landau institute, Berlin
and Rio de Janeiro the leitmotiv of the research was to use two dimensional
factorizing models as a theoretical laboratory to test ideas coming from
higher dimensions under mathematically controllable conditions.

The uniqueness claim was based on the difficulty to solve the nonlinear
bootstrap relations i.e. on the fallacious suggestion that something nonlinear
which appears difficult, and for which one could not find any solution at the
time, may be rather unique, if it has any solution at all; this kind of
attitude towards nonlinear structures (example: Schwinger-Dyson equations,
similar ideas in conformal QFT) was not new and was in each case contradicted
by the existence of infinitely many solutions. Apart from the uniqueness
ideology, the bootstrap project was founded on one extremely vague requirement
called "maximal analyticity" (motto: impose analytic properties as the
calculation proceeds until you run into contradictions), but since its ad hoc
nature was so plainly evident, non of the concrete postulates caused any
serious metaphoric confusions. Nevertheless it is clear, though not at the
time of bootstrap, that the Zeitgeist was heading away from the laborious and
precise approach of LSZ flanked by Wightman and Haag with the dispersion
theory as the observational anchor. \ 

There is one positive assertion in favor of the S-matrix bootstrap which
should be mentioned; it placed for the first time an important property into
the focus of attention which before was occasionally observed in perturbation
theory (and whose name refers to a graphical property of sums of Feynman
diagrams of a given perturbative order): the \textit{crossing property}. Among
the many analytic properties following from the spectral and causality
attributes of QFT, crossing is the only one which, although being derived from
QFT, can be formulated solely on the (complex) mass shell. For this reason it
plays an absolutely crucial role in the conceptual understanding of the
field/particle interplay as well as in the nonperturbative construction of QFT
via S-matrix properties \cite{founcor}\cite{Mu-S}.

These ideas, as well as their extension from the S-matrix to formfactors
(matrixelements of fields between ket in- and bra out-states) have led to the
first explicit construction of strictly renormalizable models of QFT within
the before mentioned infinite family of 2-dim. factorizing models
\cite{Ba-Ka}\cite{Annals}\cite{Lech}. The obtained results may not have
satisfied the S-matrix bootstrap community since their diversity represent the
contrary of what one expects of a TOE. Nevertheless, measured against the
absence of any existence proof of interacting properly renormalizable QFTs,
this is a large step towards a better understanding of nonperturbative local
quantum physics.

Although the interest in the bootstrap S-matrix program deservedly subsided
with the ascent of gauge theories and the standard model, it did not fade out
because of any incorrectness in its postulates; in fact all its S-matrix
properties (apart from maximal analyticity and the illusion of a TOE) \ had a
solid conceptual basis\footnote{The often heard objection that the principle
of "nuclear democracy" contradicts the idea of quark/gluon
confinement/invisibility is the result of a too naive understanding of both
concepts.}. But the S-matrix bootstrap, as another previous attempt at a pure
S-matrix approach, did not lead to interesting and trustworthy calculations on
actual problems of particle physics, and hence it was no match to the rich new
world of perturbative gauge theories. Such scientific explanations for the
success of a theory are of course not entirely convincing since the
sociologically much more successful ST did not contribute anything to its
self-declared raison d'\^{e}tre, namely to link gravity with the other forces,
and yet it flourished at least on a sociological level.

ST was born at the time when the bootstrap ideas with their explosive
uncontrolled nonlinear structures created were loosing their attraction.
People were looking for a setting in which nice computations could be
formulated and carried through. The phenomenologically supported calculational
opening started with the attempt of Veneziano \cite{Ven} to find
\textit{explicit realizations of the crossing property}. The construction was
generalized to arbitrarily many particles by Dolan, Horn and Schmidt in their
work on the "dual resonance model" \cite{D-H-S}. All these contributions were
thought of as a consequence of a problem which was previously studied by
Mandelstam, using a spectral representation for the 2-particle scattering
amplitude whose validity, although not being deducible from QFT, was
conjectured on the basis of generalizing established analyticity properties.
The proven crossing property, as it can be rigorously implemented in
constructions of soluble d=1+1 factorizing models, is a delicate interplay in
which one-particle poles and cuts from the scattering continuum become
intermingled. The proof by quantum field theorists used powerful techniques of
analytic functions of several variables as they emerged from the study of
Wightman, retarded and time-ordered correlation functions in conjunction with
the LSZ stationary scattering formula.

The solution which Veneziano encountered by using mathematical properties of
Gamma and Beta functions, (later was referred to as duality) leading to
infinitely many particle poles, is not the crossing symmetry of QFT but came
out of mathematics. It was first formulated for the elastic scattering
amplitude and later generalized to an arbitrary number of particles. Dual
models with varying particle content were proposed \cite{Vech}. All these dual
models by \ It is important, despite all superficial similarities, to be aware
of this distinction. Whereas all these dual models were originally constructed
"by hand", using properties of special functions, nowadays we know that all of
them originate as Mellin transforms from conformal QFT \cite{Mack}, in fact it
is not difficult to the trained eye that behind the "operator representation"
of the dual model in \cite{Vech} is the Mellin transform of the chiral
conformal theory of exponential potential of a multicomponent current theory.

Mellin transforms have no operator status on their own, the operator
properties are limited to the untransformed conformal QFT. There are as many
Mellin transforms as there are conformal QFTs; the infinitely many poles in a
Mellin transform correspond to the infinitely many terms in the global
conformal operator expansions \cite{Mack}. The requirement of positivity
imposed on the Mellin transform turns out to be extremely restrictive; the
only solution corresponds to the d=10 superstring. This is somewhat easier to
see in the string formulation (canonical quantization of the Nambu-Goto
Lagrangian). I that setting one is dealing with an infinite component field
for which in addition to the action of the Poincar\'{e} transformations which
preserve the vertical mass-spin tower structure there are vertically acting
oscillator operators which not only interconnect the different levels but also
set the mass-spin spectrum. As mentioned before the construction of such an
object was the (failed) aim of a group of people under the name dynamical
infinite component fields before ST \cite{Tod}.

This raises the question what has the Mellin transform of a conformal field
theory to do with crossing which refers to formfactors i.e. matrixelements of
local operators between bra "out" and ket "in" particle
states\footnote{Scattering amplitudes are a special case.}? The answer is:
nothing. In a way the reformulation in terms of ST removes the acuteness of
this antagonism since a ST free field has no direct relation with a scattering
amplitude. From the point of view of Regge trajectory phenomenology, duality
was quite welcome, independent of its compatibility status with respect to
crossing. On shell crossing can be shown in perturbation theory, but it is
impossible to separate out the one particle pole contributions in such a way
that crossing can be realized among the direct and exchanged one particle
contributions. Crossing is always an extremely subtle interplay between poles
and cuts i.e. between the full analytic structure of scattering amplitudes; it
holds also in those cases in which there are no bound states.

It is legitimate to ask about the intrinsic quantum physical content of the
reparametrization invariant Nambu-Goto Lagrangian, taking into account that
the reparametrization invariance makes such a system \textit{classically
integrable}. For such systems one really should not apply the standard
canonical quantization but rather abstract the Poisson bracket relations
between the infinitely many conserved charges, convert them in the spirit of
Dirac into algebraic relations for abstract operators, and represent the
latter as concrete operators in a Hilbert space. This approach was
successfully applied before to two-dimensional integrable QFT where it leads
to a Poisson algebra structure for the infinitely many conserved "charges",
which is then the point of departure for the construction of a quantum algebra.

Precisely this was Pohlmeyer's starting point \cite{Po}. As expected, the
resulting quantum theory had little to do with the kind of ST which resulted
allegedly from the canonical quantization of Nambu-Goto like Lagrangians, in
fact the two quantizations are inequivalent \cite{Bahns}\cite{Me-Re}.
Pohlmeyer's almost lifelong dedication to the real ST (described by the
integrable reparametrization-invariant interacting Nambu-Goto form), after
having made several important contributions to the foundations of QFT, is an
illustration of perseverance in the defence of something which is
mathematically and conceptually correct in the face of an ever swelling sea of
publications in which the integrability of the Nambu-Goto system was ignored
in favor of the canonical quantization of a model which is classically
equivalent and which links up with the dual model. \ 

It is hard to avoid the conclusion that a 40 years lasting domination of a
theory, which in addition to its doubtful origin in the bootstrap S-matrix
program created its own incorrect concepts, is a major calamity in the almost
80 year impressive history of particle physics. For physics the aspect of
interpretation is absolutely crucial, whereas for a mathematician this is of
minor relevance. Whether a geometric formalism refers to spacetime or to an
inner symmetry space is of no relevance, the mathematics works the same in
both cases. This explains why mathematicians are often impressed by ST. They
love theories of physicists which have a somewhat metaphoric appearance and
leaves some freedom to use it for inspiration in the pursuit of new
mathematical theorems and to get to new subjects. The only point on which they
are unrealistic is their belief that physicists are living in the best of all
times; they would not understand if we tell them that they are creating their
golden mathematical castles on our ruins.

The damage such a misinterpretation causes in particle physics is however
considerable because (as was emphasized in section 4) a profound understanding
of modular string localization is really of vital importance in any attempt to
extend gauge theory \cite{charge1}. The latter had a steep start in QFT when
its importance for getting spin s=1 fields into the framework of a
renormalizable action was recognized. But the resulting obviously incomplete
setting did not really change for almost 40 years, and it is a legitimate
question to inquire why it has sunk into stagnation.

One reason is fairly obvious, the present ghost formalism is designed to
perturbatively construct local observables in terms of the standard
perturbative pointlike setting i.e. without any new investment into a revision
of perturbation theory. All the important nonlocal (more precisely
semiinfinite) physical objects as the electrically charged fields in QED and
the question of the physical status of gluons and quarks are not described
within the present perturbative gauge formalism \cite{charge1}. In the case of
electrically charge fields there is at least a definition by hand which
formally has been known since the time of Jordan 1935 (\ref{DJM}). This
Dirac-Jordan-Mandelstam formula which describes a semiinfinite stringlike
localized charge field is easily written down but to show that this object is
renormalizable in every order is a prohibitively difficult \cite{Stein} and
required to develop a new perturbative technology. Hence its construction,
unlike that of the gauge dependent Lagrangian fields, is not part of the
standard perturbative formalism. In the Yang-Mills case there is not even a
formal proposal how semiinfinite stringlike field look in terms of Lagrangian variables.

In the case of massive vectorfields the situation is slightly better. The
coupling of BRST ghosts is consistent with both the cohomological BRST rules
and in the sense of renormalizability and power-counting (reducing the short
distance dimension of the vectormeson+BRST from 2 to 1) under one condition:
the massive vectormeson needs a scalar massive companion \cite{Du-Sch}%
\cite{Du}. This appears inches away from a crucial theorem: the locality
principle + renormalizability $\rightarrow~$the Schwinger-Higgs screening
mechanism; for a presentation of the history behind this concepts see
\cite{Swieca}. It is interesting to note that Higgs exemplified his idea in
scalar QED (which has one parameters in addition to the mass of the complex
charged field and the coupling to photons), because what he showed is
precisely the existence of a different phase in which the charge becomes
screened to zero, the complex scalar field looses half of its degrees of
freedom by becoming a copiously produced real particle (the imaginary part
serves to convert the photon into a vectormeson) which amounts to a
charge-symmetry breaking. It is not clear whether Higgs thought about his
mechanism in terms of a charged particle which was subjected to the process of
complete screening, but Schwinger definitely had this very physical idea. He
could not exemplify it in QED and hence he invented the two-dimensional
Schwinger model. Swieca proved a structural theorem linking the appearance of
massive vectormesons to charge screening \cite{Swieca}.

The most remarkable aspect of the screening is that a partially nonlocal
theory of string-localized charged fields becomes totally local in the
screened phase. Whereas the particles corresponding to the semiinfinite
string-like charged fields are \textit{infraparticles} i.e. particle-like
objects whose mass is somewhat blurred\footnote{What is perhaps more important
is the fact that there is no tensor-product structure with a clear notion of
statistics.} due to the presence of an "eternal" soft photon cloud whose
energy can in principle be pushed below any energy resolution but not stripped
of completely. The infraparticle property in momentum space is a consequence
of the interaction which leads to the DJM string localization being the best
possible. On the other hand the localization resulting from screening is
pointlike i.e. the real part of the matter field which remains after screening
is point-like.

On the eve of new LHC experiments it becomes painfully clear that these ideas
which were discussed more than 4 decades ago were not developed further and
instead the metaphoric "Mexican hat" (which is part of the two-parametric
coupling manifold of scalar QED) and more recently "God's particle" took the
stage. One expects that the generalization of the screening idea to the
Yang-Mills situation also leads to a completely local theory, however the
understanding of the nonlocal aspects in the phase corresponding to the
unscreened Yang-Mills phase is a much harder problem (see section 4).

The return of metaphoric thinking in the 70s after its banning by Bohr and
Heisenberg through the insistence in operational definition of observables in
the beginnings of QM is a Zeitgeist phenomenon, and as such it would be too
simple to see it solely as a consequence of the acceptance of ST. It is not an
isolated event which suddenly arose in the context of the reformulation of the
dual resonance model. Its seeds were already there at the time when duality
was confused with a realization of crossing. It is conceivable that this mode
of thinking will linger on even after ST has been given up as the result of
its difficulties in delivering results.

\section{A case study: the impact of superstrings in Germany}

Traditionally particle theory in Germany has been quite strong on the
conceptual and mathematical side. It has been claimed \cite{Fo} that the
discovery of quantum mechanics in a war-torn Germany was somehow related with
the fact that its probabilistic aspect was especially in tune with the gloomy
philosophical post world-war I mood expressed in Spenglers "decline of the
west". But a more convincing explanation is that the sciences were hardly
affected by the destruction of worldwar I, and the time of widespread
political terror and racial persecution was still 15 years away. The situation
after the world-war II was totally different; this time the moral and
intellectual fiber was devastated, university live came to a standstill and
some of the remaining physicists had left the country to work under less
restrictive conditions.

Taking these conditions into consideration it was a small miracle that by the
mid 50s particle theory was back on its feet in the Federal Republic. The work
of Lehmann, Symanzik and Zimmermann (LSZ) who discovered the LSZ scattering
theory and the work of Haag and Ruelle who showed that time-dependent
scattering theory follows directly from the locality and the existence of a
mass gap in the energy-momentum spectrum, were milestones in the understanding
of the subtle connection between particles and fields. Lehmann developed
together with Jost and a later important contributions by Dyson the so-called
JLD representation which in turn was the basis of a rigorous derivation of a
particle analog of the optical Kramers-Kronig dispersion which some time later
was observationally verified. It was a well-balanced construction in that it
combined highly conceptual aspects in the improvement of the subtle
particle-field relation with useful formulas for scattering amplitudes in
terms of vacuum correlations of fields. The conceptually well-founded
derivation of dispersion relations and their experimental verification maybe
considered unattractive by modern adherers of a TOE, but according to my best
knowledge it is the only project on foundational problems in the history of
particle physics which within in a reasonable amount of time was brought to a
"mission accomplished " status.

There was a second similar line of research initiated by Rudolf Haag
\cite{Haag} and taken up end enriched by Hans-Juergen Borchers and others.
Both explorations followed a similar leitmotiv, namely to get away from the
quantization parallelism to classical physics and its limitation to
perturbation theory\footnote{The insight that the perturbative series are not
even Borel convergent would have enforced this scepticism against any
non-autonomous approach, but was not available in the early days of AQFT.}.
The liberation from classical quantization "crutches" was already on the mind
of QFT Pascual Jordan, the protagonist of QFT \cite{Jordan}, but then it was
only a dream. Only in the hands of Haag and his collaborators it led to a
profound understanding of mathematical structures which are outside the range
of Lagrangian quantization as the understanding of spin\&statistics and the
superselection rules of generalized charges and their relation to global gauge
symmetries from the structure of local observables \cite{Haag} known under the
name Doplicher-Haag-Roberts (DHR) theory.

After this resounding achievements in the early 70s, there were several other
impressive contributions, all related to issues of locality of operator
algebras, localization of states and thermal states on open systems. Although
these results were not easily accessible to most particle physicist since they
use novel (albeit very appropriate) mathematics, their importance was
recognized by some, as it is obvious from the fact that the recognition of
these achievements led to 3 Max Planck medals. The important discovery of the
relevance of gauge theories and in particular of the standard model did not
require any methodological revolution since it is an extension of the
technology of renormalized perturbation theory of the 50s.

The differences in style between the LSZ kind of particle physics represented
by LSZ and that of LQP represented by the Haag "school" were bigger than that
of its scientific content. At the beginning it amounted mainly to a disparity
of personalities. LSZ, in particular their spokesman Harry Lehmann personified
the same engaging spirit as Pauli, K\"{a}ll\'{e}n, Jost and
others\footnote{There was a strong personal affinity between Pauli and Lehmann
which certainly played a role in Lehmann being able to take Wilkelm Lenz%
%TCIMACRO{\U{b4}}%
%BeginExpansion
\'{}%
%EndExpansion
s chair in Hamburg as one of Germanys youngest professors.} who always put
scientific truth ahead of sociological wellness and political correctness
within the physics community; in contrast the people in the LQP project worked
in a more withdrawn, if not to say ivory tower like setting. \ 

Lehmann's style (as LSZ spokesman) was that of critical engagement and if
necessary confrontational but only against equals, never against novices. When
Landau publically stated that there is no derivation of dispersion relation
outside perturbation theory, Lehmann contradicted in public, but after that
squabble they were on very friendly terms.

The favorite target of the LSZ critique was Heisenberg%
%TCIMACRO{\U{b4}}%
%BeginExpansion
\'{}%
%EndExpansion
s nonlinear spinor theory which for a short time was supported by Pauli. While
on a speaking tour through the US (where his presentation of the nonlinear
spinor theory) a critical encounter with Feynman led to a change of mind.
However Pauli's leaving the boat did not sink it; it did not change the
atmosphere of veneration around the Max Planck Institute in Munich, but at
least LSZ succeeded to impede its spreading to other places by preventing
young newcomers to fall prey to ill-founded but seductive claims of famous
physicists. One who came to the MPI at the end of the 50s and did not fall
prey was Jorge Andr\'{e} Swieca with whom I had the pleasure to collaborate
for more than one decade. As a result of the situation at the MPI he
contemplated leaving physics, apparently thinking that what he saw there was
representative of particle physics at that time which it wasn't, not even in Germany.

This points to a general problem in the sciences which has its origin in the
enormous reputation which comes with the Nobel prize. It is more an exception
than a rule that Nobel prize laureates are able to contribute fruitful ideas
after they received the prize up to their old age. On the one hand they live
under the pressure that this is what is expected of them, and on the other
hand their ideas meet little criticism because of the Nobel prize aura around them.

Compare this episode with the actual situation of particle physics at the MPI
Munich. With the helping hand of another Nobel prize laureate who later
entered ST and became a member of the MPI advisory board, the entire particle
theory department of the MPI became a home for ST and the TOE project. This
time there was no engaging spirit which could have taken over the torch of
public criticism from the previous LSZ generation; it was as if the personal
softspoken non-engaging manner of Haag had changed into to an accommodating
spirit of the entire LQP community.

Lehmann, like Pauli, K\"{a}ll\'{e}n and Jost often used to compress their
critique into sarcastic verdicts. Lehmann did not mince his words when,
beginning at the end of the 50s, he criticized the bootstrap S-matrix
approach, the Mandelstam representation and the dual resonance model salvaged
from the former and the ST abstracted from it. During his time in active
research he successfully prevented any of these speculative fashionable but
unfounded proposals to take a hold in German universities. Unlike his
collaborator and friend Res Jost, with whom he shared his critical views, he
never formulated his critical arguments in the form of an article; they rather
remained in the verbal realm.

Having attended as a student in Hamburg discussions among Lehmann, Jost and
K\"{a}ll\'{e}n and informal after seminar remarks, I had some problems at that
time to understand why they all were so fiercely critical about the S-matrix
bootstrap and those ideas which emerged in its orbit as the Mandelstam
representation. After all educated speculations in certain situations
contributed to the success of particle physics. Looking however with hindsight
at the kind of particle theory ideas which resulted, one understands that
their early doubts were justified.

If one has any doubts about the danger of ideas which lead to the formation of
a community of followers before they have been critically processed, one
should study the demise of particle theory in the aftermath of the S-Matrix
bootstrap, the Mandelstam representation and the later development via the
dual model towards ST. Privily Lehmann and others may have expected that the
popularity of some of those ideas was a transient fashion and over short or
long people would find back to the kind of balanced physics where spectral
representations are something to be either proved or dismissed.

During his active lifetime Lehmann used his prestige and influence in order to
prevent particle theory getting stuck in senseless mind games. Non of the
aforementioned ideas had a basis in the principles underlying particle theory;
in contrast to the derivation of dispersion relations which used proven (and
not guessed) spectral representation, the mind games of the 60s and 70s did
not make contact with physical reality, apart from grafting certain ideas
whose relation to QFT was extremely doubtful onto experimental data which two
years later were superseded by new data\footnote{The first phenomenological
support for Regge trajectories and the dual model faded away with new
scattering data arriving.} which brought that kind of phenomenology to an end.

But the LSZ community certainly underestimated the tenaciousness of ST to hang
on, they believed too much in the self-correcting power of science and
underestimated the strong influence of increasingly globalized sociological
forces. They failed to see that by succeeding to create a community which
accepted the move from the observationally contradicted phenomenological Regge
trajectory setting of strong interactions to a highly speculative setting of a
theory of quantum gravity at the Planck length, ST will have acquired a secure
place at which it is immune against any critic. Since this was already 2
decades after its beginnings, it was too late for a foundational study as the
one in this paper; the string theorists of the 80s by that time hardly knew
their own history and their knowledge of QFT was in terms of computational
recipes. Those who knew the QFT (especially the localization problematics)
sufficiently were not interested in ST; as Lehmann they thought that ST is a
transient phenomenon.

The quantum gravity coup which identified the string tension with the Planck
length made ST experimentally inaccessible and the notion that there are
theoretical principles to be met and the necessary knowledge to implement them
was already in decline. Who will not concede that a "low energy effective
approximations" of a theory, which has been proposed at distances of the
Planck length, at the present state of computational art is too dubious for
serving for an observational comparison?

Having thus reached immunity against the Damocles sword of experimental
judgement which generations of particle physicists had to live under, the only
remaining danger for the new mind game was critique from the side of physical
principles and concepts. But as was shown in a previous section, the knowledge
about conceptual structures in local quantum physics got lost with the string
community and probably even beyond; we cited as evidence in section 5 the lost
distinction between string- and point- like localization and the missed
relevance of the notion of phase space localization in issues of
correspondences and holographic projections.

After Lehmann's retirement, when as a result of gradually failing health he
had already withdrawn into private life and lost much of his political clout
in the German particle physics scene, he expresses his doubts about the LQP
community to stem the tide on many occasions. He worried about the
self-centeredness and the lack of engaging spirit about what was going on in
the particle physics mainstream as well as the active support ST received from
some even some renown particle theoreticians (including Nobel laureates).

How correct he was in his assessment became obvious to me only some years
later; three years before his death, on the occasion of one of my visits to
Hamburg, he wanted to know from me what happened at the Berlin universities on
the former DDR side in the confused aftermath of the German unification (which
as far as the universties are concerned looked more like an "Anschluss"), when
a group in ST was installed there. I found myself in the somewhat unpleasant
situation to explain to my former advisor Harry Lehmann that none of the
quantum field theorists and mathematical physicists at both of the academic
institutions in Berlin had been asked for advice or had any influence on that decision.

This is somewhat indicative that when a group of researchers who starting in
the 70s at the FU made important contributions to QFT is not consulted and not
even informed about planned important changes. Normally physicists from nearby
places who have contributed to the foundations, and hence have a lot of
expertise, would either be directly invited to participate in the selection
process for a new departments. But even Harry Lehmann, perhaps the
internationally best known German physicist (at that time already retired) was
informed. He was surprised that I, on a visit to Hamburg coming from Berlin,
could not tell him either. We wondered why an area as ST, which produced not a
single tangible physical result, was selected to present the new particle
physics at one of the most important universities in the new old capital.

At first some people, including myself, thought that there was some foul play.
But when I many years later saw that some of the best US universities hired
people of a similar line of research it became clear to me that the millennium
physics was precisely that tune of a TOE in form of ST and extra dimensions.
In fact the distinction between sciences, politics and social ideas evaporates
because there are easily recognizable cross connections which are part of the
Zeitgeist: the millennium physics corresponding to the finance markets of the
casino capitalism, and ST and the idea of a TOE relates to the trade with
derivatives. Only as far as the crash goes, the Zeitgeist in the sciences lags
somewhat behind.

Since science is an occupation of only a few, the scientific side of the
breakdown will be less spectacular; a silent end of a more than 40 years old
theories which has not led anywhere and, as we have shown in this essay, are
build on misunderstandings of local quantum physics, will be the most probable
event. This would be a pity, because the end of a theory which has captivated
so many minds for such a long time contains valuable messages; in fact the
message contained in such a failure is, in my opinion as important as that in
a successfull theory. The people who will have the greatest interest to
understand what was going on are naturally the historians and philosophers
because to explain what has been going on and what was on the mind of the
protagonists is their profession.

Human activities even those in the exact sciences were never completely
independent of the Zeitgeist. In fact if there is any sociological phenomenon
to which the frantic chase for a TOE finds it analog, it is the post-cold war
reign of globalized capitalism with its "end of history" frame of mind
\cite{Fuku} and its ideological support for insatiable greed and exploration
of natural resources. It is hard to imagine any other project in physics which
would fit the post cold war millennium spirit of power and glory and its
hegemony claims in the pursuit of a these goals better than superST: shock and
awe of a TOE against the soft conceptual power of critical thinking.

Whereas the post cold war social order has, contrary to its promises of a
better life for mankind, accentuated social differences and caused avoidable
wars and deep political divisions, the three decades long reign of the project
of a TOE in particle physics has eradicated valuable knowledge about QFT and
considerably weakened chances of finding one's way out of the present crisis.

In the beginning of this section I argued against Foreman's thesis claiming
that the probabilistic aspect of QM was a result of the "demise of the west"
gloom and doom spirit in post world-war I Germany. But on the other hand the
relation between the "end of the history" capitalism and the glorious "end of
the millennium" TOE in particle physics is something which would be hard to
negate. ST with its lavish support is an activity where this analogy is most
visible. The way in which the LHC research is publicized as the run for the
"God's particle" is another Zeitgeist-phenomenon. The theory stagnates since
almost 40 years and now an experiment is expected to decide its fate. \ 

With the death of Lehmann, Symanzik and Jost, the engaging style of research
reached an end and metaphoric arguments became increasingly accepted. The
sense for the importance of deriving observational accessible laws from
established principles as in those previously mentioned cases (in particular
the setting of dispersion relations), or for deriving results in a systematic
and mathematically controllable way (renormalization theory of QED and the
subsequent standard model) came to an end and the era of "everything goes" and
of metaphors began to take roots with young people, in particular with those
imaginative ones who were looking for excitement and adventure in particle
physics. In this way ST came to play the role of the millennium theory in this
new age.

Theoretical physicists as Klaus Pohlmeyer (see previous section), a
theoretical physicist who stubbornly criticized ST by showing how the
quantization of an integrable system as the Nambu-Goto Lagrangian is correctly
done \cite{Po}\cite{Me-Re}, became a tragic figure; as a listener in one of
his seminars it was evident to me that the correct quantum theory of the
Nambu-Goto Lagrangian was of little interest to people who found the
metaphorical treatment based on canonical quantization which totally ignored
the integrable structure much more relevant and exciting. Since the result did
not meet the prejudices of string theorists, his lifelong work (which probably
shares the fate with ST of being at odds with observations, but at least
passed the conceptual-mathematical consistency test) remained unknown. The
self-delusion of a TOE was stronger than the inner coherence of a theory of
something which was at least conceptually consistent.

Compared with LSZ, the public perception about the LQP community was that of
science in an ivory tower. Whereas the contributions of LSZ were in the center
of particle physics, the profound enrichments of LQP were more peripheral as
far as the day to day problems were concerned and only very few researchers
succeeded to contribute to such problems. Any allegation of lack of engagement
in the dispute about ST would be unfair if it would not also include
autocriticism of ones own role. The present work come much too late for having
any influence. As mentioned before its content aims more at future historians
and philosophers of physics.

The deep-rooted conceptual setting of the LQP research would have been the
best antidote against the metaphoric spirit of ST; in particular since the
Achilles heels of ST, namely localization, is the backbone of LQP. \ So how
come that German particle physics succumbed to the string lure as exemplified
in particular by the situation at the MPI in Munich and several other places
where it took root, including at Hamburg which was the epicenter of the
postwar renaissance of German particle physics.

The answer is outside the realm of pure science, it has to do with
personalities and the culture of critical engagement with controversial
subjects which enjoy the support of influential personalities, in short with
sociological aspects. The LSZ school provided the backup of a solid conceptual
platform for speculative excursions as well as for their critical evaluation.

This school (with a helping hand from Pauli, in his role as an apostate of
Heisenberg's nonlinear spinor theory) which successfully contained
Heisenberg's influence, did not exist any more when in more recent times the
theory group of the Heisenberg MP institute in Munich, with a Nobel prize
laureate in its advisory board, became a home of ST.

To make the irony of history of particle theory in Germany complete, slightly
more than one decade after his death Lehmann's chair went to a string
theorist. There was apparently strong pressure from the side of the Desy
theory on the theoretical physics department of the Hamburg University. As in
times of royal rulers when every court had to have its own buffoon or
oneirocritic, each reputable theory group at a high energy laboratory
\textit{must have} one string theorist. Unfortunately the Desy pressure was
successful and Lehmann's chair went to string theory.

With this step the circle about Lehmann's legacy at Hamburg ended in a full
turn. He started a successfull campaign against Heisenberg's nonlinear spinor
theory. At the end he was hepless against a theory which is conceptually not
less metaphoric and contradictory. His chair at the Hamburg University went to
a string theorist and the Heisenberg Institute in Munich became the center of
ST in Germany!

The present critique formulated in this article is by no means my private
intellectual hobbyhorse. At least some of the leading members of the LQP
community have independent knowledge of most of the points, namely that ST
wave functions are not string-like but point-like infinite component wave
functions which contradicts\footnote{Only a radical conceptual revision at the
end of which something entirely different that ST would emerge can assign a
possible new mathematical/physical content. But this is not in the interest of
string theorists and with an ill conceived physical motivation it is not
interesting for non-string physicists either.} the application of Feynman
inspired tube rules for transition amplitudes, as well as the strange
metaphoric idea that one can embed a chiral QFT as a one-dimensional localized
subtheory into a higher dimensional QFT as if it would be the same thing as a
one-dimensional cord in a 3 dimensional quantum mechanics. Even those few
string theorists who looked at the question of string localization
\cite{Mar}\cite{Lowe} noticed the pointlike nature and therefore could have
explained to their fellow string theorists that should change terminology
since their object is pointlike. But in order to save the string metaphor, or
because they were confused about the issue of localization they proposed
"invisible strings" i.e. stringlike objects for which only one point on the
string is visible. Quantum theory whose great achievement was the separation
between classical and metaphorical aspects from intrinsic properties in
Heisenberg's notion of "observables" was turned back from its feet to its head.

Localization and causality is the heart piece of LQP and the question arises
why, in a recent survey article \cite{Fred} in a joint book together with ST
contributions, the authors do not come forward with a profound criticism and
instead rather prefer to leave it at an easily overlooked remark that strings
are pointlike without indicating what this implies about ST. String theorists
have learned to live with such statements as long as no critical consequences
are drawn, after all some of their own folks have come up with the statement
that ST deals with strings in spacetime of which only one point is visible
\cite{Lowe}. As long as the localization point lies on a string, the world of
ST is in order.

The article missed the chance to engage the string theorists and preferred to
maintain the ambience of their ivory tower instead of showing that string
theorists are, conceptually speaking, "would be" emperors without cloth. The
idea that by abstaining from any profound scientific engagement with ST one
can create a modus vivendi of mutual tolerance is naive since in the minds of
ideologues of a TOE such behavior only underlines that their LQP opponents are
obsolete since they are just repeating that strings are localized on a
"string-point" which they know from their own folks.

To be fair, there were valiant attempts to inject a critical note in
discussions which arose in the tail of ST, but had their natural setting in
the conceptually well founded QFT. One of those was Rehren's \cite{Re}
critical evaluation of the Maldacena conjecture when he demonstrated that the
AdS-CFT correspondence is primarily a mathematical statement since the degree
of freedom preservation inevitably makes on side physically pathological. But
a mathematical theorem which listed heavily on the physical side, as the
result of a mismatch with the physical phase space degrees of freedom
\cite{Swieca} on one side of the correspondence i.e. the latter cannot link
two physical QFTs; this is not a statement about a particular model but rather
a structual property concering a (rather raddical) change of the localization
concept of abstract quantum matter. No wonder that some string theorists who
believed that these contributions are nothing more than a nuisance (since they
go against what Madacena conjectured) to their project began to refer to this
work at conferences as "the German correspondence".

\ It would be hard to deny that in the times of the Pauli, Jost, Kall\'{e}n
and LSZ \ Streitkultur, feathers were often roughened, but the critical
balance which was kept in this way was a great \textit{blessing for particle
physics}. Should social wellness be allowed to curtail conceptual clarity
through a Streitkultur?

When the question comes up why many young people were lured into the metaphors
of ST and why some of the important theoretical research institution had been
taken over by ST crew, at least a partial answer is clear; a large part of the
particle theory community, in particular the leading members of the LQP
preferred social accommodation over scientific engagement. In contrast to
their LSZ predecessors, they limited their critical standards to a
well-defined region which included their ivory tower.

In this context it is worthwhile to comment on two statements from my LQP
colleagues. One statement is that even if the the foundation of what people
are doing are metaphoric or confuse, they at least can communicate between
them and there is always a certain chance that somebody finds something
interesting. I do not know of any case which could support this point of view.
Another statement which expresses a much stronger form of non-engagement and
which protects conferences in which ST, the Maldacena conjecture, holography
and the quark-gluon plasma are stirred together is the vernacular: "many
people cannot err". I only know of cases where many people did err and the
more people the greater and more subtle was the error. When I entered particle
physics it was the S-matrix bootstrap and as time passed the errors increased
with the size of the communities defending them and with ST the subtlety of
the error and the size of the community are of such large proportion that
incorrect theory cannot any more be disposed of. Since the engagement of those
who know better is zero, the community which host such ideas will even grow
because lack of engagement especially at a place dedicated to mathematical
physics is interpreted as acceptance.

Whereas several aspects in this section (namely how ST took its roots, and in
particular the fact that the knowledge of LQP is the ideal point of departure
for a critical evaluation of superstrings) are specific for the situation in
Germany, the dominance of ST in other counties followed a similar pattern. In
most cases it started by a few renown members of the particle theory community
beating the drums for superST as a TOE. What is more surprising is that those
few, who had all the background and the intellectual capability to engage,
refrained from doing so (this includes myself).

For a long time even sceptical minds as myself believed that the problem will
take care of itself, the exact sciences represent a different mind-set from
other human activities e.g. in the political domain. When after many years
people realize that this is not what is happening they finally try to make the
best out of it, either by accommodation themselves with the ST or, if they are
retired (and do not have a nice ST neighbor) they write articles as this one.

All jokes aside, the present situation in Germany is quite serious. After the
QFT in Berlin was closed down at the beginning of this century, it is clear
that there will be no continuation of QFT in Goettingen, and it does not take
any visionary power that QFT in Hamburg, the place where LQP in Germany begun
will also end. This is a pity because different from ST, the reseach on LQP
never promised rapid success and after several decades of investment it has
reached an interesting stage with many new results which promise to
revolutionize QFT and particle theory.

\section{Trying to get back on track: a particle theory without metaphors}

The way out of the present plight is not simply a return to pre-string times.
In exact sciences which support the concept of \textit{truth }as opposed to
other more opinion-driven human activities,\textit{ }having taken a wrong
direction is not necessarily a total loss. The comprehension of why and under
what circumstances the misleading journey begun carries the chance of
revealing a deeper insight into truth than that obtained by a direct path
where the potential slips and some subtleties of the concepts were not
appreciated. But such a situation would require a more serious study of the
forgotten incomplete achievements and of those foundational insights into the
building of local quantum physics which went largely unnoticed or got lost in
the noise of a TOE. So one must find a way to break the unproductive alliance
between the metaphoric bombast of a TOE and the longed for career and fame of
our junior scientists. It is a myth that the development of the exact sciences
follows its own intrinsic laws. Fashions and manifestations of the Zeitgeist
are as influential as elsewhere.

As in any other human activity it is futile to hope for any auto-critical
attitude from people who have dedicated a significant part of their scientific
life to ST and its derivatives. A recent quotation of Churchill's war
endurance rallying cry: "never, never,...give in", which David Gross used to
rally support behind ST precisely illustrates this point. This statement is
quite an escalation from the previous "the only game in town". It is difficult
not to agree with Anderson and understand such statements for what they really
are: oubursts of arrogance which comes naturally with the conviction of having
the keys to a TOE.

With the old Streitkultur between equals having been substituted by a
propaganda about a dominant monoculture of a TOE, particle theory became more
representative of a Zeitgeist which led to the present hegemonic capitalism.
There is the not unfounded fear that the partisans of ST could succeed to
drive particle physics against a brick wall according to the tune of "ST or
nothing", the statements of some of its leading representatives sound like
this. The Achilles heel here is the new generation; are they willing to study
and extend the still incomplete conceptual framework of QFT and put new life
into the old Streitkultur, or do they prefer to follow the TOE-tunes?

There is no deficit of important problems and ideas how to solve them, which
could end the 40 year conceptual stagnation about the standard model
\cite{charge1}. QFT owes its survival and its present status as the most
successful and universal of all physical theories to the methodological
breakthrough obtained with renormalized QED. But even this impressive
achievement of the quantized setting of Maxwell's electrodynamics shows an
incompleteness which is symptomatic for the points which will be raised in
this concluding section. Whereas, apart from one interpretational issue going
under the name of the \textit{problem of the ether}\footnote{The ether
postulate led to a wrong interpretation but had no effect on the form of the
Maxwell equations (including their Lorentz symmetry) and the proof that they
have nontrivial solution.}, the classical Maxwell theory reached its
conceptual closure\ already two decades after its conception, the fate of
renormalized QED was less fortunate.

Even nowadays, more than 60 years after its discovery, its closure is still
far away. When we talk about QED as a theory, we are using words in a much
more loose sense than our forefathers about Maxwell's theory at the beginning
of last century. What really encourages to carry the conceptual problems
behind gauge theory to their closure is the realization that these ideas, even
in their present imperfect setting, have such impressive observational
predictive power.

It is unfortunate that the misconception which presents ST as being
string-localized has found a widespread acceptance as a fait accompli, because
it shows that the issue of localization in relativistic QT which could decide
about the fate of ST has been pushed to the wayside. As explained in section
4, ST is founded on the idea that the potentials $\Phi_{\mu}$ of an
n-component chiral current can be interpreted as describing the locus of a
string $X_{\mu}(\sigma)$ in a n-dimensional target space\footnote{In order
that the target space carries a unitary positive energy representation of the
Poincar\'{e} group one has to admit spinorial target indices. The unique
solution of these requirements is the d=10 superstring representation.}. This
idea which insinuates the incorrect association to a relativistic position
operator contradicts the intrinsic localization concept in QFT which was
explained in section 5. Whereas the misunderstandings around the ether at the
end of the 19th century had no effect on the mathematical structure of
Maxwell's theory (the outcome of the Michelson-Moreley experiment did not
require a change in the Maxwell-Lorentz equations but only on their
interpretation), the TOE at the turn of the millennium had a devastating still
ongoing impact on particle theory which will reach far into the future of this century.

The worst aspect is not that in ST we are confronting a theory which is
totally empty-handed with respect to observational checks because the
existence of a consistent theory which generalizes the successfull QFT would
be a remarkable achievement even if at the end it is rejected by nature. Even
the more than 50 years of its dominance can be tolerated if it would be
conceptually consistent, or if at the end its inconsistency would lead to a
conceptual enrichment as in the resolution of the ether problem a century before.

The seriousness of the confusion can be seen by looking at thousands of
publication on ST as well as semi-popular publications on string localization
in ST and QCD \cite{Witten} including the inevitable drawings of open and
closed spacetime strings in order to remove the last doubts about the
spacetime meaning of strings.

With a "never give up" intransigence on the one side and the absence of the
engaging spirit of the old Streitkultur on the side of those who know better
on the other side of the fence, one has to be a notorious optimist in order to
expect a change in the near future.

Particle physics and quantum field theory are presently in an upheaval. Many
particle theorists look to new experiments at the LHC to lift them out of the
present conceptual labyrinth which probably overburdens LHC. On the other hand
there has been a very dedicated project for almost 5 decades which less than
looking for discovering gems near the paths of great caravans has stubbornly
pursuit one aim: the understanding the inner workings of QFT, despite all its
incompleteness by far our most successful physical theory. This LQP project
still did not lead to the closure of QFT, but it recently arrived at some
remarkable results which are all related to modular localization.

After a long time of stagnation of gauge theory it became clear that its
relation with modular localization (string-localization of potentials
\cite{charge1}\cite{nonlocal}) suggests an interesting and potentially
important extension which is not limited to s=1 but applies also to higher
spin. It also suggests a complete new look at the Schwinger-Higgs screening
which is the better presentation of the Higgs mechanism. The already mentioned
derivation of the crossing property of formfactors from modular localization
is another result which together with the idea of wedge algebra generators
\cite{inter}\cite{founcor} lead to the first mathematically controlled
construction of factorizing models. Last not least it explains the relation of
modular localization with thermal properties which originally were thought of
as an intrinsic property of localization behind event horizons in curved space
time. These ideas lead to a formula for "localization entropy" and a
conjecture about a universal relation between heat bath- and
localization-caused- thermal behavior. It is precisely this progress in
modular localization which facilitates the critical evaluation of string
theory although a direct criticism on specific claims of ST was always possible.

\textit{Acknowledgement}: I am indebted to Jakob Yngvason for inviting me to
the Erwin Schroedinger Institute for mathematical physics where the last two
sections of this essay were written. The simultaneous occuring workshop on
"AdS Holography and the Quark--Gluon Plasma" confirmed the pessimistic outlook
that globalized communities, different from individuals, can perfectly ignore
known theorems and steamroll not mass supported knowledge into the ground. At
least in contemporary particle physics the old vernaculars "the end justifies
the means" and "many people cannot err" acquire an ominous meaning; the larger
the community, the easier it is for a conjecture coming with a big name to
spread without ever receiving a critical review.

\bigskip

\begin{thebibliography}{99}                                                                                               %


\bibitem {Haag}R. Haag, \textit{Local Quantum Physics}, Springer, second
edition, 1996

\bibitem {Darrigol}O. Darrigol, The origin of quantized matter fields, Hist.
Stud. Phys. Sci.\textbf{16}, (1986) 197

\bibitem {Hei}W. Heisenberg, \textit{\"{U}ber die mit der Entstehung von
Materie aus Strahlung verkn\"{u}pften Ladungsschwankungen}, Verhandlungen der
S\"{a}chsischen Akademie der Wissenschaften zu Leipzig, \textbf{86}, (1934) 317-322

\bibitem {Fu-Op}W. H. Furry and J. R. Oppenheimer, Phys. Rev. \textbf{45},
(1934) 245

\bibitem {charge1}B. Schroer, \textit{Unexplored regions in QFT and the
conceptual foundations of the Standard Model}, arXiv:1006.3543

\bibitem {Pic}R. Ferrari, L. Picasso and F. Strocchi, Comm. Math. Phys.
\textbf{35}, (1974) 25; Nuovo Cim. \textbf{39} A, (1977) 1

\bibitem {Roberts}P. Leyland, J. Roberts and D. Testard, Duality For Quantum
Free Fields, C.N.R.S. preprint July 1978, unpublished

\bibitem {Ander}Philip Anderson, in \textit{the Times Higher Education 25
August 2006}, \ http://www. timeshighereducation.co.uk/story.asp?soryCode=204995\&sectioncode=1

\bibitem {crisis}B. Schroer, \textit{String theory, the crisis in particle
physics and the ascent of metaphoric arguments,}

\bibitem {Hed}R. Hedrich, \textit{The Internal and External Problems of String
Theory - A Philosophical View}, physics/0610168

\bibitem {Woit}P. Woit, \textit{Not Even Wrong, the failure of string theory
and the continuing challenge to unify the laws of physics}, Jonathan Cape
London 2006

\bibitem {Smolin}L. Smolin, \textit{The Trouble With Physics: The Rise of
String Theory, the Fall of a Science, and What Comes Next}, Sept. 2006

\bibitem {Vech}P. Di Vecchia, \textit{The birth of string theory}, arXiv 0704.0101

\bibitem {Nambu}Y. Nambu, Lectures at the Copenhagen Symposium, 1970, unpublished

\bibitem {Goto}T. Goto, Progr. Theor. Phys. \textbf{46} (1971) 1560

\bibitem {Barut}A. O Barut and H. Kleinert, Phys. Rev. \textbf{157}, \ (1967) 1180

\bibitem {Frons}C. Fronsdal, Phys. Re. \textbf{156}, (1967) 1652

\bibitem {Dimock}\textit{ }J. Dimock, \textit{Locality in Free String Field
Theory-II}, Annales Henri Poincare 3 (2002) 613, math-ph/0102027

\bibitem {Wei}S. Weinberg, \textit{The Quantum Theory of Fields I}, Cambridge
University Press 1995

\bibitem {Tod}N.N. Bogolubov, A.A. Logunov, A.I. Oksak and I.T. Todorov,
\textit{General Principles of Quantum Field Theory}, Kluwer 1990

\bibitem {Lowe}D. A. Lowe, Phys. Lett. B 326, (1994) 223

\bibitem {Mar}E. Martinec, Class. Quant. Grav. \textbf{10}, (1993) 187

\bibitem {W97}S. Weinberg, \textit{What is Quantum Field theory, and what did
we think it is?}, arXiv:hepth/9702027

\bibitem {Fo}P. Foreman, \textit{Weimar culture, causality and quantum
theory1918-1927: Adaptation by German physicists and mathematicians to a
hostile intellectual environment. }Hist. Stud. Sci, 3:1-115\newline

\bibitem {founcor}B. Schroer, \textit{A critical look at 50 years particle
theory from the perspective of the crossing property}, arXiv:0905.4006

\bibitem {Fub}S. Fubini and G. Veneziano, Annals of Physics 63, 12 (1971),
\ E. Del Giudice, P. Di Vecchia and S. Fubini, Annals of Physics, 70, (1972) 378

\bibitem {Jor}B. Schroer, \textit{Pascual Jordan's legacy and the ongoing
research in quantum field theory}, in preparation

\bibitem {BMT}D. Buchholz, G. Mack and I. Todorov, Nucl.Phys. B, Proc. Suppl.
\textbf{5B}, (1988) 20

\bibitem {Stas}C. P. Staszkiewicz, Die lokale Struktur abelscher Stromalgebren
auf dem Kreis, Freie Universitaet Thesis Berlin 1995, unpublished

\bibitem {Longo}R. Longo and Y. Kawahigashi, Adv. Math. \textbf{206}, (2006)
729, and references therein

\bibitem {Mack}G. Mack, \textit{D-dimensional Conformal Field Theories with
anomalous dimensions as Dual Resonance Models}, arXiv:0909.1024,
\textit{D-independent representations of conformal field theories in D
dimensions via transformations to auxiliary dual resonance models. The scalar
case}, arXiv:0907:2407

\bibitem {SSV}B. Schroer, J. A. Swieca and A. Voelkel, Phys. Rev. D
\textbf{11}, (1975) 11

\bibitem {H-H-W}R. Haag, N. M. Hugenholz and M. Winnink, Commun. Math. Phys.
\textbf{5}, (1967) 215

\bibitem {Bi-Wi}J. Bisognano and E. Wichmann, J. Math. Phys. \textbf{16},
(1975) 985

\bibitem {Wigh}R. F. Streater and A. Wightman, \ \textit{PCT, Spin and
Statistics and All That}, \ Benjamin, New York 1964

\bibitem {Annals}B. Schroer, Ann. Phys. \textbf{295}, (1999) 190

\bibitem {B-G-L}R. Brunetti, D. Guido and R. Longo, \textit{Modular
localization and Wigner particles}, Rev.Math.Phys. \textbf{14}, (2002) 759

\bibitem {F-S}L. Fassarella and B. Schroer, J. Phys. \textbf{A 35}, (2002) 9123

\bibitem {M-S-Y}J. Mund, B. Schroer and J. Yngvason, \textit{String-localized
quantum fields and modular localization}, CMP\textbf{\ 268} (2006) 621, math-ph/0511042

\bibitem {Su}S. J. Summers, \textit{Tomita-Takesaki Modular Theory}, math-ph/0511034

\bibitem {Mu}J. Mund, Ann. Henri Poincar\'{e} \textbf{2}, (2001) 907

\bibitem {Lech}G. Lechner, \textit{An Existence Proof for Interacting Quantum
Field Theories with a Factorizing S-Matrix}, math-ph/0601022

\bibitem {L-R}Y. Kawahigashi, R. Longo, U. Pennig and K.-H. Rehren, Commun.
Math. Phys.\textbf{271}, (2007) 375

\bibitem {Mu2}J. Mund, Commun. Math. Phys. \textbf{294}, (2010) 505, arXiv:0902.4434

\bibitem {A-S}B. Schroer and N. K. Nielsen, Nuclear Phys. B\textbf{120},
(1977) 62

\bibitem {LRT}P. Leyland, J. Roberts and D. Testard, Duality For Quantum Free
Fields, C.N.R.S. preprint July 1978, unpublished

\bibitem {Stein}O. Steinmann, Ann. Phys. (NY) \textbf{157}, (1984) 232

\bibitem {B-N}F. Bloch and A.Nordsiek, Phys. Rev. \textbf{52}, (1937) 54

\bibitem {infra}B. Schroer, Fortschr. Physik \textbf{143}. (1963) 1526

\bibitem {YFS}D. Yenni, S. Frautschi and H. Suura, \textit{The infrared
divergence phenomena and high energy processes}, Ann. of Phys. \textbf{13},
(1961) 370-462

\bibitem {scree}J. A. Swieca, Phys. Rev. D \textbf{13}, (1976) 312

\bibitem {D-R}S. Doplicher and J.E. Roberts, Commun. Math. Phys. 131 (1990) 51

\bibitem {Witten}E. Witten, \textit{CURRENT SCIENCE} \textbf{81}, NO. 12,
(2001) 25

\bibitem {Du-Sch}M. Duetsch and B. Schroer, \textit{Massive Vector Mesons and
Gauge Theory}, .Phys. A33 (2000) 4317

\bibitem {Du}M. D\"{u}tsch, J. M. Gracia-Bondia, F. Scheck and J. C. Varilly,
\textit{Quantum gauge models without classical Higgs mechanism}, arXiv:1001.0932

\bibitem {Sei-Wi}N. Seiberg and E. Witten, arXiv:hep-th9607163

\bibitem {Jae}C. Gerard and C. Jaekel, Prog. Math. \textbf{251}, (2007) 125

\bibitem {po-int}B. Schroer, \textit{Positivity and Integrability}
(Mathematical Physics at the FU-Berlin), hep-th/0603118

\bibitem {Buch}D. Buchholz, I. Ojima, \textit{Spontaneous Collapse of
Supersymmetry}, Nucl.Phys. B498 (1997) 228, hep-th/9701005

\bibitem {inter}B. Schroer, \textit{Localization and the interface between
quantum mechanics, quantum field theory and quantum gravity I (the two
antagonistic localizations and their asymptotic compatibility)}
arXiv:0912.2874, \ \textit{Localization} .....\textit{gravity II} ((The search
of the interface between QFT and QG) arXiv:0912.2886

\bibitem {Ka-Wi}R. Kaehler and H.-W. Wiesbrock, JMP \textbf{42}, (2000) 74,
and references to prior work therein

\bibitem {Re}K.-H. Rehren, \textit{A Proof of the AdS-CFT Correspondence}, In:
Quantum Theory and Symmetries, H.-D. Doebner et al. (eds.), World Scientific
(2000), pp. 278, hep-th/9910074, \ M. Duetsch, K.-H. Rehren, \textit{A comment
on the dual field in the AdS-CFT correspondence}, Lett.Math.Phys. 62 (2002)
171, \ K.-H. Rehren, \textit{QFT Lectures on AdS-CFT}, hep-th/0411086

\bibitem {Suss}L. Susskind, \textit{The Cosmic Landscape: String Theory and
the Illusion of Intelligent Design, }Little, Brown (2005)

\bibitem {Schele}B. Schellekens, Rep. Prog. Phys. \textbf{71}, (2008) 07220

\bibitem {Teg}M. Tegmark, arXiv:0709.4024

\bibitem {Capra}\textit{F. Capra, The Tao of Physics: An Exploration of the
Parallels Between Modern Physics and Eastern Mysticism}, Shambhala
Publications of Berkeley, California 1975

\bibitem {STW}B. Schroer, T.T. Truong and P. Weiss, Phys. Lett. B \textbf{63},
(1976) 422

\bibitem {KTTW}M. Karowski, H.-J. Thun, T.T. Truoung and P. Weiss, Phys. Rev.
Lett. B \textbf{67}, (1977) 321

\bibitem {KW}M. Karowski and P. Weiss, Nuccl. Phys. B \textbf{139}, (1978) 445

\bibitem {Mu-S}J. Mund and B. Schroer, \textit{A generalized KMS condition and
its relation to the crossing property}, in preparation

\bibitem {Ba-Ka}H. Babujian and M. Karowski, Int. J. Mod. Phys. \textbf{A1952}%
, (2004) 34, and references therein to the beginnings of the
bootstrap-formfactor program

\bibitem {Ven}G. Veneziano, Nuovo Cim. A \textbf{57}, (1968) 190

\bibitem {BEG}J. Bros, H. Epstein and V. Glaser, Com. Math. Phys. \textbf{1},
(1965) 240

\bibitem {Po}K. Pohlmeyer, Phys. Lett. \textbf{119 B}, (1982) 100

\bibitem {Bahns}D. Bahns, J. Math. Phys. \textbf{45}, (2004) 4640

\bibitem {Me-Re}C. Meusburger and K.-H. Rehren, Commun.Math.Phys.
\textbf{237}, (2003) 69, \ arXiv:math-ph/0202041

\bibitem {Swieca}B. Schroer, \textit{Particle physics in the 60s and 70s and
the legacy of contributions by J. A. Swieca}, Eur. Phys. J. H. \textbf{35},
(20010) 53

\bibitem {}(2010) arXiv:0712.0371

\bibitem {Ni-Sc}N. K. Nielsen and B. Schroer, Nucl.Phys. \textbf{B127}, (1977) 493

\bibitem {BMS}B. Schroer, \textit{BMS symmetry, holography on null-surfaces
and area proportionality of "light-slice" entropy}, arXiv:0905.4435

\bibitem {Vir}M. A. Virasoro, Phys. Rev. \textbf{177}, (1969) 2309

\bibitem {D-H-S}R. Dolen, D. Horn and C. Schmid, Phys. Rev. \textbf{166},
(1968) 1768

\bibitem {Jens}J. Mund, J. Math. Phys. \textbf{44}, (2003) 2037

\bibitem {Jordan}P. Jordan, \textit{The Present State of Quantum
Electrodynamics}, in \textit{Talks and Discussions of the Theoretical-Physical
Conference in Kharkov} (May 19.-25., 1929) Physik.Zeitschr.XXX, (1929) 700

\bibitem {Fuku}F. Fukuyama, \textit{The end of history and the last man}, Free
Press 1992

\bibitem {Fred}K. Fredenhagen, K.-H. Rehren and E. Seiler, Springer Lecture
Notes Phys. \textbf{721} (2007) 61, \ arXiv:hep-th/0603155

\bibitem {Mal}J. A. Maldacena, Adv. Theor. Math. Phys. \textbf{2}, (1998) 231

\bibitem {nonlocal}B. Schroer, \textit{An alternative to the gauge theory
setting}, arXiv:1012.0013
\end{thebibliography}
\end{document}